\documentclass[10pt,journal,compsoc]{IEEEtran}
\usepackage{booktabs}
\usepackage{multirow}
\usepackage{graphicx}
\usepackage{subcaption}
\usepackage{hyperref}
\usepackage{amssymb}
\usepackage{amsmath}
\usepackage{amsthm}
\usepackage{mathrsfs}
\newtheorem{theorem}{Theorem}
\newtheorem{definition}{Definition}%
\newtheorem{assumption}{Assumption}%
\newtheorem{corollary}{Corollary}

\ifCLASSOPTIONcompsoc
  \usepackage[nocompress]{cite}
\else
  \usepackage{cite}
\fi
\ifCLASSINFOpdf
\else
\fi
\begin{document}
%
\title{A Circuit Domain Generalization Framework for Efficient Logic Synthesis in Chip Design}
%
%
%
%

\author{
Zhihai Wang$^{\dagger}$, Lei Chen$^{\dagger}$, Jie Wang$^{\ast}$,~\IEEEmembership{Senior Member,~IEEE,} Xing Li, Yinqi Bai,\\ Xijun Li, Mingxuan Yuan, Jianye Hao,~\IEEEmembership{Member,~IEEE,} \\ Yongdong Zhang,~\IEEEmembership{Senior Member,~IEEE,} and Feng Wu,~\IEEEmembership{Fellow,~IEEE}
\IEEEcompsocitemizethanks{
\IEEEcompsocthanksitem Z. Wang, J. Wang, Y. Bai, Y. Zhang, F. Wu are with: a) CAS Key Laboratory of Technology in GIPAS, University of Science and Technology of China, Hefei 230027, China; b) Institute of Artificial Intelligence, Hefei Comprehensive National Science Center, Hefei 230091, China. E-mail: \{zhwangx,byq000324\}@mail.ustc.edu.cn, \{jiewangx,zhyd73,fengwu\}@ustc.edu.cn.
\IEEEcompsocthanksitem L. Chen, X. Li, X. Li, M. Yuan, J. Hao are with Huawei Noah's Ark Lab. E-mail: \{lc.leichen, li.xing2, xijun.li,Yuan.Mingxuan,haojianye\}@huawei.com.
\IEEEcompsocthanksitem J. Hao is with College of Intelligence and Computing, Tianjin University.
}
\thanks{Manuscript received April 19, 2005; revised August 26, 2015. This work was done when the first author was an intern at Huawei Noah's Ark Lab. $^{\dagger}$Equal contribution. $^{\ast}$Corresponding author.}}

%
%

\markboth{Journal of \LaTeX\ Class Files,~Vol.~14, No.~8, August~2015}%
{Shell \MakeLowercase{\textit{et al.}}: Bare Demo of IEEEtran.cls for Computer Society Journals}
%



\IEEEtitleabstractindextext{%
\begin{abstract}
Logic Synthesis (LS) plays a vital role in chip design---a cornerstone of the semiconductor industry. A key task in LS is to transform circuits---modeled by directed acyclic graphs (DAGs)---into simplified circuits with equivalent functionalities. To tackle this task, many LS operators apply transformations to subgraphs---rooted at each node on an input DAG---sequentially. However, we found that a large number of transformations are \textit{ineffective}, which makes applying these operators highly time-consuming. In particular, we notice that the runtime of the Resub and Mfs2 operators often dominates the overall runtime of LS optimization processes. To address this challenge, we propose a novel data-driven LS operator paradigm, namely PruneX, to reduce ineffective transformations. The major challenge of developing PruneX is to learn models that well generalize to unseen circuits, i.e., the \textit{out-of-distribution (OOD) generalization} problem. Thus, the major technical contribution of PruneX is the novel \textit{circuit domain generalization} framework, which learns domain-invariant representations based on the transformation-invariant domain-knowledge. To the best of our knowledge, PruneX is \textit{the first} approach to tackle the OOD problem in LS operators. We integrate PruneX with the aforementioned Resub and Mfs2 operators. Experiments demonstrate that PruneX significantly improves their efficiency while keeping comparable optimization performance on industrial and very large-scale circuits, achieving up to $3.1\times$ faster runtime. 

\end{abstract}

\begin{IEEEkeywords}
Chip Design, Logic Synthesis, Deep Learning, Domain Generalization
\end{IEEEkeywords}}

\maketitle

\IEEEdisplaynontitleabstractindextext

%
\IEEEpeerreviewmaketitle

\begin{figure*}[t]
    \centering
    \begin{subfigure}{0.9\textwidth}
        \includegraphics[width=\textwidth]{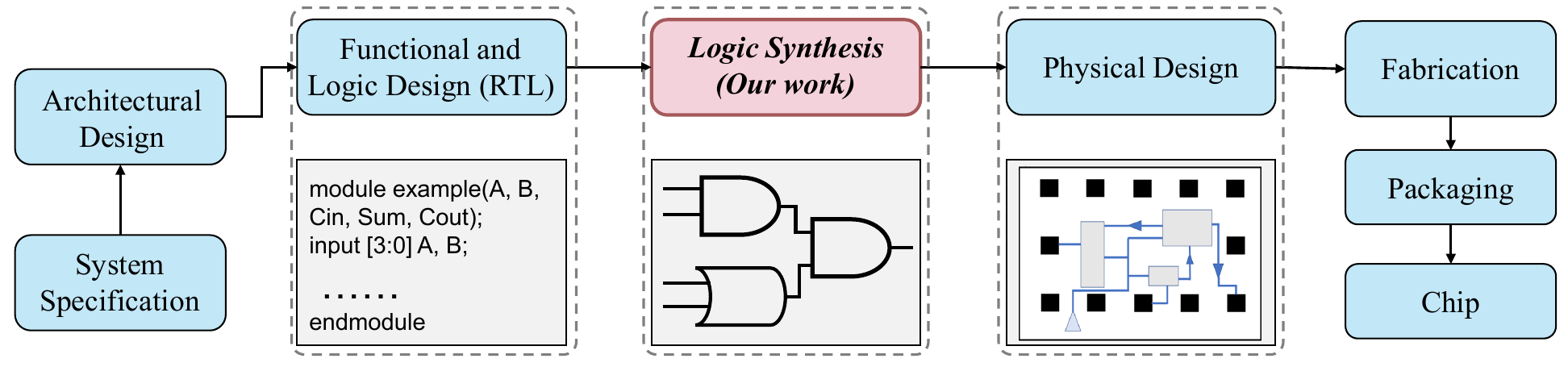}
        \caption{The chip design workflow}
        \label{fig:chip_design_flow}
    \end{subfigure}
    \begin{subfigure}{0.45\textwidth}
        \includegraphics[width=\textwidth,height=0.45\textwidth]{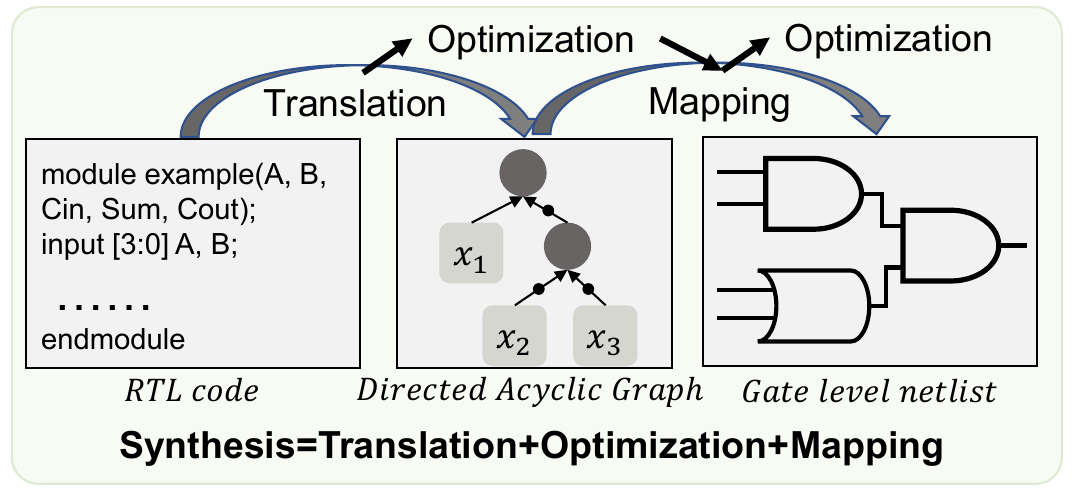}
        \caption{Logic synthesis}
        \label{fig:illustration_ls}
    \end{subfigure}
    \begin{subfigure}{0.45\textwidth}
        \includegraphics[width=\textwidth,height=0.45\textwidth]{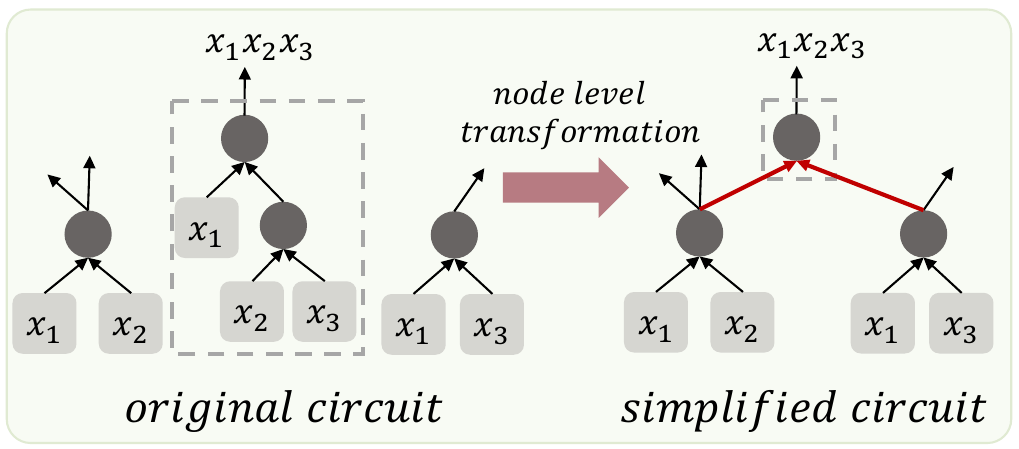}
        \caption{Circuit optimization}
        \label{fig:illustration_circuit_optimization}
    \end{subfigure}    
    \caption{(a) We illustrate the chip design workflow, and we focus on Logic Synthesis (LS) in this paper. (b) LS converts the behavioral description of a circuit to a transistor-level implementation, which generally consists of circuit optimization and technology mapping \cite{technology_mapping, ren2023machine}. (c) Circuit optimization aims to simplify an input circuit without changing its functionality.}
    \label{fig:illustration_ls_co}
\end{figure*}

\IEEEraisesectionheading{\section{Introduction}\label{sec:introduction}}
    \IEEEPARstart{C}hip design is a cornerstone of the worldwide semiconductor industry, promoting the development of an extensive market of electronic devices, such as cellular phones, personal computers, smart automobiles, etc.
    Logic Synthesis (LS) is one of the most important modules in chip design \cite{review_ai4lo,aisyn} as illustrated in Fig. \ref{fig:chip_design_flow}. Specifically, LS transforms a behavioral-level description of a design into an optimized gate-level circuit as illustrated in Fig. \ref{fig:illustration_ls}. 
    In brief, LS is the ``compiler'' in chip design. 
    A key task in LS is Circuit Optimization (CO),
    which aims to transform an input circuit into a simplified circuit with equivalent functionality and reduced size and/or depth as shown in Fig. \ref{fig:illustration_circuit_optimization}. Thus, it is crucial to well tackle the CO task, as it can significantly improve the Quality of Results, i.e.,  various metrics that evaluate the quality of designed chips, such as the area, delay, and performance of the chips \cite{de2021fast, rewrite, review_ai4lo}.
                  
\begin{figure*}[t]
    \centering
    \includegraphics[width=0.9\textwidth]{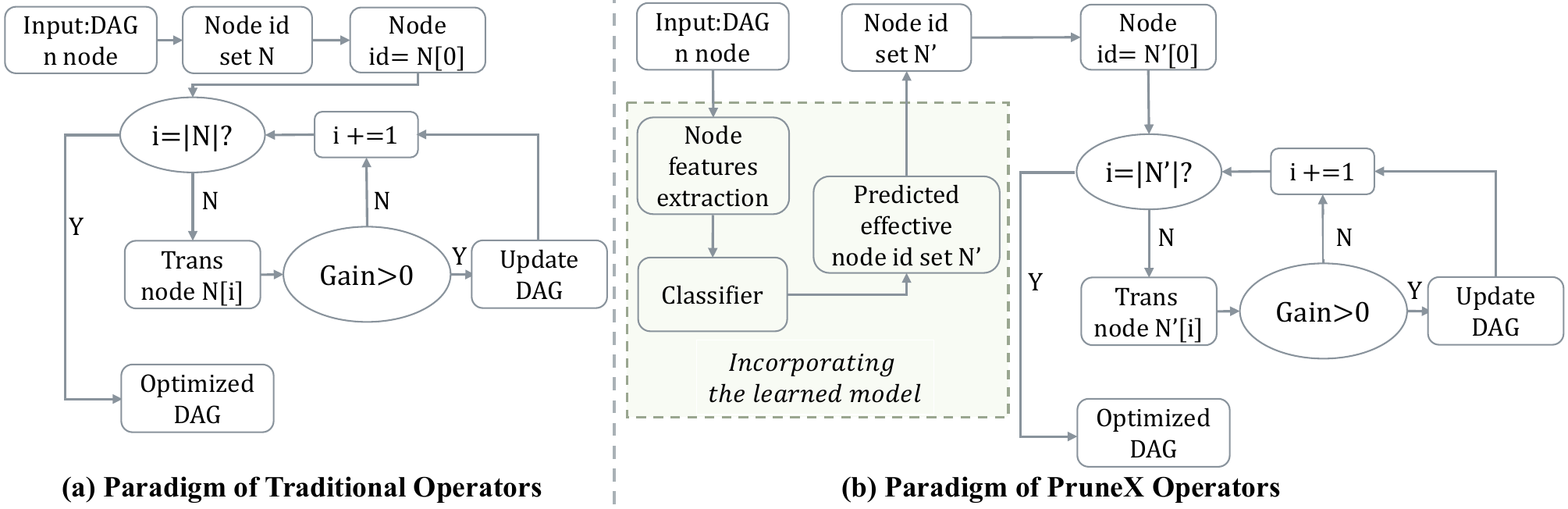}
    \caption{We illustrate the paradigm of many traditional operators and PruneX operators. Many commonly used operators follow the same paradigm while differing primarily in the specific methods for node-level transformations. Our PruneX incorporates learned classifiers into these operators to reduce the number of ineffective transformations. We denote the node-level transformation by ``Trans node N[i]'' in the figure.}
    \label{fig:paradigm_operators}
\end{figure*}

    However, the CO task can be extremely hard to tackle as it is a $\mathcal{NP}$-hard problem \cite{syn_textbook, farrahi1994complexity, aisyn}. To approximately tackle the CO task, many existing LS frameworks \cite{abc, mvsis}, including the open-source state-of-the-art LS framework known as ABC \cite{abc}, have developed a rich set of LS operators, such as Mfs2 \cite{mfs2}, Resub \cite{resub}, Rewrite \cite{rewrite, rewrite2}, Refactor \cite{refactor, rewrite2}, etc. Specifically, given an input circuit modeled by a directed acyclic graph (DAG), many commonly used LS operators apply transformations to 
    subgraphs rooted at each node---that is, the node-level transformation---sequentially for all nodes on the DAG. We illustrate a unified paradigm of these LS operators as shown in Fig. \ref{fig:paradigm_operators}. 
    
    However, we found an important problem leading to inefficient LS, that is, a large number of node-level transformations are \textit{ineffective}, making applying these operators highly time-consuming.      
    In particular, we notice that applying the Resub \cite{resub} and Mfs2 \cite{mfs2} operators take much longer runtime, roughly ranging from $30\times$ to $70\times$, than the other operators (see Fig. \ref{fig:motivation_ineffective_node}). Moreover, we found that the runtime of the two operators often dominates the overall runtime of LS optimization processes---accounts for approximately $79\%$ of the overall runtime (see Appendix \ref{appendix:motivation_time_analysis}). Thus, the runtime of the two operators acts as a bottleneck to the efficiency of LS, and inefficient LS 
    may significantly increase the Time to Market \cite{neto2021read, time_to_market_logic_synthesis, time_to_market_logic_synthesis2}, i.e., the overall duration for developing and commercializing new chips.  
    Moreover, inefficient LS could significantly degrade the Quality of Results (see Section \ref{exps:improving_qor}). For example, chip designers often have to reduce the number of times of using time-consuming operators to optimize large-scale circuits in LS, which may significantly increase the area and delay of chips.  

    To promote efficient LS, we propose a novel data-driven LS operator paradigm (see Fig. \ref{fig:paradigm_operators}), namely PruneX, which can significantly improve 
    the efficiency of LS operators by learning to reduce a large number of ineffective node-level transformations.    
    Specifically, PruneX learns a generalizable classifier to predict nodes with ineffective transformations (ineffective nodes\footnote{We apply an X operator to a given circuit, and then we denote those nodes with effective
    (ineffective) node-level transformations by effective (ineffective) nodes.}) accurately on unseen circuits, and avoids applying transformations to these ineffective nodes.
    An appealing feature of PruneX is that it is applicable to many commonly used LS operators---which follow the paradigm illustrated in Fig. \ref{fig:paradigm_operators}---to significantly improve their efficiency, thus possibly reducing the Time to Market. 
    
    The major challenge of developing PruneX is how to learn models that can well generalize to unseen circuits, that is, the out-of-distribution (OOD) generalization problem across circuits in LS operators. The major reason for the OOD generalization problem in LS is the large distribution shift across different circuits (see Fig. \ref{fig:generalization_motivation_tsne_visu_sample_batch}). As a result, PruneX could significantly degrade the optimization performance compared to the default operator, as it could reduce many effective transformations when failing to classify effective nodes accurately on unseen circuits. Thus, it is crucial for PruneX to tackle the OOD generalization problem across circuits in LS operators to achieve faster runtime while keeping comparable optimization performance. 

    To enable generalization, the major technical contribution of PruneX is the novel \textbf{c}ircuit d\textbf{o}main \textbf{g}eneralization (COG) framework (see Fig. \ref{fig:cog}) to learn domain-invariant representations for generalizable classifiers based on the transformation-invariant domain knowledge.
    To the best of our knowledge, COG is \textit{the first} data-driven method to well formulate and tackle the OOD generalization problem across circuits in LS operators, which is  
    critical for the success of data-driven LS algorithms.  
    Specifically, COG first formulates the OOD problem as a circuit domain generalization task by 
    formulating each circuit as an environment for collecting a dataset about the aforementioned classification task.    
    Then, COG proposes to learn domain-invariant node representations by aligning node embeddings across circuits from different domains based on the transformation-invariant domain knowledge---a node-level transformation is significantly associated with the local subgraph rooted at the node, regardless of the circuit to which it pertains.
    

    We integrate PruneX with the two aforementioned most time-consuming operators, i.e., the Resub \cite{resub} and Mfs2 \cite{mfs2} operators, among commonly used operators. Extensive experiments demonstrate that PruneX significantly and consistently improves their efficiency while keeping comparable optimization performance on three challenging benchmarks, achieving up to $3.1\times$ faster runtime. The challenging benchmarks include industrial and very large-scale circuits. Thus, our PruneX has the potential to save thousands of hours for developing new chips. Moreover, we conduct experiments to demonstrate that applying PruneX operators twice can significantly improve the optimization performance while achieving faster runtime. Note that a one percent improvement in the optimization performance may yield substantial economic value.     

    We summarize our major contributions as follows. (1) We found an important problem that leads to inefficient LS, i.e., many LS operators apply a large number of \textit{ineffective} node-level transformations (see Fig \ref{fig:motivation_ineffective_node}). (2) To promote efficient LS, we propose an effective data-driven LS operator paradigm, namely PruneX, which is applicable to many LS operators to significantly improve their efficiency.
    (3) The major technical contribution of PruneX is the novel circuit domain generalization framework to learn domain-invariant representations based on the transformation-invariant domain knowledge. 
    (4) To the best of our knowledge, PruneX is \textit{the first} data-driven method to tackle the OOD generalization problem in LS operators, which is critical for the success of data-driven LS algorithms.
    (5) Experiments demonstrate that PruneX significantly and consistently improves the efficiency of LS operators on three challenging benchmarks, including industrial and very large-scale circuits.

\section{Related Work}
    \subsection{Machine Learning for Logic Synthesis}
    As chip complexity has grown exponentially with the development of semiconductor technology, using machine learning (ML) to assist the automated chip design workflow has been an active topic of significant interest in recent years \cite{place_nature, huang2021machine, survey_gnn4eda, neto2021read, lai2022maskplace, lai2023chipformer}.
    As shown in Fig. \ref{fig:chip_design_flow}, the chip design workflow consists of many stages, such as high-level synthesis, logic synthesis, placement, routing, testing, verification, etc \cite{huang2021machine, ren2023machine}. In most of the stages in the workflow, recent studies have demonstrated significant improvement by using ML methods compared with traditional methods, including high-level synthesis \cite{xppe, hls_iccad18, hls_dac13}, logic synthesis \cite{neto2021read, review_ai4lo, lsoracle}, and placement \cite{place_nature,lai2022maskplace, lai2023chipformer,autodmp,cheng2022the}.
    
    In this paper, we focus on using machine learning to promote efficient logic synthesis (LS), which plays a vital role in efficient chip design and can yield substantial economic value \cite{fawcett1994synthesis, neto2021read}. 
    Existing research on machine learning for LS can be roughly divided into three categories \cite{review_ai4lo, ren2023machine}.
    First, \cite{dsoai, cadence_cerebrus, hosny2020drills, grosnit2022boils} use machine learning to tune the optimization flow of LS operators.   
    Second, \cite{neto2021read, kirby2019congestionnet, zhou2019primal} use machine learning to predict key metrics after physical design and leverage the prediction to guide LS optimization. 
    Third, \cite{neto2021slap, neto2019improving} use machine learning to improve decision-making in traditional LS methods. Different from existing work, our work uses machine learning to improve the paradigm of traditional LS operators to promote efficient LS. An appealing feature of our PruneX is that it is applicable to many LS operators---which follow the paradigm illustrated in Fig. \ref{fig:paradigm_operators}---to significantly improve their efficiency. 

    \subsection{Generalizable Prediction in Chip Design Workflow}
    Prior research has investigated the utilization of machine learning (ML) techniques to develop generalizable congestion prediction models within the chip design workflow \cite{congestionnet, LHNN, cross_graph, yang2022versatile}. 
    Nevertheless, our work differs from previous studies in two fundamental aspects. \textbf{First}, we address dissimilar input data. Prior research mainly focuses on the physical design stage with circuits represented by gate-level netlists or designed layouts, while we focus on the logic synthesis stage with circuits represented by Boolean networks. The dissimilarity in input data poses significant challenges in directly applying their methods to our setting. To the best of our knowledge, our work is the first data-driven method to well tackle the out-of-distribution (OOD) generalization problem across circuits in LS operators, which is critical for the success of data-driven LS algorithms.
    \textbf{Second}, we employ different methodologies for learning models.
    Generally, they propose problem-specific graph neural network architectures and learn the models via supervised learning. In contrast, our PruneX formulates the OOD generalization problem across circuits as a novel circuit domain generalization task, which offers promising avenues for future research on prediction tasks in chip design.   
    Moreover, based on a key observation called the transformation-invariant domain knowledge, our PruneX further proposes to learn domain-invariant representations for enhanced generalization capabilities. 

\section{Background}\label{sec:background}
    \subsection{Logic Synthesis (LS)}\label{background:ls}
    Driven by Moore’s law, the chip design complexity has grown exponentially \cite{khailany2020accelerating, lopera2021survey, huang2021machine, place_nature, ren2023machine}. Thus, the chip design workflow has incorporated multiple Electronic Design Automation (EDA) tools to synthesize, simulate, test, and verify different
    circuit designs efficiently and reliably. These EDA tools automatize the chip design workflow as shown in Fig. \ref{fig:chip_design_flow}. A LS tool---which aims to transform a behavioral description of a design into an optimized gate-level circuit implementation---is one of the most important modules in the EDA tools. In general, LS consists of pre-mapping optimization, technology mapping, and post-mapping optimization \cite{drills, ren2023machine}. \textbf{In this paper, we define Circuit Optimization (CO) by both pre-mapping optimization and post-mapping optimization.} First, in the pre-mapping optimization phase, logic optimization operators, such as Rewrite \cite{rewrite}, Resub \cite{resub}, and Refactor \cite{refactor}, are applied to an input circuit to optimize the circuit. Then, in the technology mapping phase, the optimized logic circuit is mapped to the available technology library, e.g., a standard-cell netlist \cite{braytontechnology} or k-input lookup-tables \cite{technology_mapping}. Finally, post-mapping optimization operators, such as Mfs2 \cite{mfs2}, are applied to the mapped circuit to further optimize it.

    \subsection{Circuit Optimization in LS}\label{background:co}
    Circuit Optimization (CO) in LS aims to transform an input circuit into a simplified circuit with equivalent functionality and reduced size and/or depth. Thus, well tackling the CO task can significantly save the hardware resources required to design a specific chip \cite{de2021fast}. As circuit representations in the pre-mapping and post-mapping optimization phases are different, we first discuss circuit representations in the two phases. Then, we provide details on LS operators.

    \subsubsection{Circuit Representation}\label{bg:circuit_representation}
    In the LS stage, a circuit is usually modeled by a Boolean network. In this paper, we use the terms Boolean network and circuit interchangeably. A Boolean network is a directed acyclic graph (DAG), where nodes correspond to Boolean functions and directed edges correspond to wires connecting these functions. A Boolean function takes the form $f:\mathbf{B}^{n}\to \mathbf{B}$, where $\mathbf{B}=\{0,1\}$ denotes the Boolean domain. 
    Given a node, its \textit{fanins} are nodes connected by incoming edges of this node, and its \textit{fanouts} are nodes connected by outgoing edges of this node. The \textit{primary inputs (PIs)} are nodes with no fanin, and the \textit{primary outputs (POs)} are nodes with no fanout. The \textit{size} of a circuit denotes the number of nodes in the DAG. The \textit{depth (level)} of a circuit denotes the maximal length of a path from a PI to a PO in the DAG. 
    The size and depth of a circuit are proxy metrics for the area and delay of the circuit, respectively.    

    Common types of DAGs for CO include And-Inverter Graphs (AIGs) for pre-mapping optimization \cite{rewrite, rewrite2} and K-Input Look-Up Tables (K-LUTs) for post-mapping optimization \cite{technology_mapping,mfs2}.
    In the pre-mapping optimization phase, an AIG is a DAG containing four types of nodes: the constant, PIs, POs, and two-input And (And2) nodes. A graph edge is either complemented or not. A complemented edge indicates that the signal is complemented. 
    In the post-mapping optimization phase, a K-LUT is a DAG with nodes corresponding to Look-Up Tables and directed edges corresponding to wires. A Look-Up Table in a K-LUT is a digital memory that implements the Boolean function of the node.

    \subsubsection{Commonly Used LS Operators}
    A rich set of LS operators have been developed to tackle the CO task in the pre-mapping and post-mapping optimization phases \cite{abc}. In this paper, we focus on commonly used LS operators on large-scale industrial circuits---that is, the Resub \cite{resub}, Mfs2 \cite{mfs2}, Rewrite \cite{rewrite}, and Refactor \cite{refactor} operators---which often act as a bottleneck to the efficiency of the LS optimization processes (see Section \ref{sec:efficiency_analysis}). We notice that these LS operators follow the same paradigm as shown in Fig. \ref{fig:paradigm_operators}. Specifically, they apply transformations to subgraphs rooted at each node---that is, the node-level transformation---sequentially for all nodes on an input DAG. Note that the major differences among these operators lie in the node-level transformation mechanism. 

\begin{figure*}[t]
    \centering
    \begin{subfigure}{0.9\textwidth}
        \centering
        \includegraphics[width=0.32\textwidth, height=0.32\textwidth]{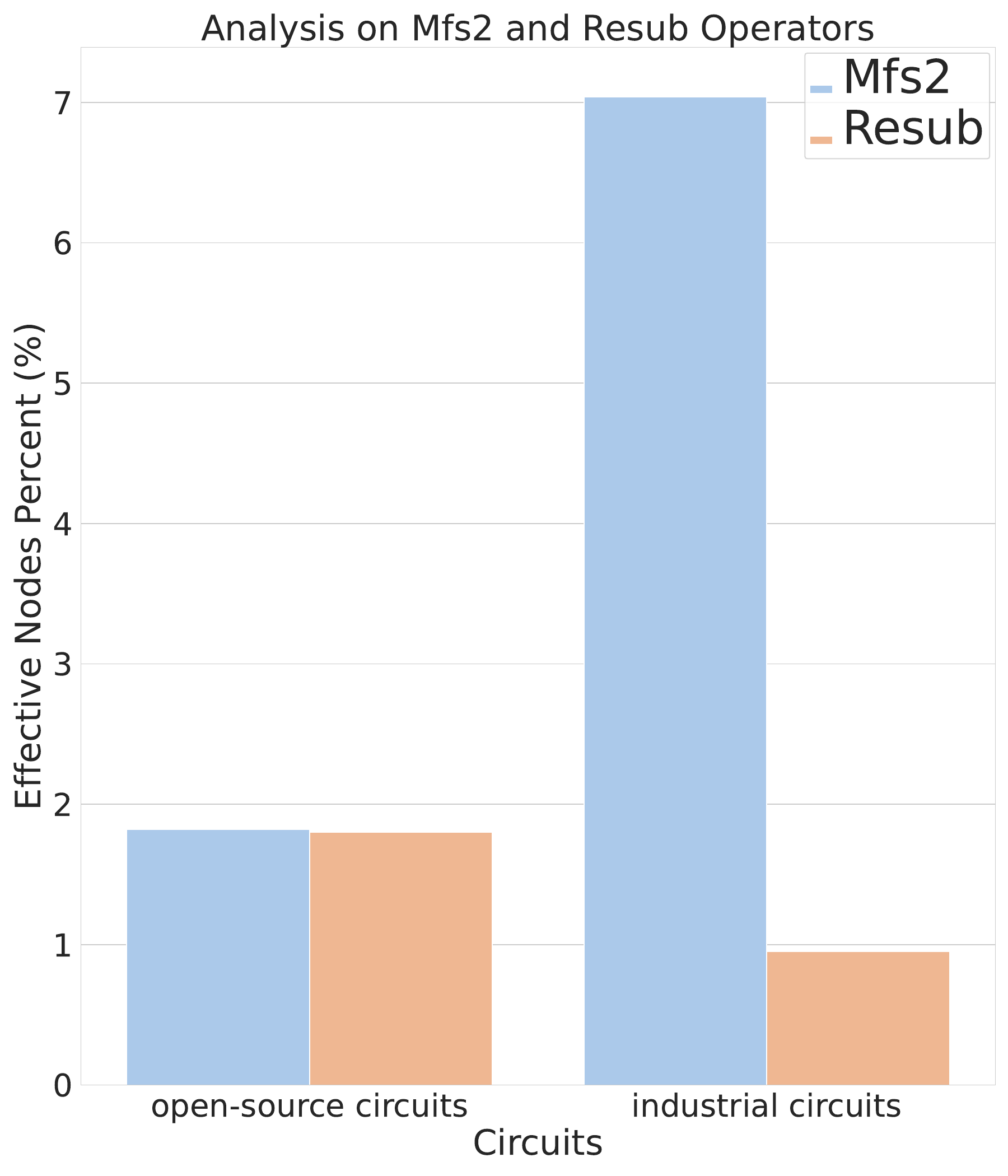}
        \includegraphics[width=0.32\textwidth, height=0.32\textwidth]{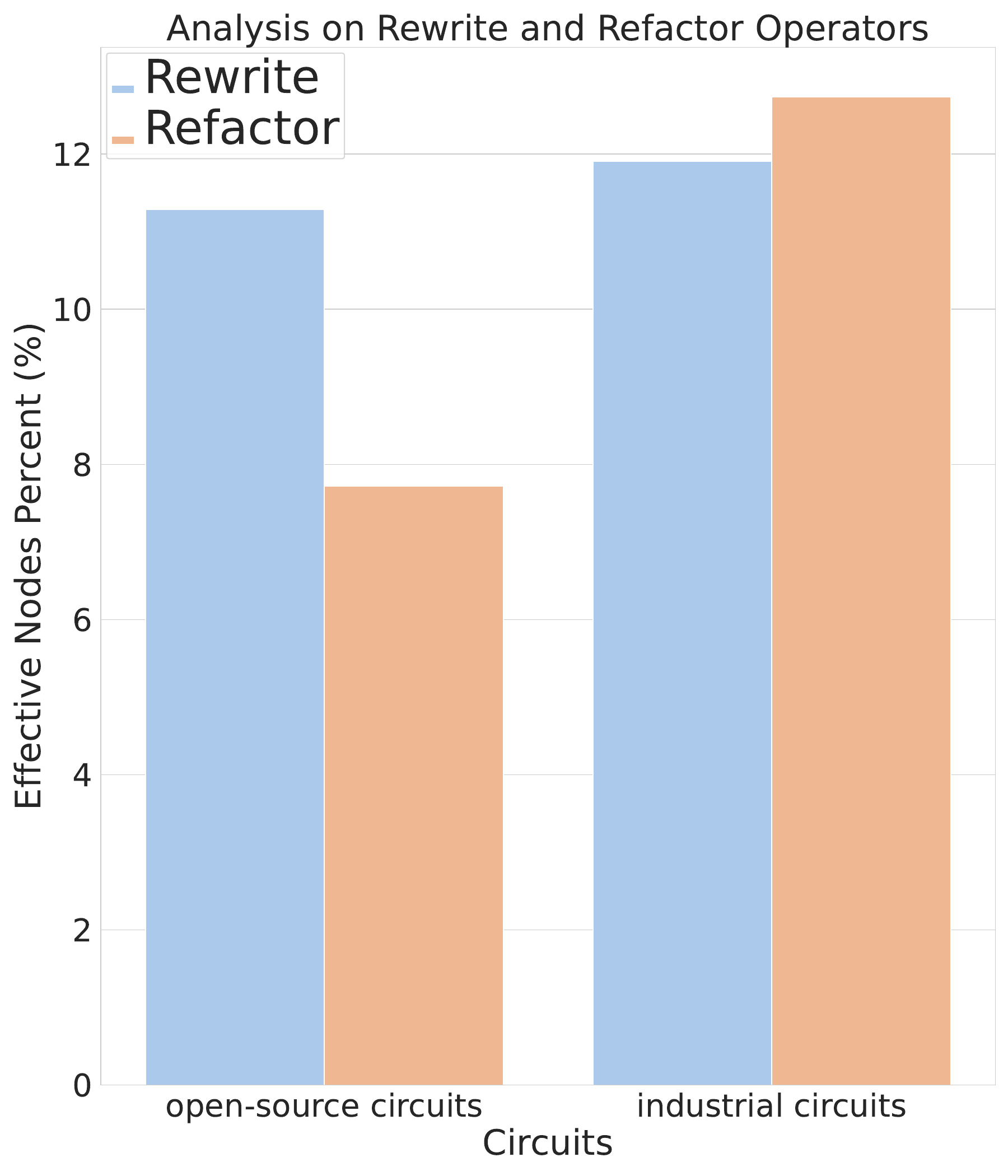}
        \includegraphics[width=0.32\textwidth, height=0.32\textwidth]{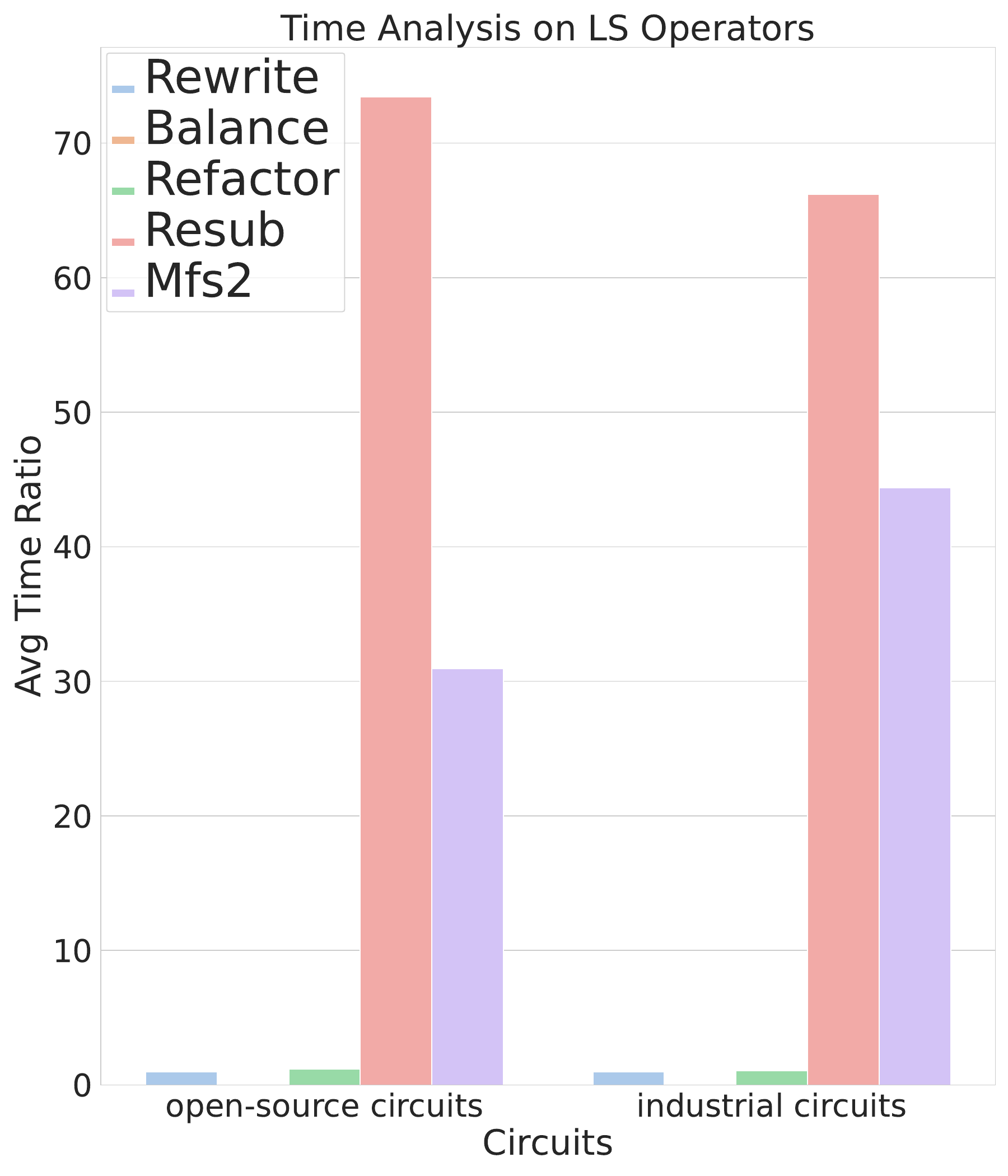}
        \vspace{-2mm}
        \caption{We analyze the percentage of effective nodes and runtime of LS operators.}
        \label{fig:motivation_ineffective_node}
    \end{subfigure}
    \vspace{1.5mm}
    \begin{subfigure}{0.9\textwidth}
        \centering
        \includegraphics[width=0.32\textwidth,height=0.25\textwidth]{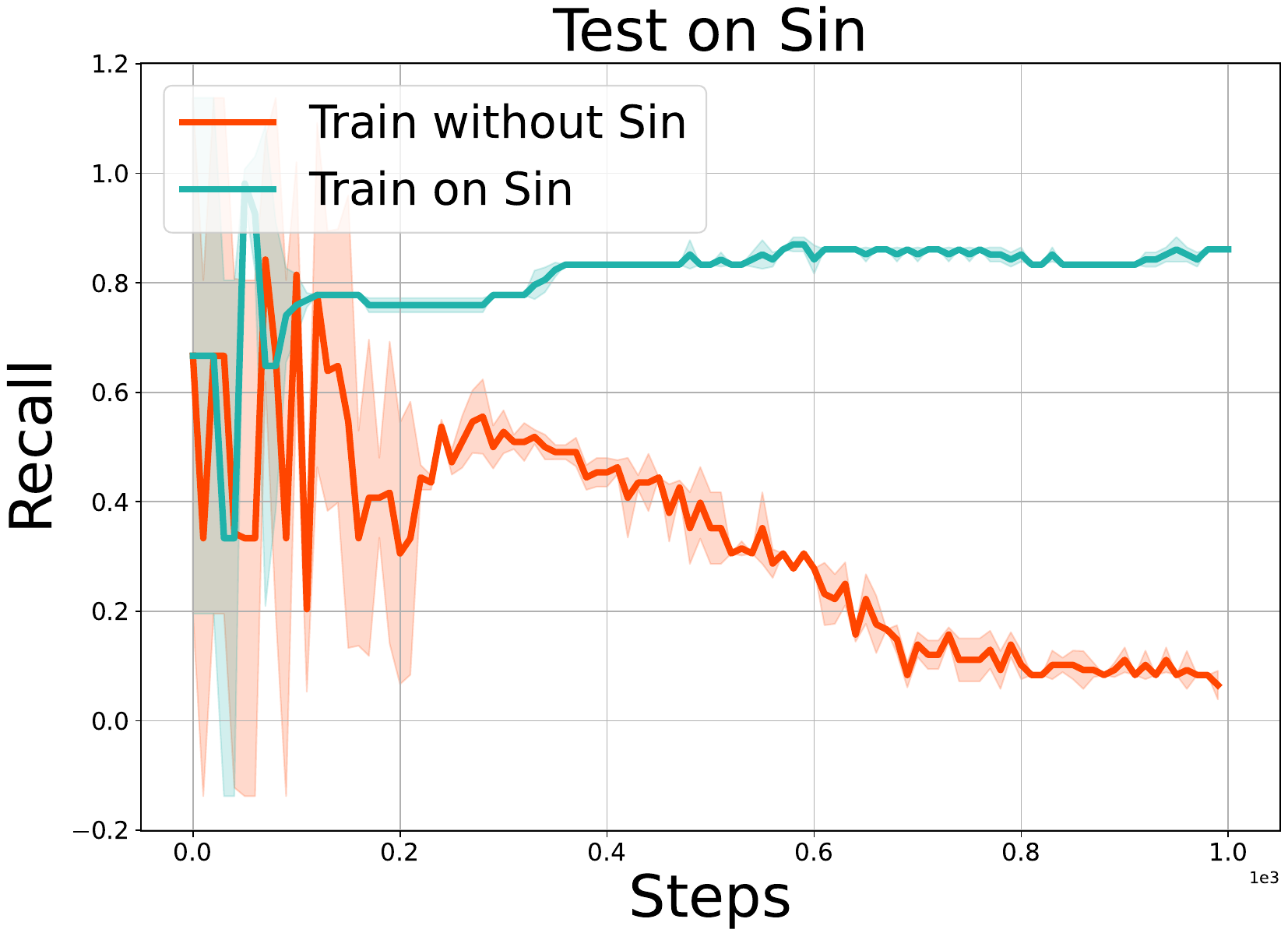}
        \includegraphics[width=0.32\textwidth,height=0.25\textwidth]{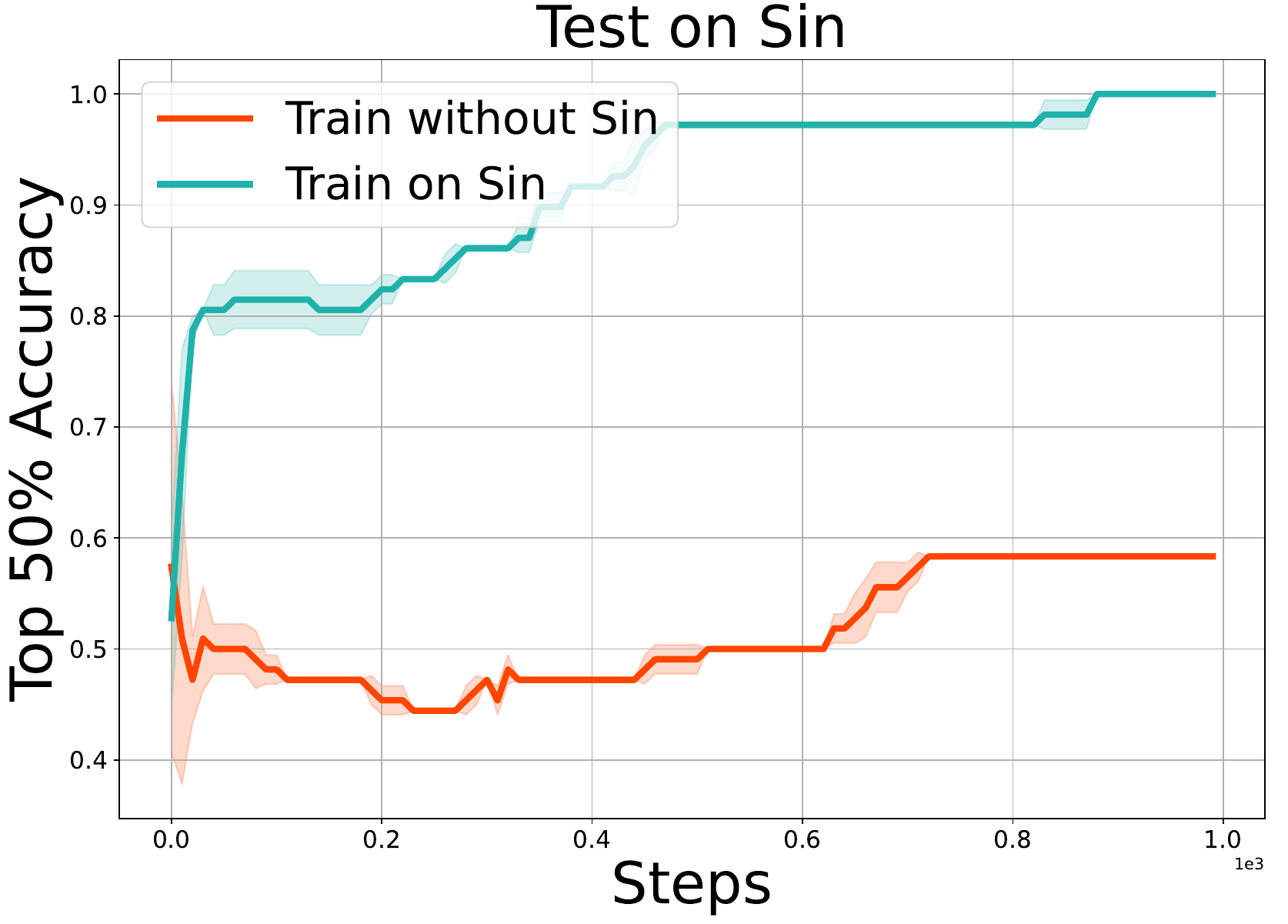}
        \includegraphics[width=0.32\textwidth,height=0.25\textwidth]{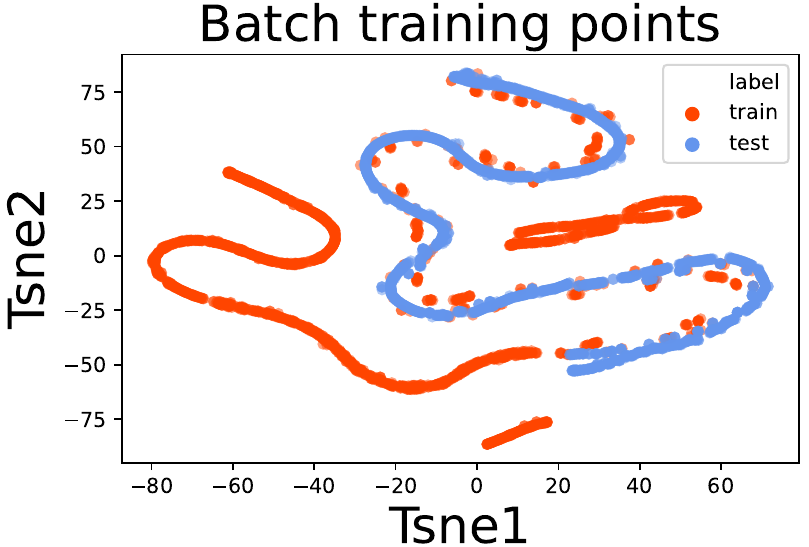}
        \vspace{-1mm}
        \caption{We analyze the out-of-distribution generalization challenge in LS.}
        \label{fig:generalization_motivation_tsne_visu_sample_batch}
    \end{subfigure}
    \vspace{-2.5mm}
    \caption{(a) The results demonstrate that a large number of transformations are ineffective (exceeding $87\%$) in the four operators. Moreover, the results show that applying the Resub and Mfs2 operators are the most time-consuming among commonly used operators. (b) We train and evaluate models using a dataset with Sin (green) and without Sin (red). The results show that models trained without Sin (red) performs poorly on Sin. Furthermore, we visualize the data points from the dataset without Sin (red) and Sin circuit (blue), respectively. The results suggest that distribution shift indeed exists.}
    \label{fig:motivation}
\end{figure*}

\section{Motivating Results}\label{sec:motivating_results}
    \subsection{Ineffective Node-Level Transformations Problem}
    As shown in Fig. \ref{fig:motivation_ineffective_node}, the Mfs2 and Resub operators apply a large number of Ineffective Node-Level Transformations (exceeding $93\%$) on open-source and industrial circuits. In addition, the results in Fig. \ref{fig:motivation_ineffective_node} show that a large number of node-level transformations applied by the Rewrite and Refactor operators are ineffective as well (exceeding $87\%$). We defer more results to Appendix \ref{appendix:motivation_int_problem}.

    \subsection{Efficiency Analysis of LS Operators}\label{sec:efficiency_analysis}
    We evaluate the runtime of applying five LS operators, which are commonly used in industrial settings, to open-source and industrial circuits. Specifically, the five operators include Rewrite \cite{rewrite, rewrite2}, Balance \cite{balance, rewrite2}, Refactor \cite{refactor, rewrite2}, Resub \cite{resub}, Mfs2 \cite{mfs2}. The results in Fig. \ref{fig:motivation_ineffective_node} show that applying the Resub and Mfs2 operators are the most time-consuming among commonly used operators. 
    Specifically, we report the ratio of the runtime of these operators to that of the Rewrite operator.
    The results demonstrate that applying the Resub and Mfs2 operators take much longer runtime, roughly ranging from $30\times$ to $70\times$, than the Rewrite operator.    
    Moreover, we further demonstrate that the runtime of the two operators often dominates the overall runtime of LS optimization processes---accounts for approximately $79\%$ of the overall runtime (see Appendix \ref{appendix:motivation_time_analysis}). 
    Therefore, the runtime of applying the Resub and Mfs2 operators acts as a bottleneck to the efficiency of LS.     
    
    Moreover, we found that applying the Rewrite and Refactor operators to very large-scale circuits are time-consuming as well (see Appendix \ref{appendix:motivation_time_analysis}).
    Nevertheless, our PruneX is also applicable to the Rewrite and Refactor operators, as these operators follow the same paradigm as shown in Fig. \ref{fig:paradigm_operators}. We provide more discussion on how to apply PruneX to the Rewrite and Refactor operators in Appendix \ref{appendix:discuss_apply_to_rewrite}.
    We provide results and details in Appendix \ref{appendix:motivation_time_analysis}.

    \subsection{Out-of-Distribution Generalization Problem in LS}
    To evaluate whether there exists distribution shift across different circuits, we use the Mfs2 operator to collect a training dataset with all circuits from the EPFL benchmark \cite{amaru2015epfl} and another one with all circuits except the Sin circuit. We use EnsembleMLP---a simple learning baseline using supervised learning---to train models on these two datasets, and evaluate the model on Sin. The results in Fig. \ref{fig:generalization_motivation_tsne_visu_sample_batch} show that it is challenging for models trained on circuits without Sin to generalize to the unseen Sin circuit. 
    
    Furthermore, we visualize the batch data from the training dataset without Sin and the Sin circuit as shown in Fig \ref{fig:generalization_motivation_tsne_visu_sample_batch}. The results show that the data distributions from the training and testing datasets are different. 
    Moreover, we empirically show that the data distributions from different circuits are similar but different in Appendix \ref{appendix:motivation_ood}. Due to the distribution shift, it is extremely challenging to learn models that can well generalize to unseen circuits, i.e., the OOD generalization problem across circuits in LS operators. 
    
    In addition, we empirically demonstrate that the optimization performance of the PruneX operator significantly degrades as the prediction recall decreases in Appendix \ref{appendix:recall_qor}. Therefore, it is critical to develop data-driven methods that can well tackle the OOD generalization problem for achieving high prediction recall and thus comparable optimization performance with default operators.

\section{Methods}\label{sec:methods}
    To promote efficient Logic Synthesis (LS), we first provide a detailed description of our proposed PruneX operator paradigm in Section \ref{method:prunex}, which learns classification models to predict ineffective nodes in LS. However, directly learning classification models in LS faces the out-of-distribution (OOD) generalization challenge (see Fig. \ref{fig:generalization_motivation_tsne_visu_sample_batch}).
    To address this challenge, we then present our proposed \textbf{c}ircuit d\textbf{o}main \textbf{g}eneralization (COG) framework to learn generalizable and scalable classification models in LS in Section \ref{method:cog}. 

    \subsection{A Novel Data-Driven LS Operator Paradigm}\label{method:prunex}
        As shown in Fig. \ref{fig:motivation_ineffective_node}, we found that a large number of node-level transformations in many LS operators are ineffective, which makes applying these operators highly time-consuming. To address this challenge, we propose a novel data-driven LS operator paradigm, namely PruneX, which incorporates learned classifiers into these operators to improve their efficiency. Specifically, PruneX consists of two phases: the offline and online phases.

        \textbf{The Offline Phase: Data Collection and Model Learning} In this phase, we aim to collect a training dataset $\mathcal{D}$ from multiple existing logic circuits $\{\mathcal{C}_i\}_{i=1}^{N}$ given a LS operator, and learn a classification model from the dataset.
        We first provide a definition of the Circuit Dataset as follows. 
        \begin{definition}\label{def:circuit_dataset}
            (\textbf{Circuit Dataset}) Let $\{\mathcal{C}_i\}_{i=1}^{N}$ denote a set of $N$ circuits. For each circuit $\mathcal{C}_i$, let $\mathcal{X}_i$ denote its nonempty input space and $\mathcal{Y}_i$ an output space. Let $\{\textbf{x}^i_j,y^i_j\}_{j=1}^{n_i}$ denote the data from the circuit $\mathcal{C}_i$. We define a \textbf{Circuit Dataset} by the data $\mathcal{D}_i = \{\textbf{x}^i_j,y^i_j\}_{j=1}^{n_i}$.
        \end{definition}
        In general, a logic circuit $\mathcal{C}$ is modeled as a directed acyclic graph (DAG), where nodes represent logic gates and directed edges represent wires connecting the gates. 
        Let X denote a LS operator, such as Mfs2 \cite{mfs2}, Resub \cite{resub}, Rewrite \cite{rewrite}.         
        To generate the Circuit Dataset, we apply the X operator to optimizing the circuit, which applies node-level transformations to each node on the DAG. For each node in the circuit, we generate a data pair $(\textbf{x},y)$, where $\textbf{x}$ denotes the node, and $y$ denotes the label. Specifically, if the node-level transformation is effective at the node $\textbf{x}$, then $y=1$. Otherwise, $y=0$. Given the X operator and $N$ circuits, we generate a dataset $\mathcal{D} = \{\mathcal{D}_i\}_{i=1}^{N} = \{ \{\textbf{x}_j^i,y_j^i\}_{j=1}^{n_i} \}_{i=1}^{N}$. 
        
        Given the generated dataset $\mathcal{D}$, we learn a binary classifier $f: \mathcal{X} \to \mathcal{Y}$, where $\mathcal{X}$ denotes the input space of nodes, and $\mathcal{Y}$ denotes $\{0,1\}$. Specifically, we formulate the prediction task as a binary classification task. The optimization objective for each data pair $(\textbf{x},y)$ takes the form of
        \begin{align}\label{obj:binary_classification}
            l(f(\textbf{x}),y) = -y\log(f(\textbf{x}))-(1-y)\log(1-f(\textbf{x})).
        \end{align}
        That is, the output of our learned model for an input node represents the probability that the node is an effective node. 

        Note that the Ineffective Node-Level Transformations (INT) problem in many LS operators---that is, the number of effective nodes is far fewer than ineffective nodes---leads to an extreme \textit{class imbalance} in the training dataset. This severe imbalance poses a substantial challenge to the classification task \cite{focal_loss, rota2017loss}. To tackle this problem, we leverage the Focal Loss \cite{focal_loss}, which has been shown successful in addressing class imbalance for object detection tasks. We provide more details on the Focal Loss in Appendix \ref{appendix:focal_loss}.

        To learn the classifier $f$, we propose a simple learning-based baseline, namely EnsembleMLP, which uses a multi-layer perceptron \cite{goodfellow2016deep, taud2018multilayer} with ensemble learning \cite{zhou2021ensemble} to parameterize the model, and trains the model by averaging the loss in (\ref{obj:binary_classification}) over all training samples. We provide more details about EnsembleMLP in Appendix \ref{appendix:ensemblemlp}.    
        
        \textbf{The Online Phase: Incorporating Learned Models into LS Operators} As shown in Fig. \ref{fig:paradigm_operators}, PruneX uses the learned classifier to predict ineffective nodes on \textit{unseen} circuits, and avoid transformations on these nodes to accelerate the X operator in the online phase. PruneX learns an individual classifier model for each given operator, as the input circuit representations for different operators are different (see Section \ref{bg:circuit_representation} and Appendix \ref{appendix:node_features} for details). 
        
        Note that the inference time of the learned model should be short, otherwise the time saved by avoiding transformations cannot compensate for the inference time. Thus, we formulate the task as a one-shot prediction task rather than a sequential prediction task. That is, PruneX only performs model inference once on all nodes, and selects the predicted effective nodes to apply node-level transformations (see Fig. \ref{fig:paradigm_operators}). As a result, the model takes notably shorter inference time in comparison to the transformation time in an operator.
        Please refer to Appendix \ref{appendix:model_inference_time} for discussion and results. 

        In addition, we found that the learned classifier suffers from negative bias due to the aforementioned imbalance problem (see Appendix \ref{appendix:exp_imbalance}). Thus, 
        $0.5$ is an inappropriate threshold for evaluating whether a sample is positive, while determining an appropriate threshold is challenging. To tackle this problem, we view the prediction task as a scoring task, and select those nodes with top $k$ scores as predicted positive samples. That is, we view an output $y$ of the model $f$ as the score of a node $\textbf{x}$. Note that the higher the score, the higher the probability that the node is an effective node. Consequently, we can control the trade-off between runtime and Quality of Results (QoR) by adjusting the hyperparameter $k$. That is, 
        as the value of $k$ decreases, the runtime reduces, but this reduction may come at the cost of a compromised Quality of Results (QoR) due to the decreased number of nodes for applying transformations. Please see Appendixes \ref{appendix:recall_qor} and \ref{appendix:num_trans_runtime} for details. 
        
        \textbf{Discussion on More Advantages} An appealing feature of PruneX is that it is applicable to many commonly used LS operators, and can significantly improve their efficiency. More specifically, PruneX is applicable to LS operators that follow the paradigm shown in Fig. \ref{fig:paradigm_operators}, such as Mfs2 \cite{mfs2}, Resub \cite{resub}, Rewrite \cite{rewrite}, Refactor \cite{refactor}, and Mfs3 \cite{mfs3}.
        Moreover, PruneX carries the potential to achieve high prediction accuracy by incorporating machine learning into LS, as machine learning has achieved great success in vision prediction tasks such as image classification \cite{resnet, vit}.
        

\begin{figure*}[t]
    \centering
    \includegraphics[width=0.98\textwidth]{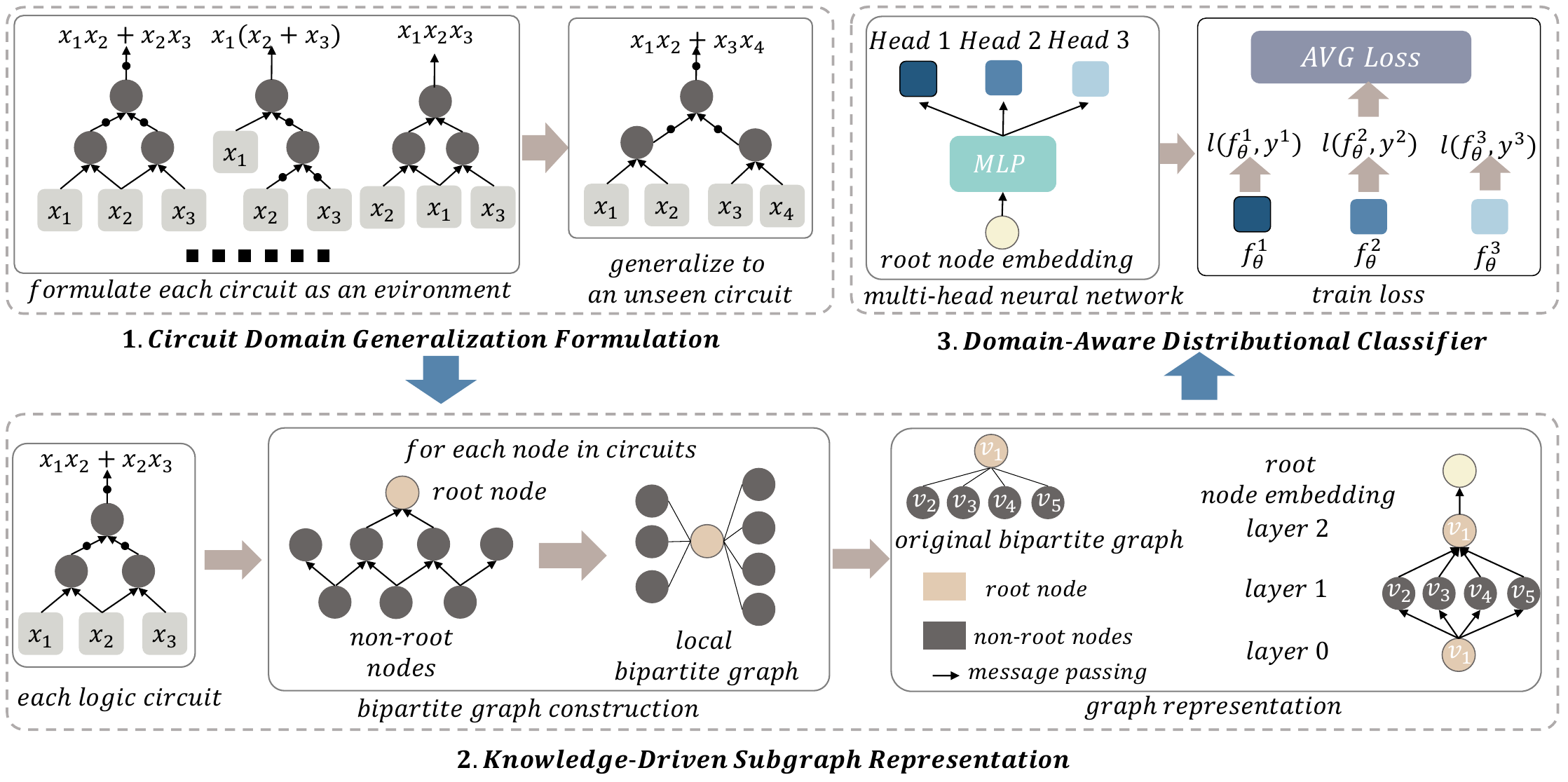}
    \vspace{-2mm}
    \caption{We illustrate our circuit domain generalization (COG) framework. Our COG consists of three main components to learn classification models that can well generalize to unseen circuits. Please see Section \ref{method:cog} for details.}
    \label{fig:cog}
\end{figure*}

    \subsection{A Circuit Domain Generalization Framework for Learning Classification Models in LS}\label{method:cog}
        In this subsection, we provide a detailed description of our proposed \textbf{c}ircuit \textbf{d}omain \textbf{g}eneralization (COG) framework, which serves as the model learning module in our PruneX to learn generalizable and scalable classifiers. Specifically, COG consists of three main components. (1) COG formulates the learning task as a circuit domain generalization task in Section \ref{method:cog_formulation}. (2) COG proposes a knowledge-driven subgraph representation learning method to learn domain-invariant representations in Section \ref{method:kdlgr}. (3) COG proposes a domain-aware distributional classifier in Section \ref{method:ds}.
 
        \subsubsection{A Circuit Domain Generalization Formulation}\label{method:cog_formulation} 
        Due to the distribution shift between training and testing Circuit Datasets as shown in Fig. \ref{fig:generalization_motivation_tsne_visu_sample_batch}, learning classification models that well generalize to unseen circuits is challenging, i.e., the out-of-distribution (OOD) generalization problem in LS. To address this challenge, we propose to formulate the OOD generalization problem across circuits as a circuit domain generalization task by 
        formulating each circuit as an environment for collecting a Circuit Dataset about the aforementioned classification task. Under this formulation, a natural idea is to formulate each Circuit Dataset as an estimation of a Circuit Domain. This allows us to incorporate prior knowledge about the domain into model learning to enhance the generalization capability of our models. 


        
        However, unlike the traditional domain generalization setting in computer vision \cite{shen2021towards,wang2022generalizing}, 
        the sample sizes of different Circuit Domains in our setting can be quite variable, leading to two undesirable challenges. First, the sizes of some circuits are small, with only about $100$ nodes. This makes the number of data points from these Circuit Datasets small, resulting in a few shot learning challenge \cite{wang2020generalizing}. Second, the sizes of different circuits range from $100$ to $100,000$ nodes, resulting in an extreme imbalance of sample sizes across different Circuit Datasets. 
        
        To address these challenges, we further propose a circuit aggregation mechanism, which is based on the domain knowledge that circuits with similar functionality are likely to possess similar distributions. Specifically, we aggregate Circuit Datasets from circuits with similar functionality into a hyper-circuit dataset. Thus, we aggregate $N$ Circuit Datasets into multiple hyper-circuit datasets, and formulate each hyper-circuit dataset as an estimation of a Circuit Domain. Before discussing details on the circuit aggregation mechanism, we first introduce the formal definition of a Circuit Domain as follows.
        \begin{definition}\label{defition_circuit_domain}
            (\textbf{Circuit Domain}) Let $\{\mathcal{D}_i\}_{i=1}^{N} $ denote $N$ Circuit Datasets. We propose a \textit{circuit aggregation mechanism} $g$ that maps the $N$ Circuit Datasets to $M$ hyper-circuit datasets ($M \leq N$). That is, $g(\{\mathcal{D}_i\}_{i=1}^N) = \{\mathcal{D}_k^{\prime}\}_{k=1}^{M}$, where $\mathcal{D}_k^{\prime}$ denotes the data from the $k$-th aggregated dataset. 
            For each $\mathcal{D}_k^{\prime}$, we assume that its data $\{ \textbf{x}^k_j,y^k_j \}_{j=1}^{n_k}$ are sampled from an underlying distribution $\mathbb{P}^k_{XY}$. We define a \textbf{Circuit Domain} by the underlying distribution $\mathbb{P}^k_{XY}$.
            
        \label{def:circuit_domain}
        \end{definition}
        Moreover, we introduce the definition of our circuit domain generalization as follows.
        \begin{definition}\label{defition_circuit_domain_generalization}
            (\textbf{Circuit Domain Generalization}) We are given $M$ aggregated datasets $\{\mathcal{D}_k^{\prime}\}_{k=1}^{M}$ from 
            $M$ source Circuit Domains $\mathbb{P}_{XY}^1,\dots,\mathbb{P}_{XY}^M$, where $\mathbb{P}_{XY}^i\neq\mathbb{P}_{XY}^j$, $1 \leq i\neq j\leq M$. The goal of circuit domain generalization is to learn a robust and generalizable predictive function $f:\mathcal{X}\to\mathcal{Y}$ from the $M$ source domains, which achieves a minimum prediction error on an unseen testing circuit sampled from $\mathbb{P}_{XY}^{t}$, i.e.,
            $\min\limits_f \mathbb{E}_{(\textbf{x},y)\sim\mathbb{P}_{XY}^{t}} [l(f(\textbf{x}), y)]$ 
        \label{def:circuit_domain_generalization}
        \end{definition}
        Under the circuit domain generalization formulation, we aim to leverage prior knowledge about the Circuit Domain to learn models that can well generalize to unseen circuits. To the best of our knowledge, COG is \textit{the first} to formulate the OOD generalization problem across circuits as a circuit domain generalization task, which is critical for the success of our COG and carries the potential to motivate future remarkable research on this OOD generalization problem. 
        
        \textbf{Discussion on the Circuit Aggregation Mechanism} It remains challenging to instantiate the idea of aggregating circuits with similar functionality for the following two reasons. First, a standardized metric for quantifying the similarity of circuit functionality is currently absent in the literature. In this paper, we propose to aggregate circuits based on the high-level functionality of circuits, such as the arithmetic, control, and memory functionality. Second, the availability of functionality information for circuits may be limited or unavailable. To alleviate this problem, we make some theoretical analysis as follows, and the theoretical results show two more basic principles for the circuit aggregation mechanism $g$ to achieve a small generalization error bound. (1) The sample sizes of different Circuit Domains should be close. (2) It is undesirable to aggregate all Circuit Datasets into one mixed domain. Finally, we propose a simple method to aggregate circuits based on their functionality and sample sizes (see Appendix \ref{appendix:domain_construction_mapping} for details). 
    
        \textbf{Theoretical Analysis} 
        In general, the OOD generalization problem is infeasible unless we make some assumptions on the existence of some
        statistical invariances across training and testing domains \cite{wang2022generalizing, shen2021towards}. 
        Based on the domain knowledge of the OOD generalization problems across circuits, we make the following assumption, which is commonly used in traditional domain generalization literature \cite{blanchard2011generalizing,muandet2013domain,wang2022generalizing}.
        \begin{assumption}\label{assumption1}
            The training Circuit Domains $\mathbb{P}_{X Y}^1, \ldots, \mathbb{P}_{X Y}^M$ are i.i.d. realizations from a hyper-distribution $\mathscr{P}$ and any possible testing Circuit Domain $\mathbb{P}_{X Y}^t$ follows the same hyper-distribution. 
        \end{assumption}
        The assumption usually holds in practice for our OOD problem across circuits. We defer detailed discussion on this assumption to Appendix \ref{appendix:discuss_assum1}.        
        Based on Assumption \ref{assumption1}, the learned classifier should include the Circuit Domain information $\mathbb{P}_{XY}$ into its input so that it can well generalize to unseen target Circuit Domains. In our case, the functional relationship $\mathbb{P}_{Y \vert X}$ across different Circuit Domains is stable, as 
        whether a node-level transformation is effective is stable across different circuits. Thus, the marginal distribution $\mathbb{P}_{X}$ contains the whole domain-specific information. As a result, the prediction of the classifier takes the form of $y = f(\mathbb{P}_{X},\textbf{x})$. Given the model $f$, its average risk over all possible target Circuit Domains takes the form of 
        \begin{align}\label{obj:ground_truth}
            \mathcal{R}(f) = \mathbb{E}_{\mathbb{P}_{XY}^t\sim \mathscr{P}} \mathbb{E}_{(\textbf{x},y)\sim \mathbb{P}_{XY}^t}[l(f(\mathbb{P}_{X}^t,\textbf{x}),y)].
        \end{align}
        Unfortunately, the expectation is intractable as the hyper-distribution $\mathscr{P}$ is unknown. Nevertheless, we can estimate (\ref{obj:ground_truth}) using finite Circuit Domains following $\mathscr{P}$, and finite $(\textbf{x},y)$ samples following each Circuit Domain distribution under Assumption \ref{assumption1}. Specifically, our empirical risk estimation objective takes the form of 
        \begin{align}\label{obj:empirical}
            \hat{\mathcal{R}}(f) = \frac{1}{M}\sum_{k=1}^{M} \frac{1}{n_k} \sum_{j=1}^{n_k} l(f(\hat{\mathbb{P}}_{X}^k,\textbf{x}_j^k),y_j^k),
        \end{align}
        where $M$ denotes the number of training Circuit Domains, $n_k$ denotes the sample size of the $k$-th Circuit Domain, and $\hat{\mathbb{P}}_{X}^k$ denotes the empirical estimation of the $k$-th Circuit Domain $\mathbb{P}_{X}^k$. 
        To evaluate how well our estimation is, we show that there is a generalization error bound between (\ref{obj:ground_truth}) and (\ref{obj:empirical}), which is inspired by previous work \cite{blanchard2011generalizing, muandet2013domain}.
        \begin{theorem}\label{obj:bound_estimation}
        Under some mild and reasonable assumptions,  we can conclude that with probability at least $1-\delta$
        \begin{align}
        & \sup _{f \in B_{\bar{K}}\left(r\right)}\lvert \mathcal{R}(f)-\hat{\mathcal{R}}(f) \rvert^2 \nonumber \\
        \leq & c_1+c_2 \frac{\log 2\delta^{-1}}{M}\sum_{k=1}^M \frac{1}{n_k}+c_3 \frac{\log \delta^{-1}}{M}+\frac{c_4}{M}.
        \end{align}
        where $c_1, c_2, c_3, c_4$ are constants, $B_{\bar{K}}(r)$ denotes the ball of radius $r$ of an Reproducing Kernel Hilbert Space (RKHS) $\mathcal{H}_{\bar{K}}$, $M$ is the number of training Circuit Domains and $n_k$ is the sample size of the $k$-th Circuit Domain. 
        \end{theorem}

        Here, RKHS is a Hilbert space of functions where the evaluation of a function at any point can be represented as an inner product with a kernel function. Please refer to Appendix \ref{appendix:proof of error bound} for detailed proof. Theorem \ref{obj:bound_estimation} shows that the generalization error bound depends on both $M$ and $
        n_k$, thus depending on the circuit aggregation mechanism $g$. Due to the aggregation mechanism, the number of training Circuit Domains $M$ and the sample size of the $k$-th domain $n_k$ are variable, and the total number of samples $n$ is fixed. This is quite different from existing domain generalization work \cite{blanchard2011generalizing,muandet2013domain,wang2022generalizing}. 
        Based on Theorem \ref{obj:bound_estimation}, we show the following Corollaries to provide valuable theoretical insights into the circuit aggregation mechanism $g$. 
        
        \begin{corollary}\label{proof:corollary3}
        If the number of Circuit Domains $M$ and the total number of samples $n$ are fixed, the generalization error bound reaches its minimum when $n_k = \frac{n}{M}  \text{ for } k = 1, 2, \dots, M.$
        \end{corollary}
        \begin{corollary}\label{proof:corollary1}
         Under some mild conditions, using domain-wise training circuit datasets (i.e., ${M} > 1$) will result in a smaller generalization error bound than just pooling them into one mixed dataset (i.e., ${M} = 1$).
        \end{corollary}
        As a result, we easily conclude the aforementioned two principles for the circuit aggregation mechanism $g$ inspired by Corollaries \ref{proof:corollary3} and \ref{proof:corollary1}. We defer more discussion and detailed proof to Appendixes \ref{appendix:corollary3_proof} and \ref{appendix:corollary1_proof}.

    \subsubsection{Knowledge-Driven Subgraph Representation}\label{method:kdlgr}
        Under the formulation in Section \ref{method:cog_formulation}, we are given $M$ aggregated Circuit Datasets for learning a classifier that can well generalize to unseen circuits.
        
        \textbf{Covariate Shift Assumption} In our case, we assume the data exhibits covariate shift, i.e., the marginal distributions $\mathbb{P}_X$ on different circuits vary, but the functional relationship $\mathbb{P}_{Y \vert X}$ across different circuit domains is stable. The assumption holds in our problem as whether a node-level transformation is effective is stable across different circuits, i.e., the label generation mechanism keeps unchanged. Note that the covariate shift assumption is commonly seen in many OOD generalization problems from computer vision and natural language process as well \cite{muandet2013domain,shen2021towards}.         
        
        Under the covariate shift assumption, previous work \cite{muandet2013domain, albuquerque2019adversarial} has theoretically and/or empirically shown that \textit{learning domain-invariant representations} can well generalize to unseen domains. Thus, we propose a novel knowledge-driven subgraph representation (KDSR) learning method to learn domain-invariant representations based on the transformation-invariant domain knowledge. Our key observation called the transformation-invariant domain knowledge is that the node-level transformation mechanism is invariant across circuits for a given LS operator. More specifically, whether a node-level transformation is effective is significantly associated with the local subgraph rooted at the node, irrespective of the circuit to which it pertains. We defer more discussion on the transformation-invariant domain knowledge to Appendix \ref{appendix:discuss_transformation_invariant}.
        

        Based on the transformation-invariant domain knowledge, KDSR effectively aligns node embeddings across circuits from different domains by focusing on the constructed subgraphs rooted at each node in the DAGs to learn node embeddings. In general, the construction of a subgraph in LS operators follows heuristic rules, and the subgraph comprises the root node and a restricted number of its neighboring nodes. For further node embedding alignment, we transform the subgraph into a bipartite graph by modeling the root node and non-root nodes as two classes of nodes (see Fig. \ref{fig:cog}). 
        As a result, transformed bipartite graphs at each node share a high degree of structural similarity, which is beneficial for node embedding alignment. Moreover, we incorporate the semantic information of nodes, i.e., functionality information, into node features to learn discriminative embeddings (see Appendix \ref{appendix:node_features} for node features).
                    
        To encode the bipartite graph, we leverage graph neural networks (GNNs) \cite{hamilton2020graph}, which have been widely applied to applications with graph-structured inputs \cite{duvenaud2015convolutional, kipfsemi, shi2023lmc}. Specifically, we propose to leverage a graph convolutional neural network (GCNN) \cite{gori2005new, kipfsemi, gasse2019exact}. Our GCNN takes as input the bipartite graph $\textbf{x}=(\mathbf{T}, \mathbf{C}, \mathbf{A})$, where $\mathbf{T}\in \mathbb{R}^{1 \times c}$ denotes the feature matrix of the root node, $\mathbf{C}\in \mathbb{R}^{m \times c}$ denotes the feature matrix of the non-root nodes, and $\mathbf{A}\in \mathbb{R}^{1 \times m}$ denotes the adjacency matrix of the graph. In detail, our bipartite graph is a fully-connected graph. That is, $\mathbf{A}_{1j}=1$ for all $j\in\{1,2,\dots,m\}$. We manually design the node features to contain its basic and functionality information (see Appendix \ref{appendix:node_features}). Due to the bipartite structure of the input graph, our GCNN model $g_{\phi}$ performs a single graph convolution, in the form of two interleaved half-convolutions. Specifically, we break down our graph convolution into two successive passes, i.e., one from the root node to the non-root nodes and one from the non-root nodes to the root node. The passes take the form of
        \begin{align*}
            & \mathbf{h}(\mathbf{c}_i) \gets \mathbf{f}_{\mathcal{C}}(\mathbf{c}_i, \mathbf{g}_{\mathcal{C}} (\mathbf{c}_i, \mathbf{t}_1)), \\
            & \mathbf{h}(\mathbf{t}_1) \gets \mathbf{f}_{\mathcal{T}}(\mathbf{t}_1, \frac{1}{m} \sum_{i=1}^{m}  \mathbf{g}_{\mathcal{T}}(\mathbf{h}(\mathbf{c}_i),\mathbf{t}_1)),
        \end{align*}
        for all $i\in \{1,\dots,m\}$, where $\mathbf{f}_{\mathcal{C}},\mathbf{g}_{\mathcal{C}}, \mathbf{f}_{\mathcal{T}},$ and $\mathbf{g}_{\mathcal{T}}$ are two-layer perceptrons with relu activation functions, $\mathbf{t}_1$ denotes the root node feature, and $\mathbf{c}_i$ denotes the $i$-th non-root node feature. Following this graph-convolution layer, we obtain a bipartite graph with the same topology as the input, but with the root node embedding $\mathbf{h}(\mathbf{t}_1)=g_{\phi}(\textbf{x},\mathbf{t}_1)$, which contains rich information from the non-root nodes for discriminative and generalizable classification.

        \textbf{Discussion on Advantages of KDSR}       
        In contrast to employing GNNs directly for learning node embeddings over the global DAG, our KDSR concentrates on the utilization of constructed subgraphs (i.e., bipartite graphs) rooted at each node to learn node embeddings, leading to two major advantages. First, KDSR leverages the transformation-invariant domain knowledge for node embedding learning, which is significant for generalization to unseen circuits. Second, KDSR focuses on small-scale subgraphs rather than large global graphs, enabling efficient parallel training and high scalability to very large-scale circuits. 

    \subsubsection{Domain-Aware Distributional Classifier}\label{method:ds}
        Based on the node embeddings given by the GCNN model in Section \ref{method:kdlgr}, 
        we further propose a domain-aware distributional classifier (DADC), which well incorporates the domain-specific information into parameterized models. 
        Specifically, we parameterize the DADC via a multi-head neural network, where each head $f_{\theta}^{k}$ learns a classifier under the corresponding training domain. The multi-head neural network is a shared neural network architecture with $M$ heads branching off independently as shown in Fig \ref{fig:cog}. Thus, each head of the multi-head network contains its corresponding domain-specific information.
        The optimization objective for our proposed DADC takes the form of        \begin{align}\label{obj:binary_classification_for_each_head}
            \mathcal{\hat{R}}(\theta) = \frac{1}{M} \sum_{k=1}^{M} \frac{1}{n_k} \sum_{j=1}^{n_k} l(f_{\theta}^k(g_{\phi}(\textbf{x}_j^k, \mathbf{t}_1)), y_j^k),
        \end{align}
        where $l$ denotes the cross-entropy loss in (\ref{obj:binary_classification}), $f_{\theta}^k(\cdot)$ denotes the output of the $k$-th head, and $g_{\phi}(\textbf{x}_j^k, \mathbf{t}_1)$ denotes the node embedding given by the GCNN model mentioned in Section \ref{method:kdlgr}.
        Please see Appendix \ref{appendix:details_cog} for more details. 

        However, it is unclear how to apply DADC to unseen testing circuits, as the testing Circuit Domain information is unavailable. For simplicity, we use the \textit{mean} of the $M$ head values to approximate the classification values under testing circuits. A major advantage of DADC is to learn a distributional representation of classifiers, which can well capture the uncertainty of classification values to enhance its robustness against the distribution shift \cite{pets, tqc, cmbac}. 

\section{Experiments}\label{sec:exps}

    We conduct extensive experiments to evaluate PruneX with COG (PruneX-COG). Specifically, our experiments have four main goals: 
    (1) to demonstrate that PruneX-COG can accurately predict effective nodes, and  significantly improve the runtime of Logic Synthesis (LS) operators with comparable optimization performance (i.e., Quality of Results, QoR); (2) to show that the effectiveness of PruneX-COG on industrial circuits and very large-scale circuits (up to twenty million nodes); 
    (3) to demonstrate that PruneX-COG can not only achieve faster runtime but also improve the QoR; (4) to present a detailed ablation study of PruneX-COG. For reproducibility, we release our code in the repository \url{https://github.com/MIRALab-USTC/AI4LogicSynthesis-PruneX}.
    
    \textbf{Benchmarks}  
    We evaluate PruneX-COG on two widely-used public benchmarks and one industrial benchmark from Huawei HiSilicon\footnote{HiSilicon is a Chinese fabless semiconductor company wholly owned by Huawei. Please see \url{https://www.hisilicon.com/en/}.}. These benchmarks consist of 69 circuits in total, including very large-scale circuits with up to twenty million nodes. In terms of open-source benchmarks, we use the EPFL \cite{amaru2015epfl} and IWLS \cite{albrecht2005iwls} benchmarks, which are commonly used in previous work on LS \cite{drills, review_ai4lo, rai2021logic, aisyn}. 
    In terms of the industrial benchmark, we use 27 circuits from Huawei HiSilicon. We defer more details to Appendix \ref{appendix:three_benchmarks}.    

    \textbf{Experimental setup} 
    Throughout all experiments, we use ABC \cite{abc} as the backend LS framework, which is a state-of-the-art open source LS framework, and is widely used in research of machine learning for LS \cite{drills,aisyn, review_ai4lo, lsoracle, neto2021read, ren2023machine}. 
    In this paper, we apply PruneX to the Resub \cite{resub} and Mfs2 \cite{mfs2} operators to demonstrate that our method is applicable to many LS operators. Note that applying the two operators take the longest runtime compared to other commonly used operators as shown in Fig. \ref{fig:motivation_ineffective_node}.
    We train our method with ADAM \cite{adam} using the PyTorch \cite{torch}. 
    We provide more details in Appendix \ref{appendix:exp_setup}. 
    
    \subsection{Evaluation Metrics and Evaluated Methods}
    Throughout all experiments, we evaluate our method in two separate phases, i.e., the offline and online phases.
    
    In the offline phase, we evaluate the prediction recall of the effective nodes. 
    We empirically show that the QoR improves with increased prediction recall in Appendix \ref{appendix:recall_qor}. Thus, it is important to achieve high recall for comparable QoR with the default operators. 
    Specifically, we present details as follows. (1) \textbf{Evaluation metrics} Due to the severe imbalance of positive and negative samples in the classification task, the learned classifier will suffer from negative bias. Therefore, 0.5 is an inappropriate threshold to predict whether a node is positive, while determining an appropriate threshold is challenging.
    To alleviate this problem, PruneX view the prediction task as a scoring task, and predict nodes with top $k$ scores to be positive. Under this prediction, we define a \textbf{top $k$ accuracy metric} by the fraction of true positive nodes that are predicted to be positive, i.e., recall.
    We defer details on this metric to Appendix \ref{appendix_top_k_acc}. In this paper, we mainly use the top $50\%$ accuracy metric, and \textbf{use the terms top $k$ accuracy and prediction recall interchangeably}.
    (2) \textbf{Evaluated methods} In the offline phase, we evaluate two baselines and our PruneX-COG. PruneX-COG is our proposed circuit domain generalization framework (see Section \ref{method:cog}). The baselines include Random, which randomly predict a score between $[0,1]$ for each node, and EnsembleMLP, which is our proposed simple learning-based baseline.
    EnsembleMLP uses a simple multi-layer perceptron to predict scores for nodes (see Appendix \ref{appendix:ensemblemlp}). 
    
    In the online phase, we evaluate the efficiency and QoR of PruneX-COG. Specifically, we present details as follows. (1) \textbf{Evaluation metrics} In terms of efficiency, we use the runtime metric. In terms of QoR, we mainly use the size, i.e., the number of nodes of optimized circuits, which has a significant impact on the chip area. Moreover, we also use the depth (i.e., level) of optimized circuits, which is a proxy metric for the delay of the designed chip. Throughout all experiments, we found that our method achieves the same optimization performance in terms of the depth with that of the default operators on most circuits. Thus, we report results in terms of the size in the main text, and provide results in terms of the depth in Appendix \ref{appendix:exps}. (2) \textbf{Evaluated methods} We evaluate the Default and PruneX-COG in the online phase. Default denotes the default operators (i.e., Resub and Mfs2) in ABC \cite{abc}. PruneX-COG denotes new operators that apply our method to the Default operators. We set the top $k$ hyperparameter as top $50\%$ for all experiments unless mentioned otherwise.
    
\begin{figure*}[t]
    \centering
    \begin{subfigure}{0.9\textwidth}
        \includegraphics[width=0.32\textwidth, height=0.255\textwidth]{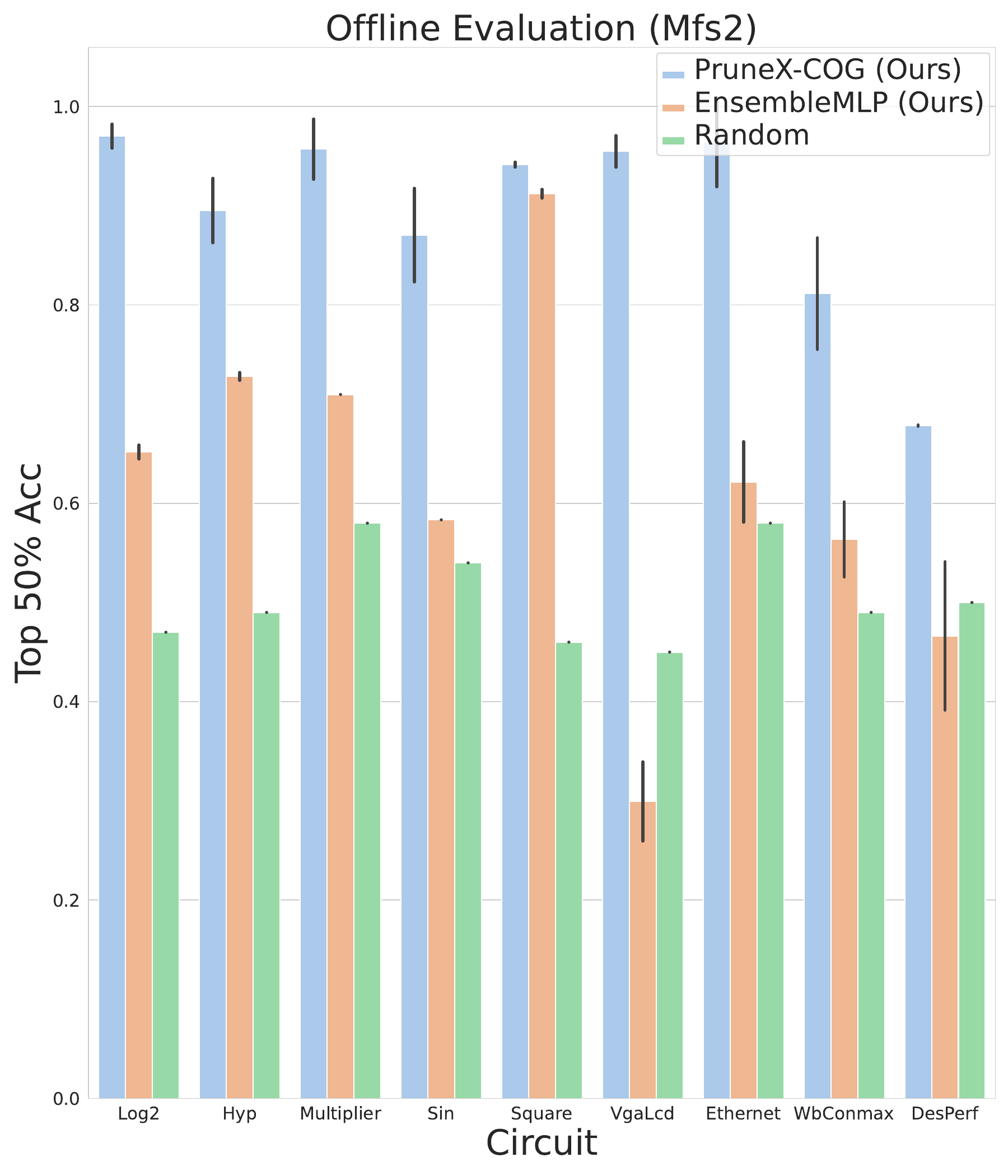}
        \includegraphics[width=0.32\textwidth, height=0.255\textwidth]{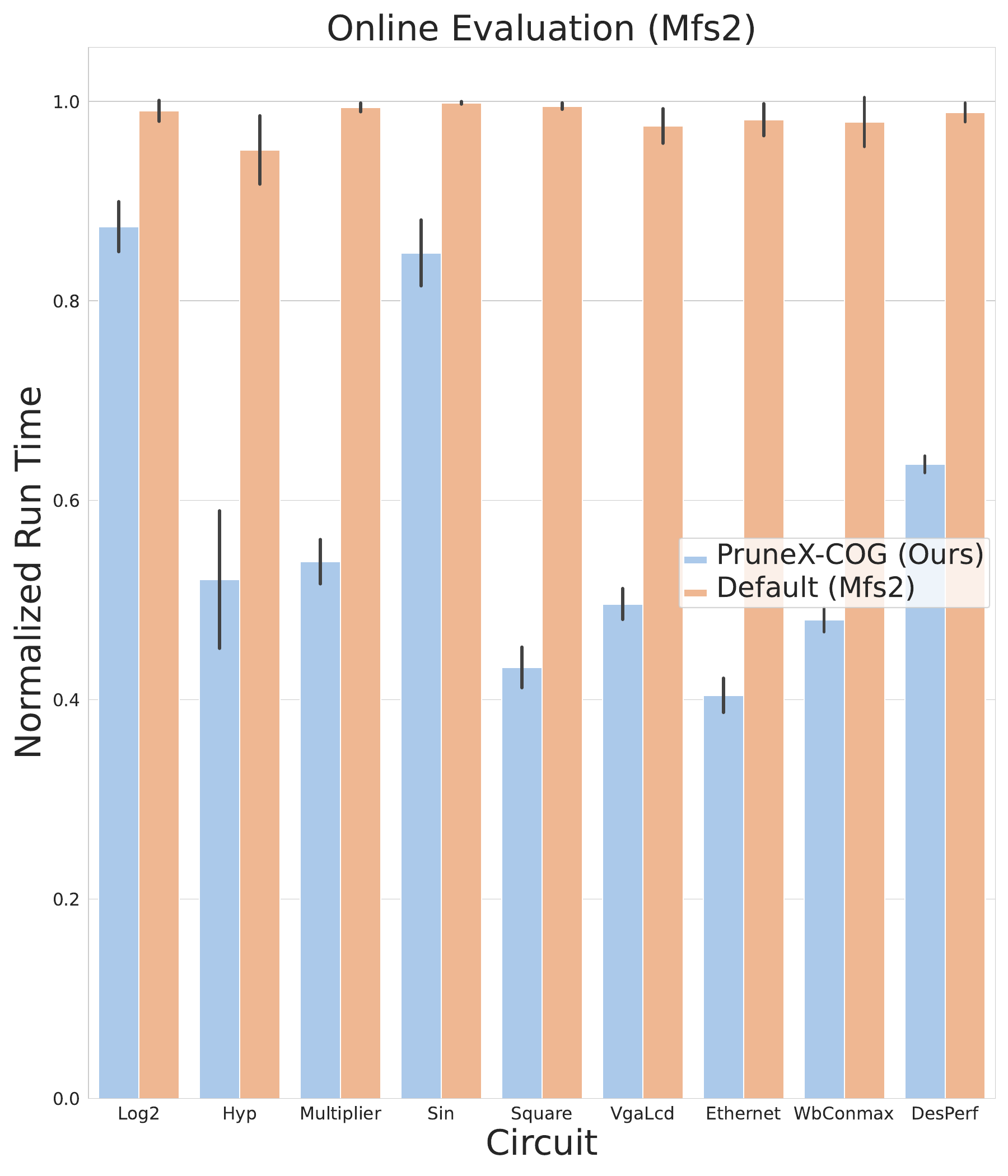}
        \includegraphics[width=0.32\textwidth, height=0.255\textwidth]{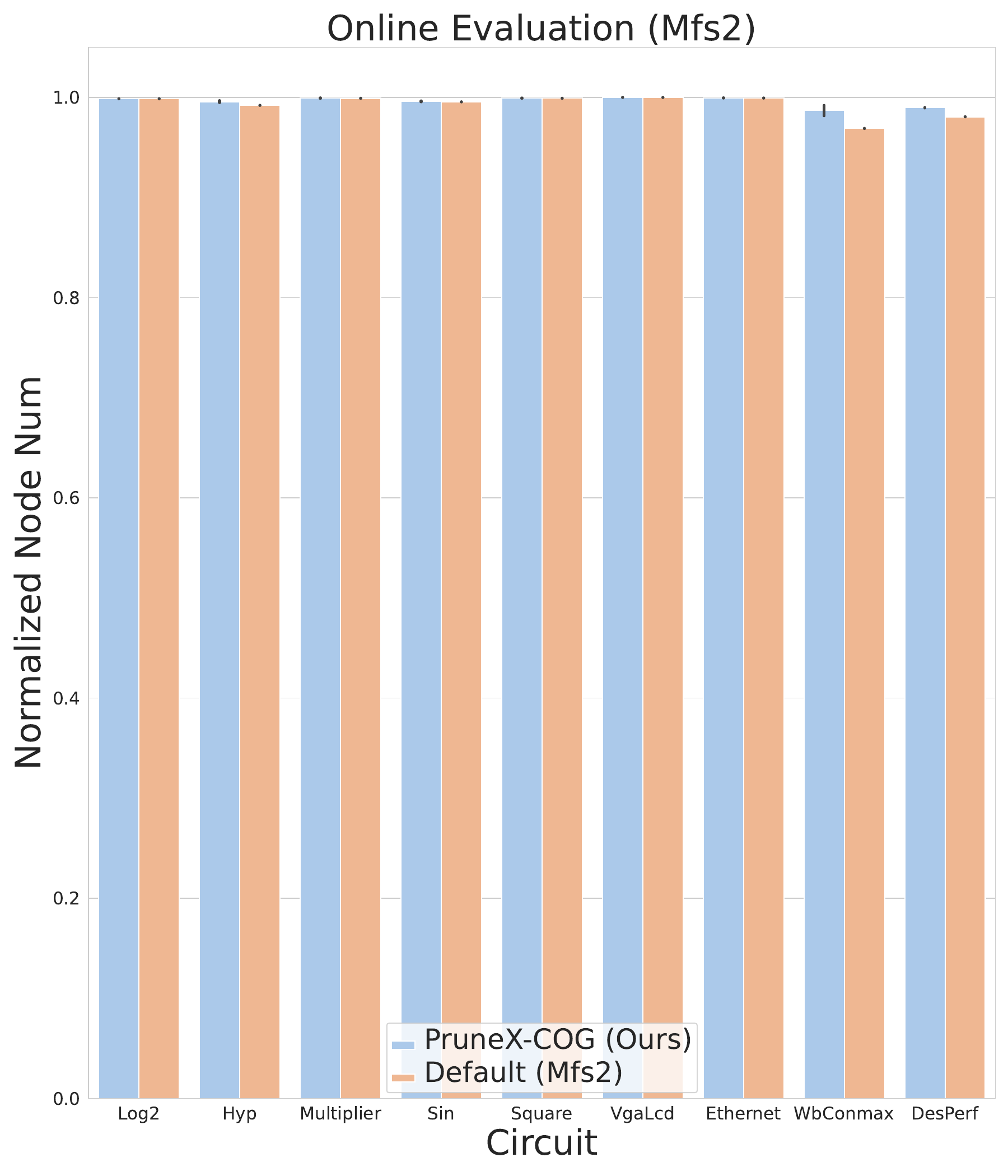}
        \vspace{-1.5mm}
        \caption{We evaluate our method under Evaluation Strategy 1 on the Mfs2 operator.}
        \label{fig:offline_evaluation_mfs2_EPFL_IWLS_acc_time_node}
    \end{subfigure}
    \begin{subfigure}{0.9\textwidth}
        \includegraphics[width=0.32\textwidth, height=0.255\textwidth]{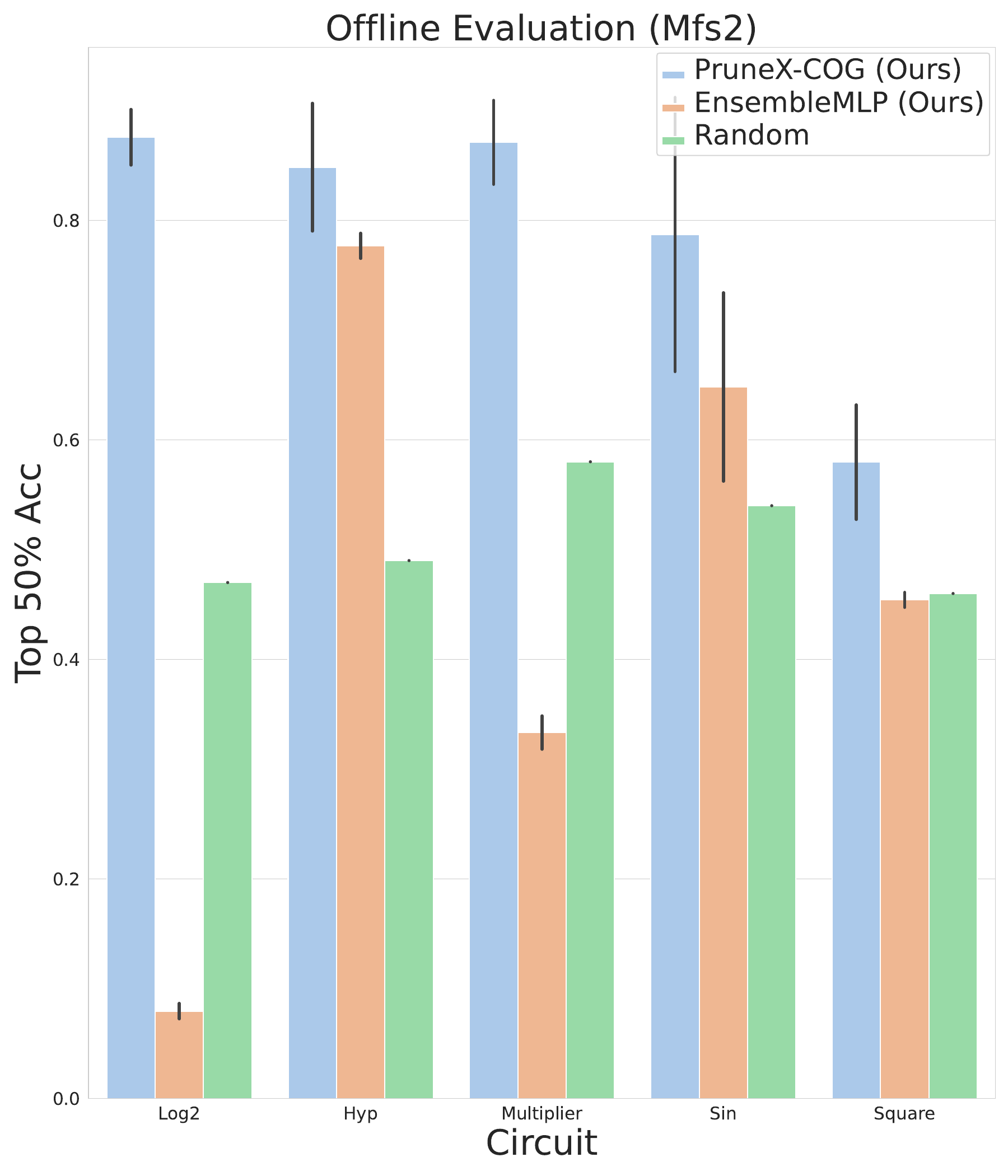}
        \includegraphics[width=0.32\textwidth, height=0.255\textwidth]{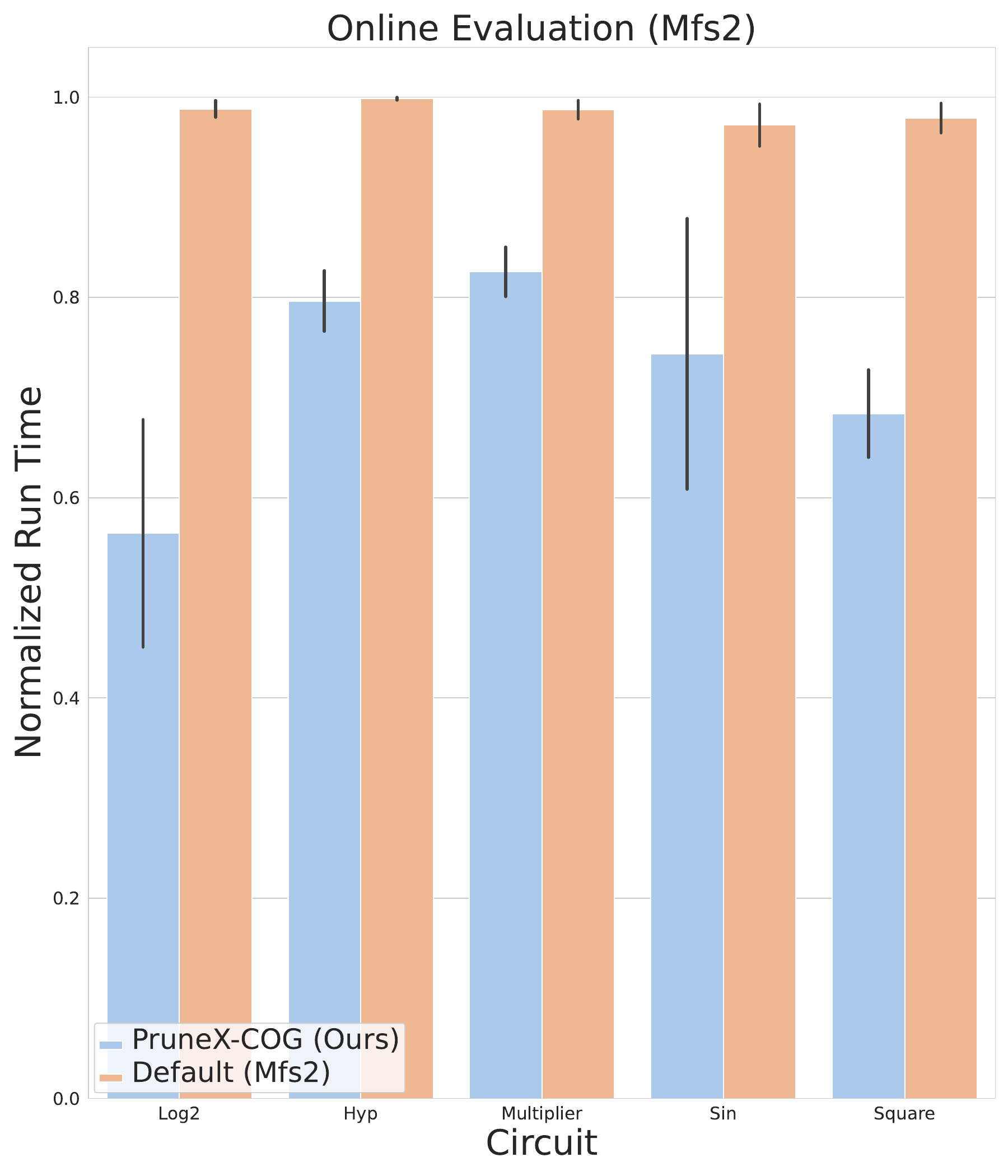}
        \includegraphics[width=0.32\textwidth, height=0.255\textwidth]{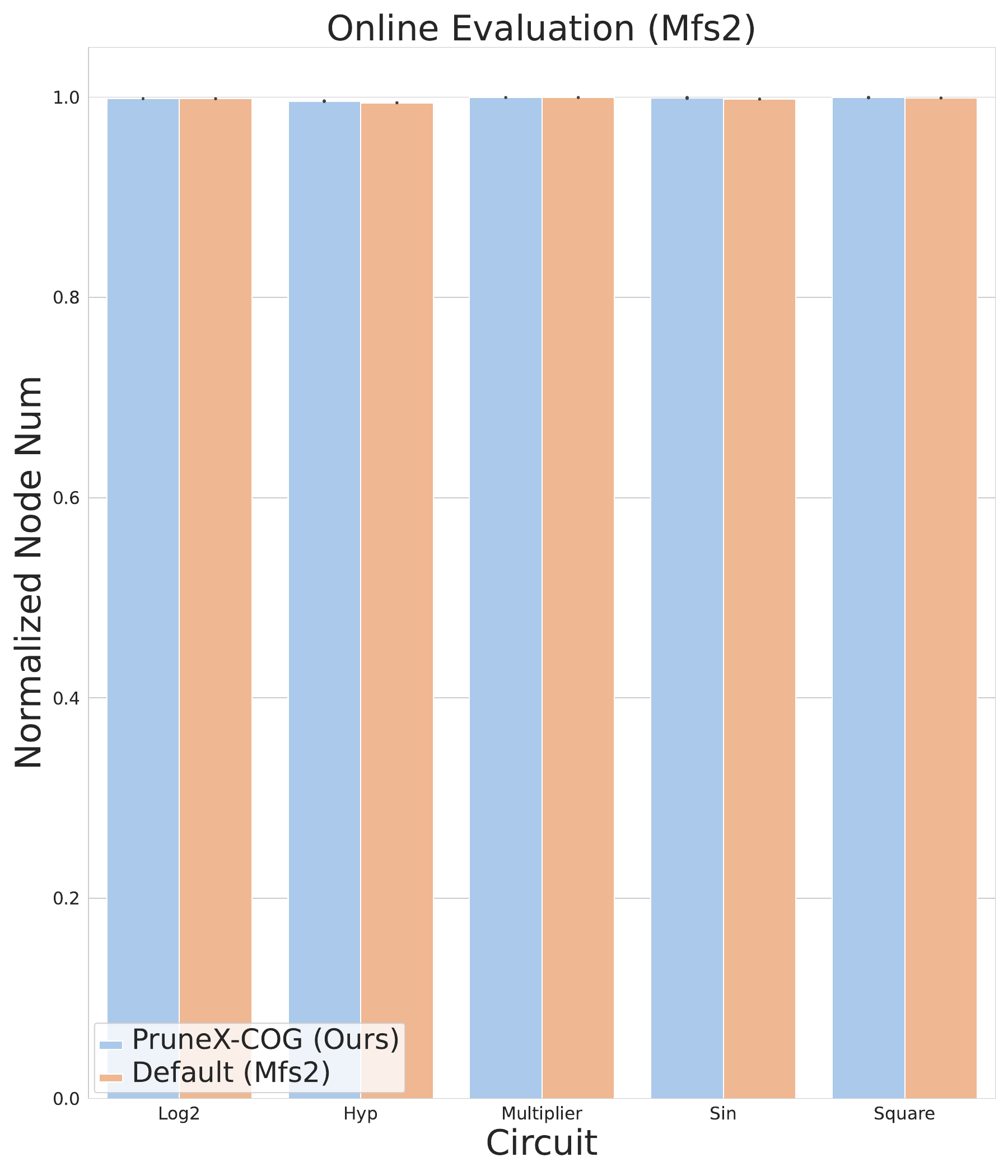}
        \vspace{-1.5mm}
        \caption{We evaluate our method under Evaluation Strategy 2 on the Mfs2 operator.}
        \label{fig:offline_evaluation_mfs2_iwls_to_epfl_acc_time_node}
    \end{subfigure}
    \begin{subfigure}{0.9\textwidth}
        \includegraphics[width=0.32\textwidth, height=0.255\textwidth]{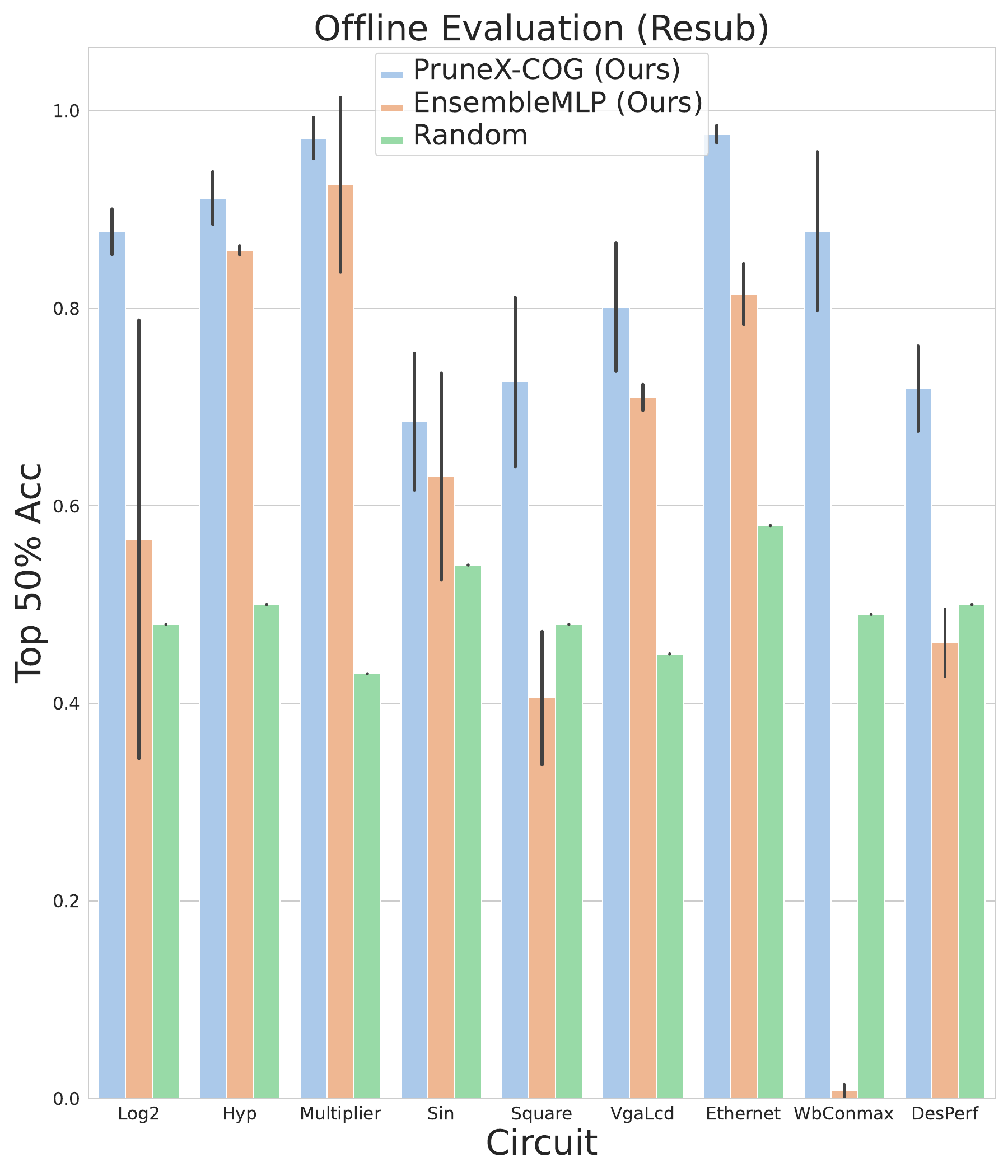}
        \includegraphics[width=0.32\textwidth, height=0.255\textwidth]{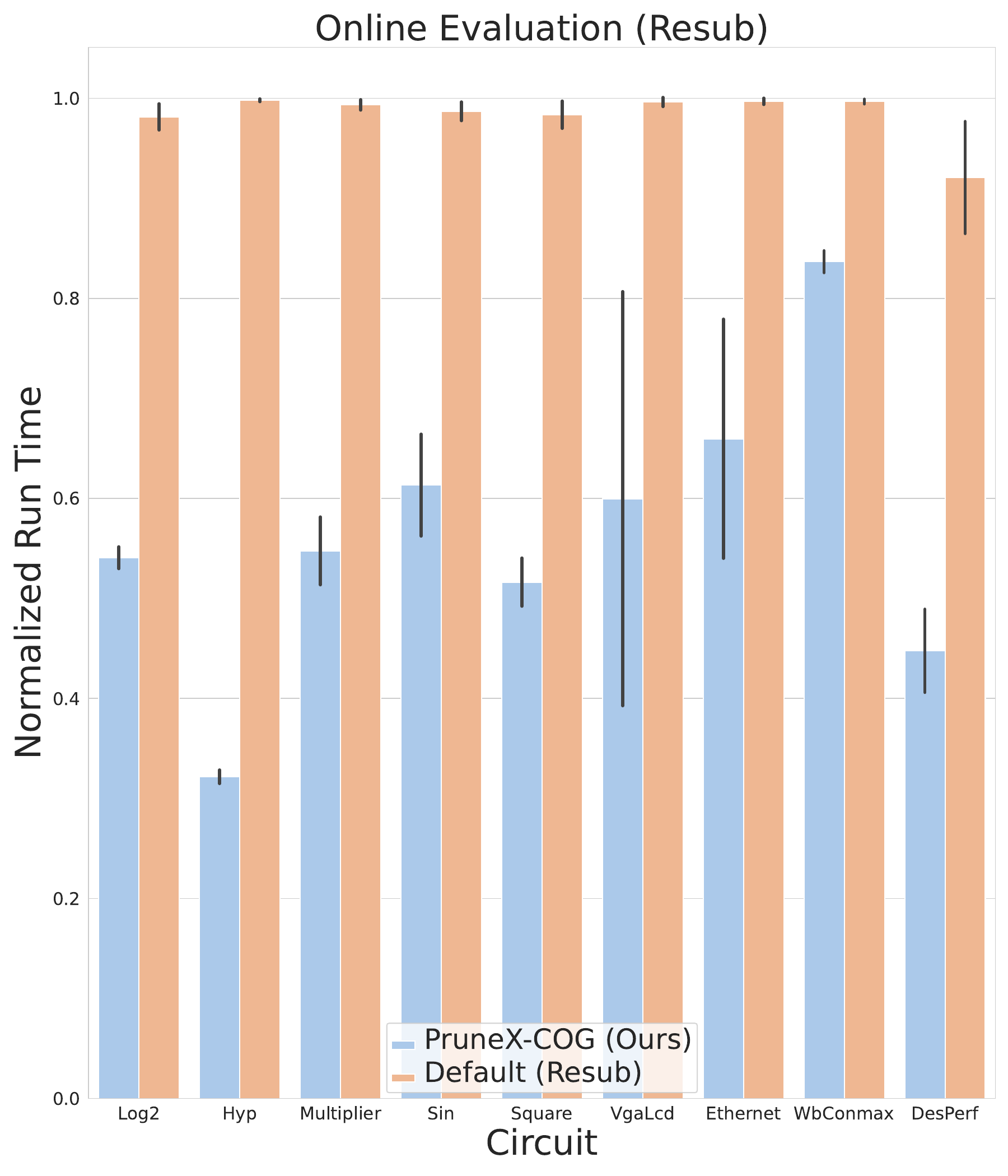}
        \includegraphics[width=0.32\textwidth, height=0.255\textwidth]{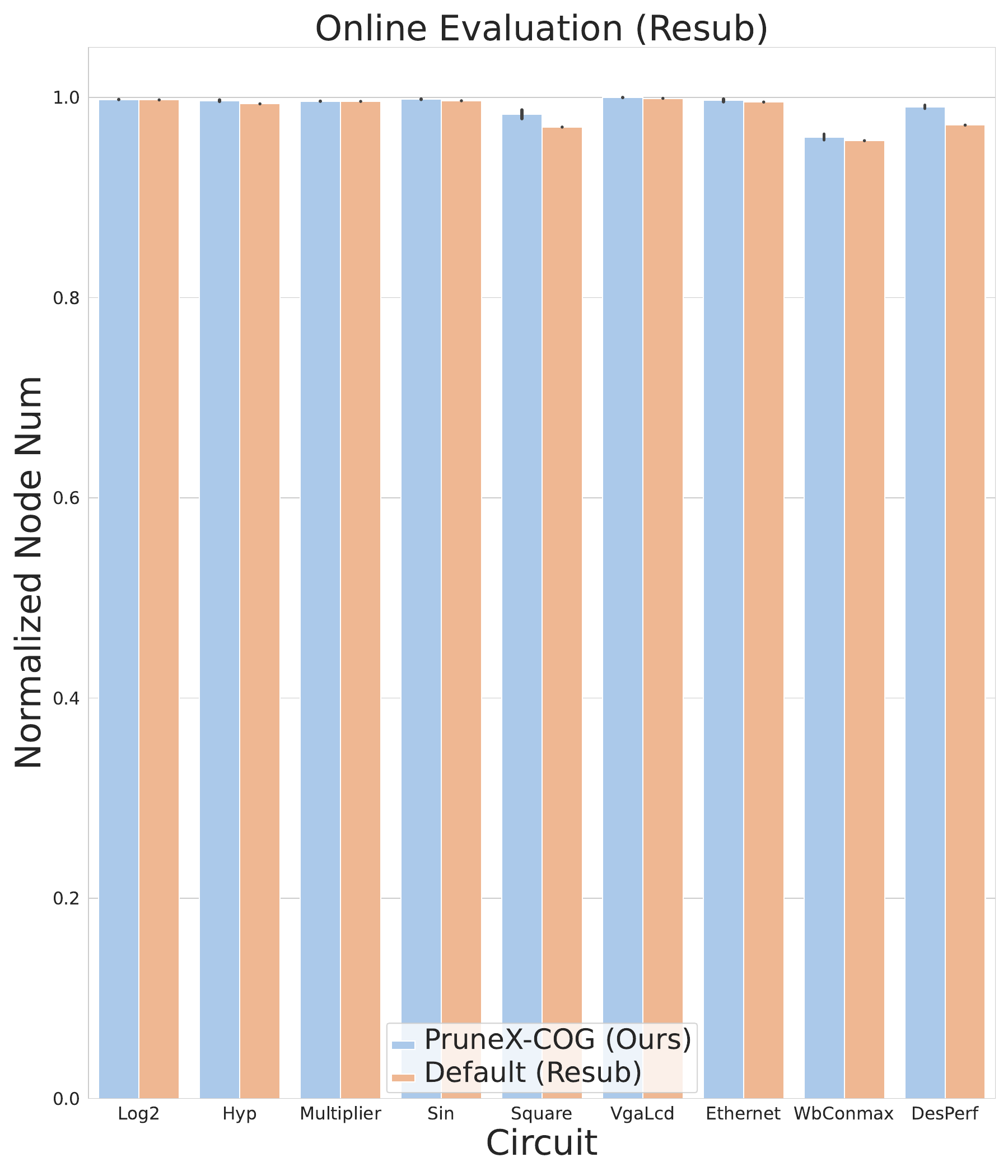}
        \vspace{-1.5mm}
        \caption{We evaluate our method under Evaluation Strategy 1 on the Resub operator.}
        \label{fig:offline_evaluation_resub_epfl_iwls_acc_time_node}
    \end{subfigure} 
    \begin{subfigure}{0.9\textwidth}
        \includegraphics[width=0.32\textwidth, height=0.255\textwidth]{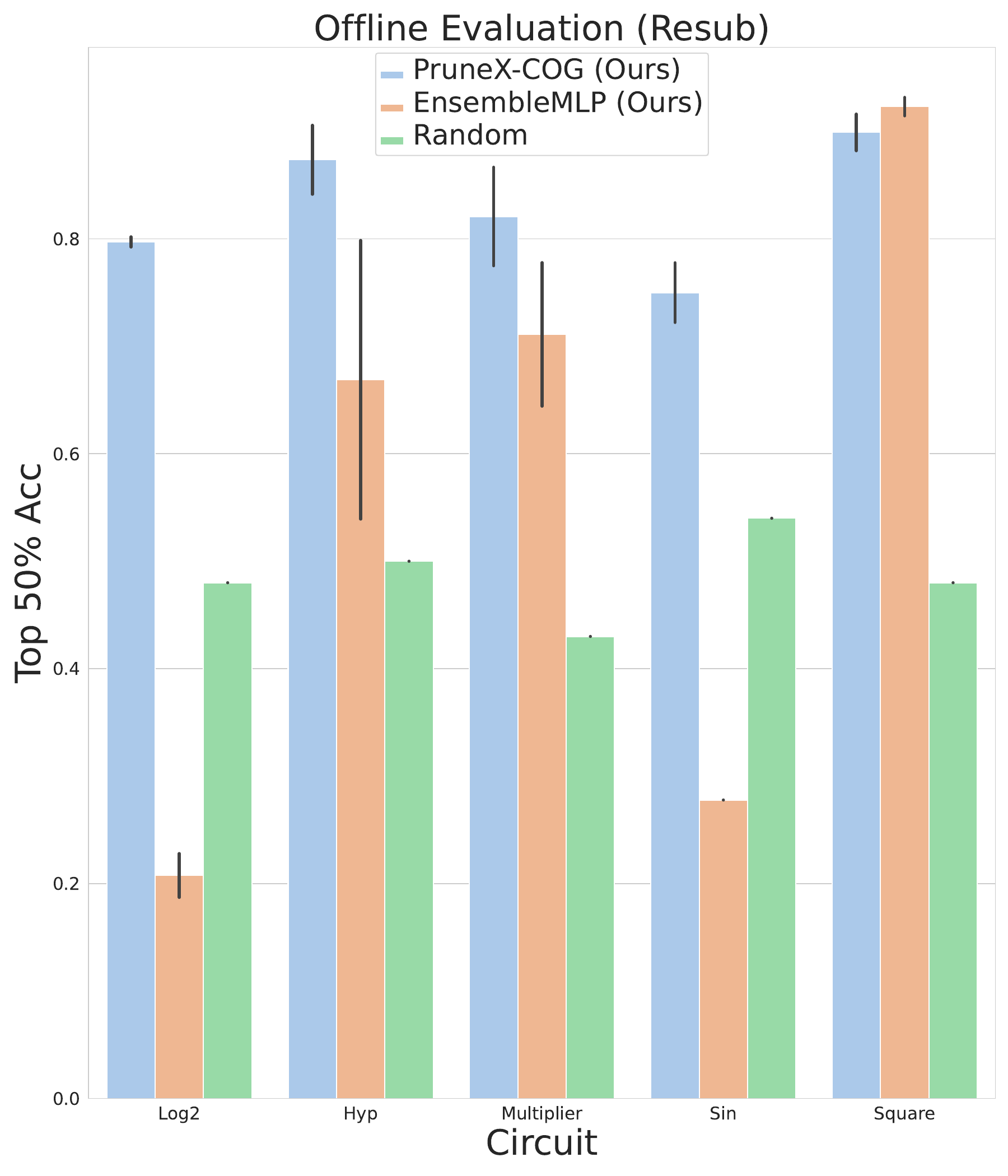}
        \includegraphics[width=0.32\textwidth, height=0.255\textwidth]{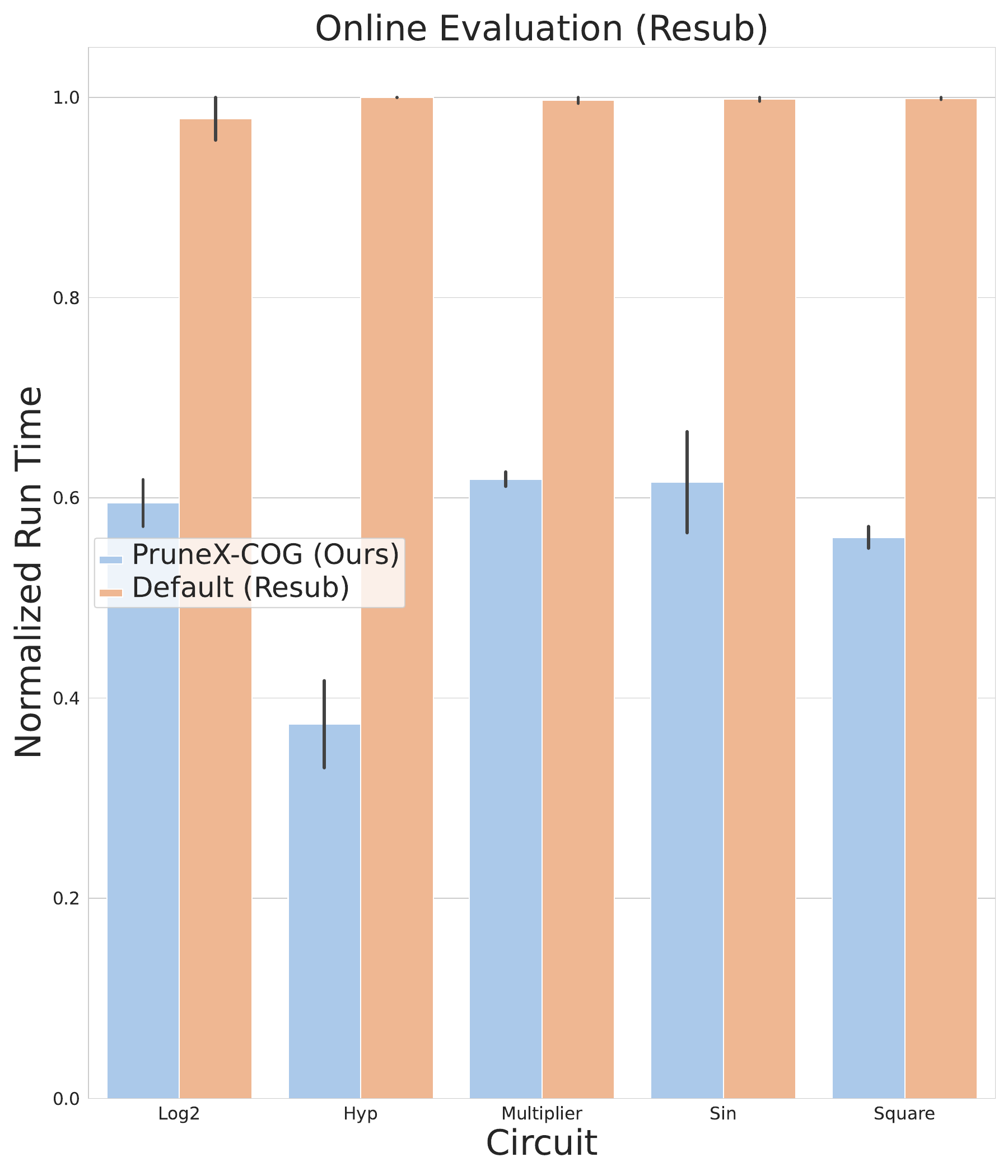}
        \includegraphics[width=0.32\textwidth, height=0.255\textwidth]{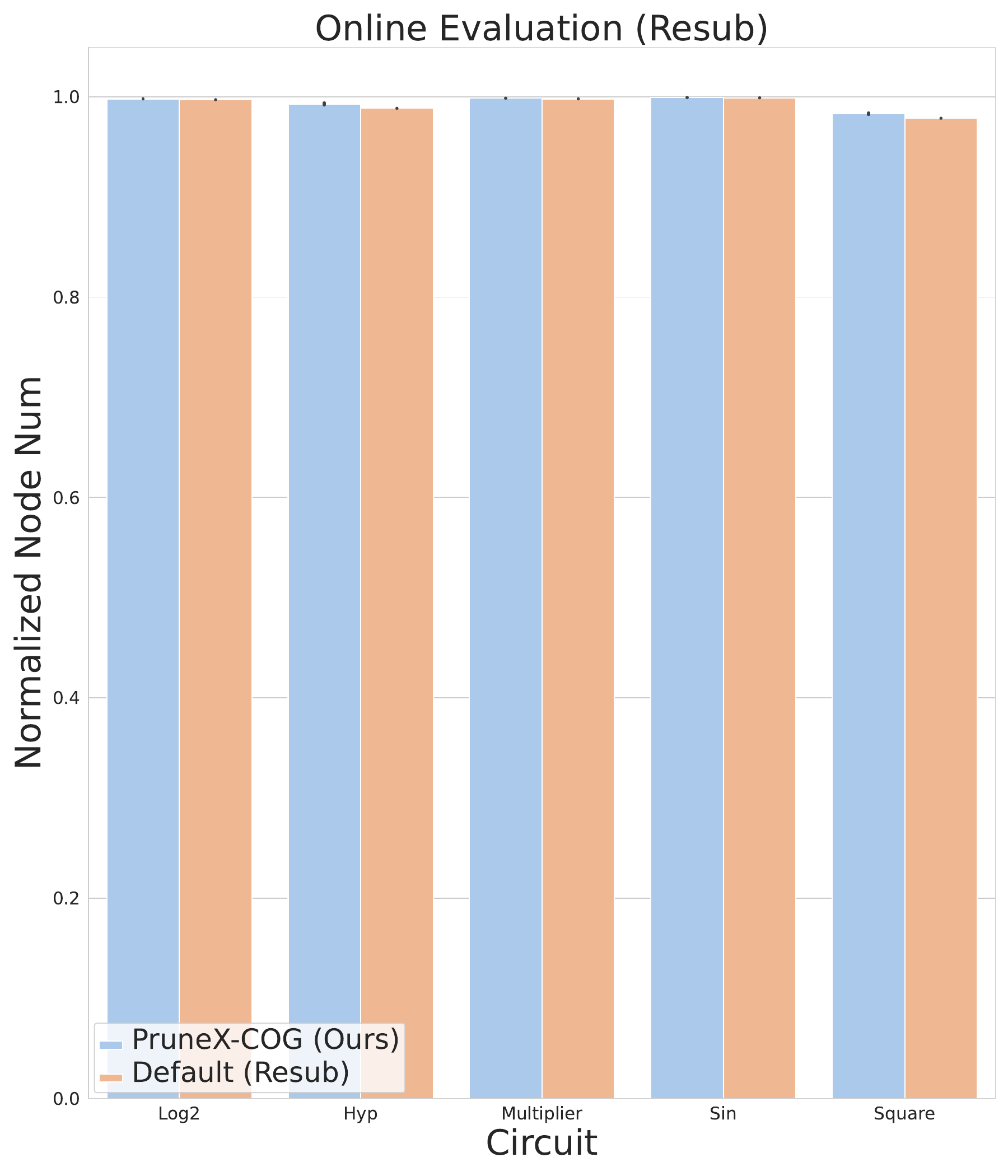}
        \vspace{-1.5mm}
        \caption{We evaluate our method under Evaluation Strategy 2 on the Resub operator.}
        \label{fig:offline_evaluation_resub_iwls_to_epfl_acc_time_node}
    \end{subfigure}
    \vspace{-2.5mm}
    \caption{The results demonstrate that our PruneX-COG achieves significant prediction recall improvement (Left, $\uparrow$), runtime reduction (Middle, $\downarrow$), and marginal QoR (size) degradation (Right, $\downarrow$). The normalized runtime (node number) denotes the ratio of the runtime (node number) to that of applying the Default operator to testing circuits.}
    \label{fig:main_results1}
\end{figure*}

    \subsection{Evaluation on Open-Source Benchmarks}\label{results:open_source}
        In this subsection, we evaluate the offline prediction recall, online runtime, and online optimization performance on the open-source EPFL \cite{amaru2015epfl} and IWLS \cite{albrecht2005iwls} benchmarks. Due to limited space, we defer more detailed results to Appendix \ref{appendix:exps_open_source}. Specifically, we design two evaluation strategies, which are inspired by previous work and real industrial scenarios.
        
        \textbf{Evaluation Strategy 1: Generalization in Single Benchmark} Inspired by the leave-one-domain-out cross-validation strategy commonly used in previous literature \cite{wang2022generalizing, grannite}, we design nine leave-one-out datasets for evaluation based on the EPFL and IWLS benchmarks. Specifically, given a benchmark, we construct a dataset by setting one circuit as the testing dataset, and the other circuits as the training dataset. For example, we construct a Log2 dataset by setting Log2 as the testing dataset, and the other in the EPFL as the training dataset. 
        In this paper, we focus on testing on circuits that are time-consuming to optimize. For the EPFL benchmark, we use Log2, Hyp, Multiplier, Sin, and Square as the testing circuit, respectively. For the IWLS benchmark, we use Des Perf, Ethernet, Wb Conmax, Vga Lcd as the testing circuit, respectively. Please refer to Appendix \ref{appendix:datasets_open_source} for more details about the datasets.
        
        \textbf{Evaluation Strategy 2: Generalization from the IWLS to EPFL}
        In real industrial scenarios, we usually train a model on   circuits, hoping that the trained model can generalize to many unseen circuits. Thus, we design the second evaluation strategy. Specifically, we set the circuits from the IWLS as the training dataset, and the five hard-to-optimize circuits from the EPFL, i.e., Log2, Hyp, Multiplier, Sin, and Square, as the testing dataset. Compared with the first strategy, it is more challenging to achieve good generalization performance under the second strategy due to the larger distribution shift between the training and testing datasets. Due to limited space, please see Appendix \ref{appendix:datasets_open_source} for details.   
               
        For the \textbf{offline evaluation} on the Mfs2 operator, Figs. \ref{fig:offline_evaluation_mfs2_EPFL_IWLS_acc_time_node} and \ref{fig:offline_evaluation_mfs2_iwls_to_epfl_acc_time_node} show that PruneX-COG significantly improves the prediction recall compared with the Random baseline in both Evaluation Strategies. Specifically, PruneX-COG achieves $37\%$ and $28\%$ recall improvement on average under the Evaluation Strategy 1 and 2, respectively.
        Furthermore, PruneX-COG achieves the prediction recall surpassing $90\%$ on most testing circuits, indicating it can maintain applying most of the effective node-level transformations. Moreover, the results show that EnsembleMLP, i.e., our proposed simple learning-based baseline, struggles to consistently achieve high prediction recall on all circuits, demonstrating that the OOD generalization problem across circuits in LS is challenging.
        
        For the \textbf{online evaluation} on the Mfs2 operator, Figs. \ref{fig:offline_evaluation_mfs2_EPFL_IWLS_acc_time_node} and \ref{fig:offline_evaluation_mfs2_iwls_to_epfl_acc_time_node} show that PruneX-COG significantly improves the runtime 
        compared with the Default Mfs2 operator, achieving $40.96\%$ and $26.66\%$ improvement on average under the Evaluation Strategy 1 and 2, respectively. Moreover, PruneX-COG achieves marginal degradation in terms of the QoR. Specifically, PruneX-COG achieves $0.36\%$ and $0.07\%$ degradation on average in terms of the size of circuits under the Evaluation Strategy 1 and 2, respectively. In addition, PruneX-COG 
        does not degrade the depth of circuits (see Appendix \ref{appendix:exps_open_source} for results). 
        Overall, the results demonstrate that PruneX-COG significantly improves the efficiency of the Mfs2 operator while keeping comparable QoR.
            
        For the evaluation on the Resub operator, Figs. \ref{fig:offline_evaluation_resub_epfl_iwls_acc_time_node} and \ref{fig:offline_evaluation_resub_iwls_to_epfl_acc_time_node} show that PruneX-COG achieves significant prediction recall improvement, runtime improvement, and marginal size degradation, which are consistent with the evaluation results on the Mfs2 operator. Specifically, 
        PruneX-COG improves the recall by $35\%$ on average. PruneX-COG reduces the runtime by $54.22\%$ and degrades the size by $0.35\%$ on average. In particular, PruneX-COG achieves $3.1\times$ faster runtime on the Hyp circuit under Evaluation Strategy 1.
        The results suggest that our method is applicable to many LS operators, which can significantly improve their efficiency while keeping comparable QoR.
        
\begin{figure*}[t]
    \centering
    \begin{subfigure}{0.9\textwidth}
        \includegraphics[width=0.32\textwidth, height=0.255\textwidth]{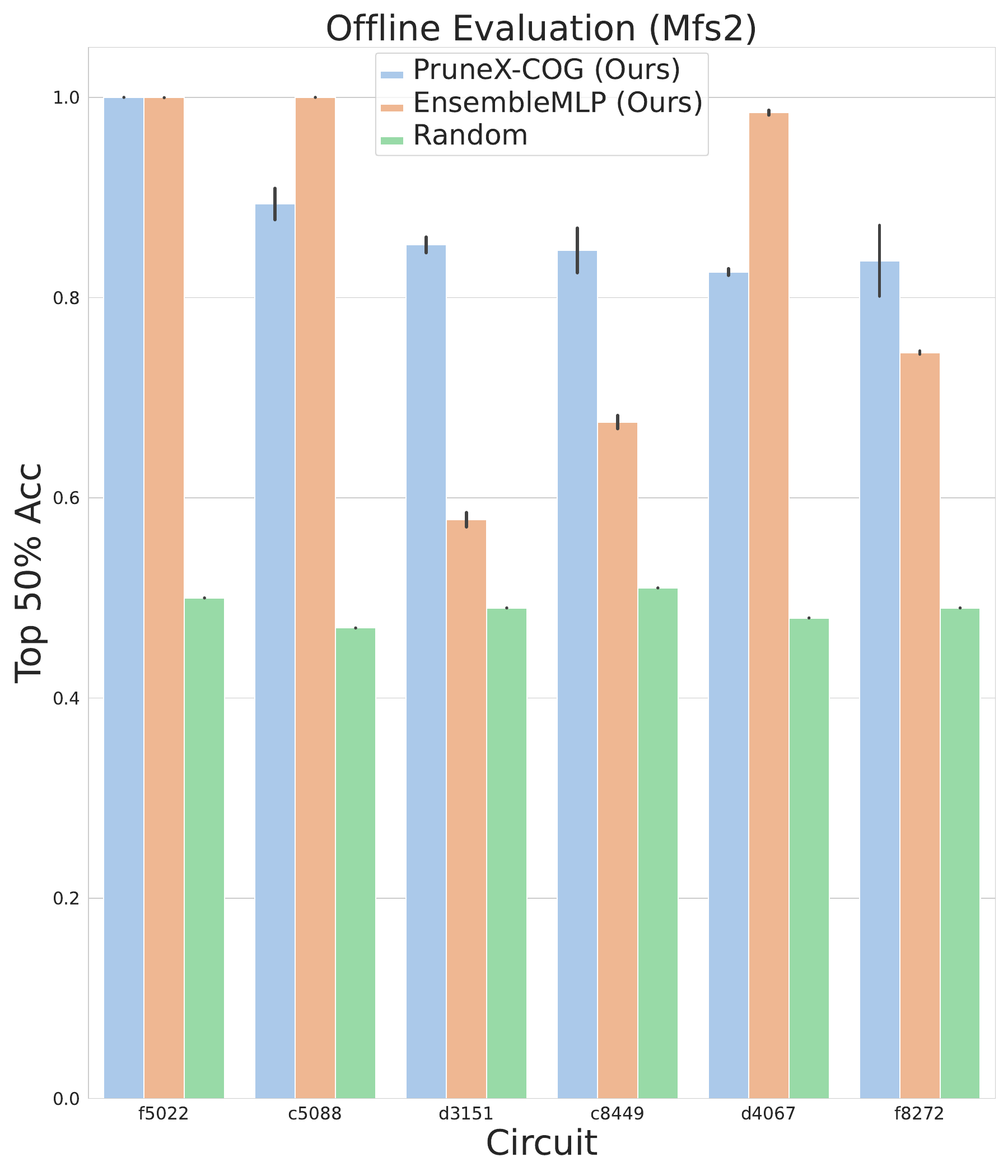}
        \includegraphics[width=0.32\textwidth, height=0.255\textwidth]{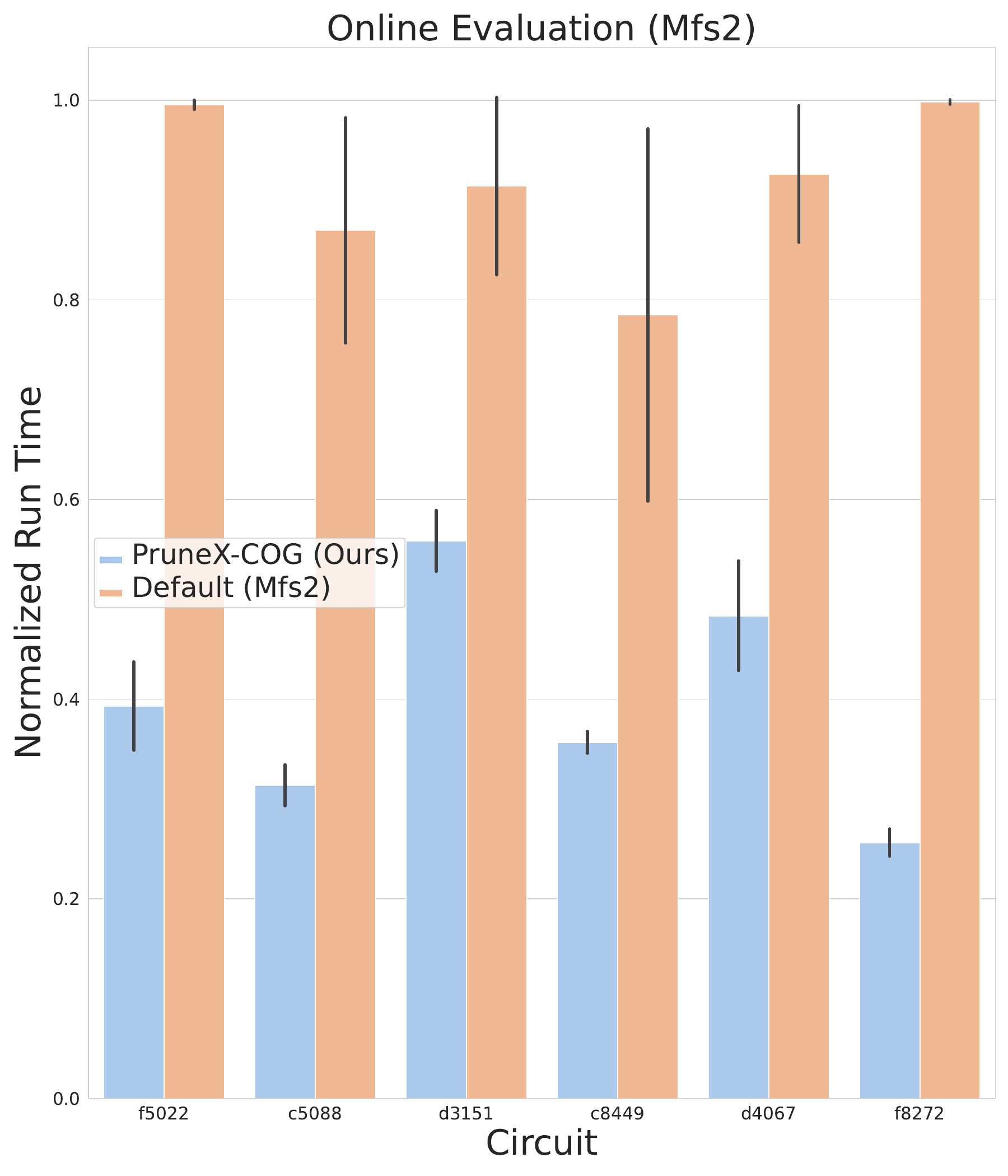}
        \includegraphics[width=0.32\textwidth, height=0.255\textwidth]{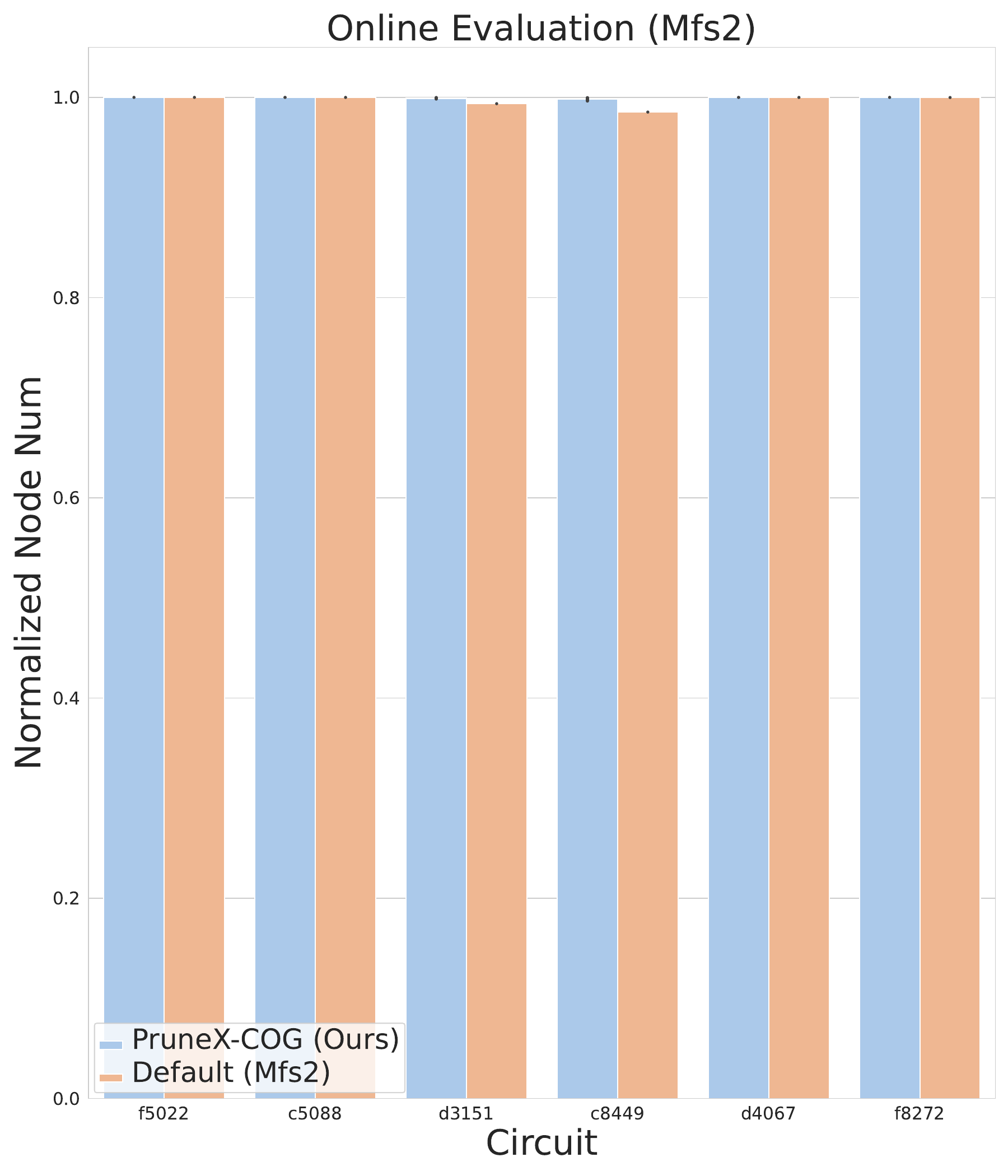}
        \vspace{-1.5mm}
        \caption{We evaluate our method on the industrial benchmark and Mfs2 operator.}
        \label{fig:offline_evaluation_mfs2_real_acc_time_node}
    \end{subfigure}
    \begin{subfigure}{0.9\textwidth}
        \includegraphics[width=0.32\textwidth, height=0.255\textwidth]{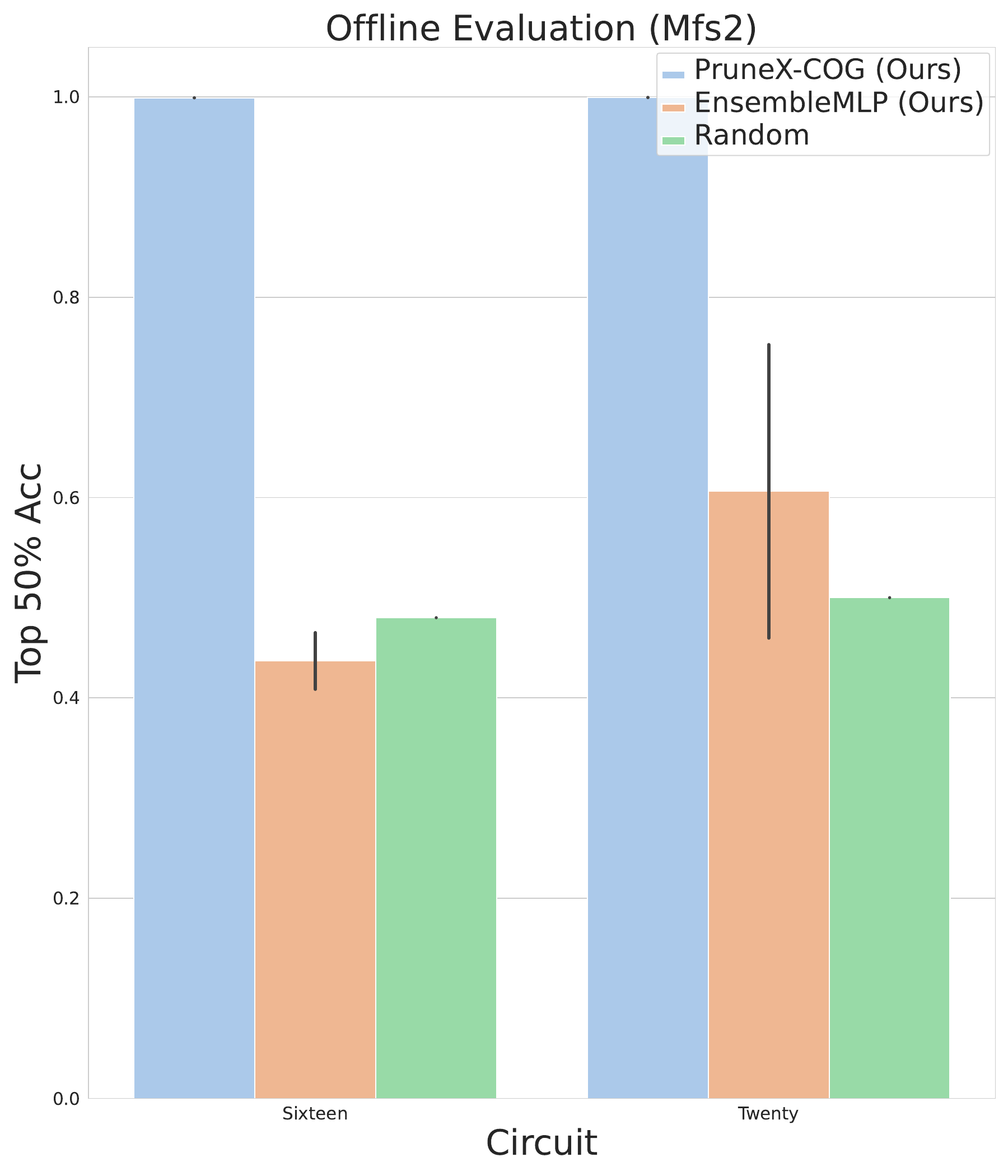}
        \includegraphics[width=0.32\textwidth, height=0.255\textwidth]{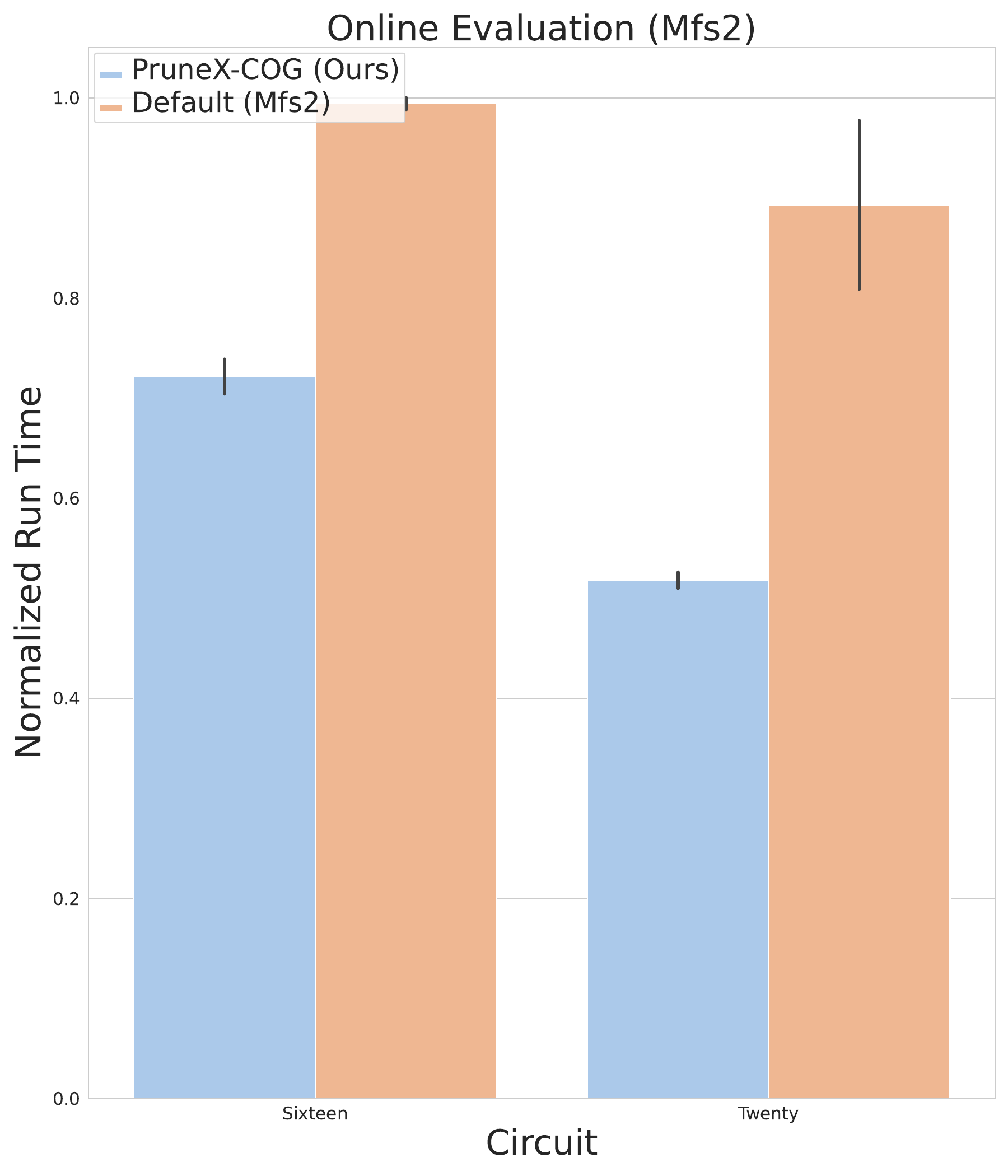}
        \includegraphics[width=0.32\textwidth, height=0.255\textwidth]{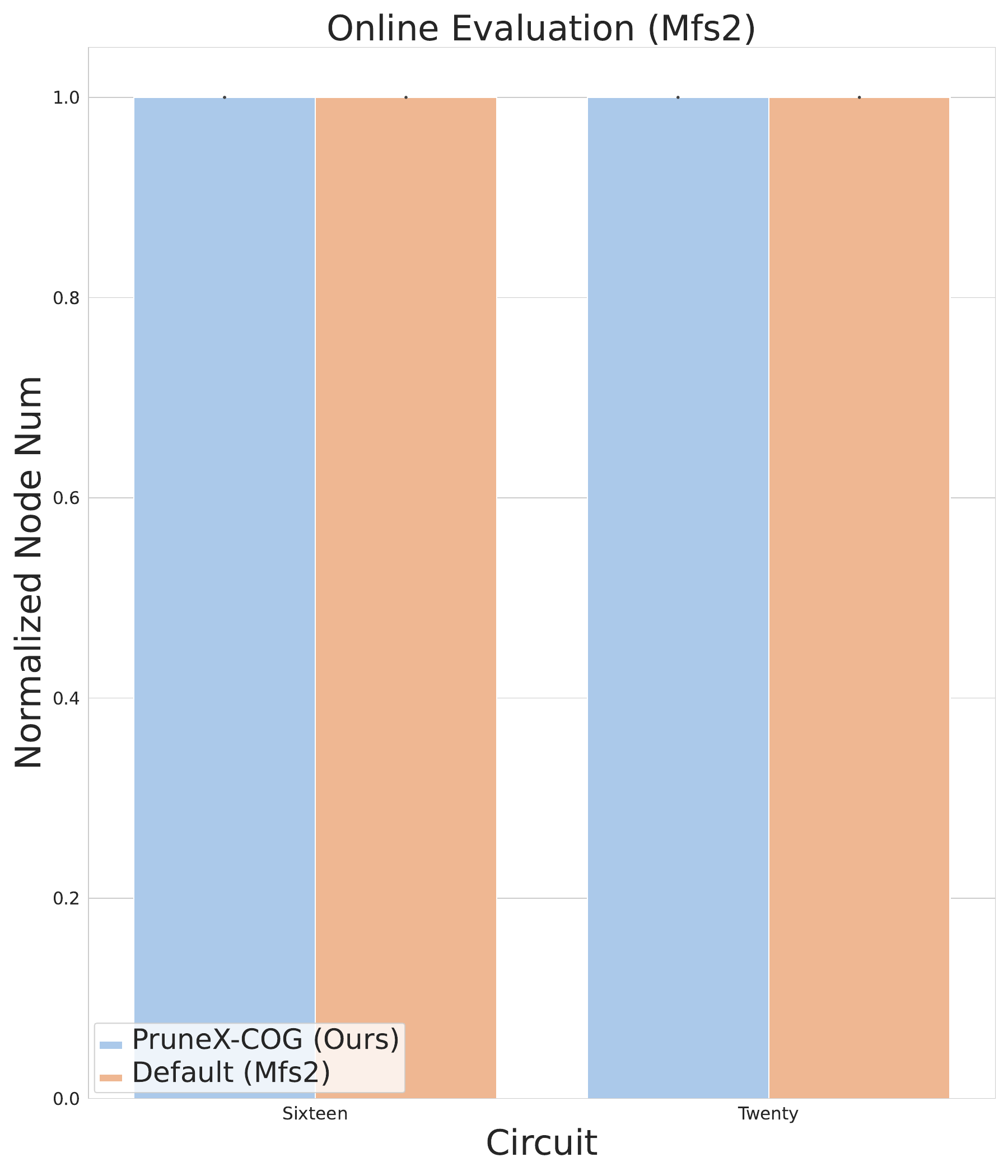}
        \vspace{-1.5mm}
        \caption{We evaluate our method on the very large-scale circuits and Mfs2 operator.}
        \label{fig:offline_evaluation_mfs2_epfl_hard_acc_time_node}
    \end{subfigure}
    \begin{subfigure}{0.9\textwidth}
        \includegraphics[width=0.32\textwidth, height=0.255\textwidth]{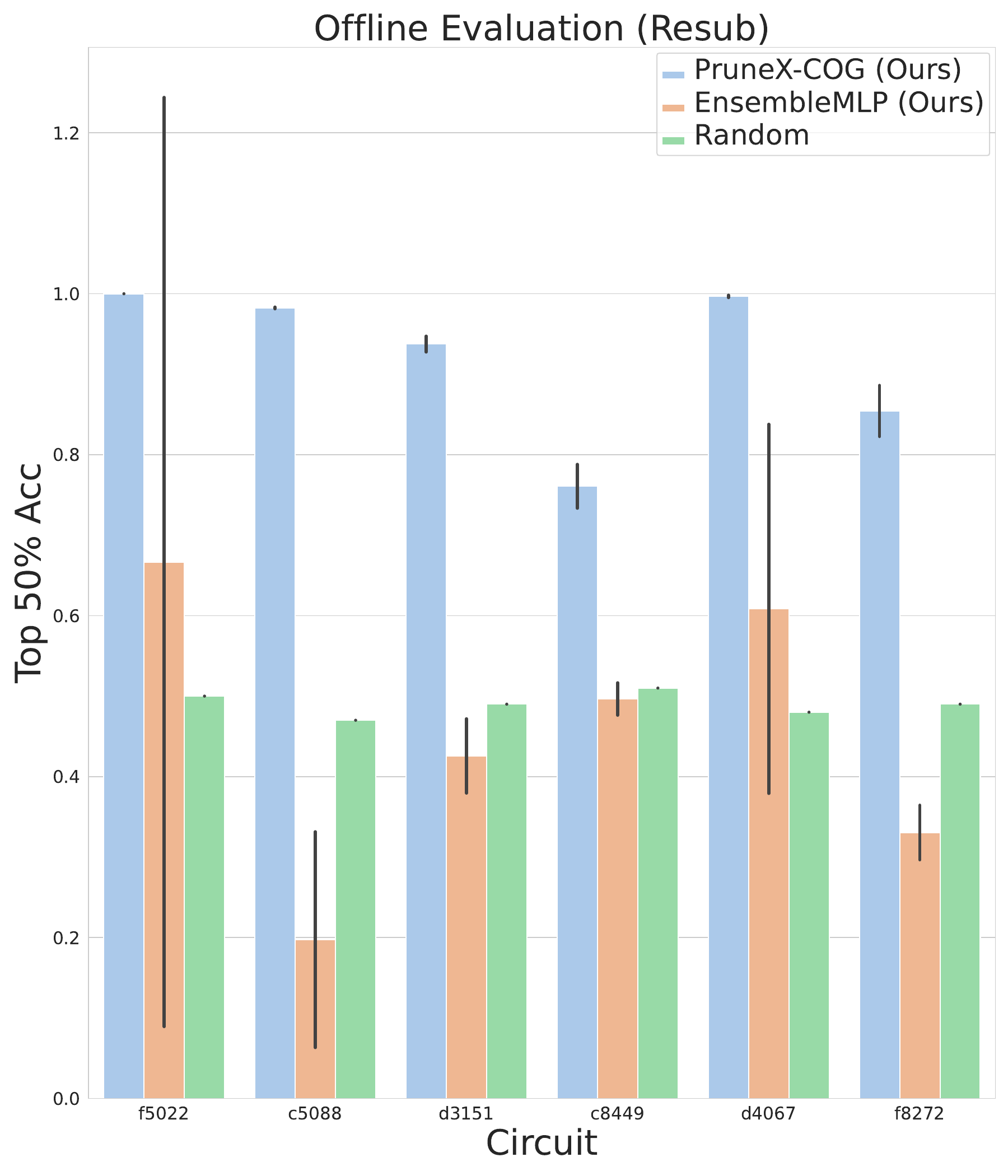}
        \includegraphics[width=0.32\textwidth, height=0.255\textwidth]{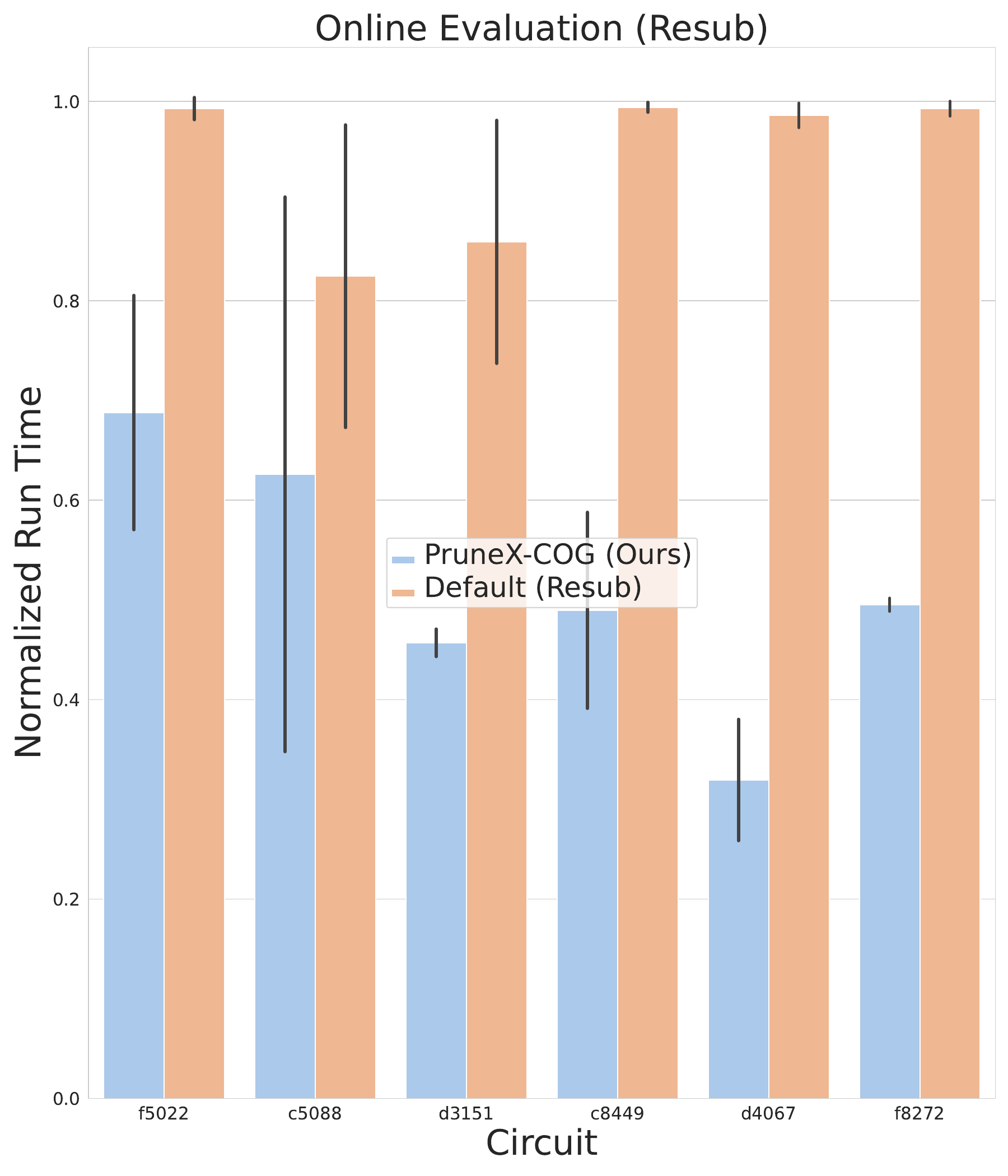}
        \includegraphics[width=0.32\textwidth, height=0.255\textwidth]{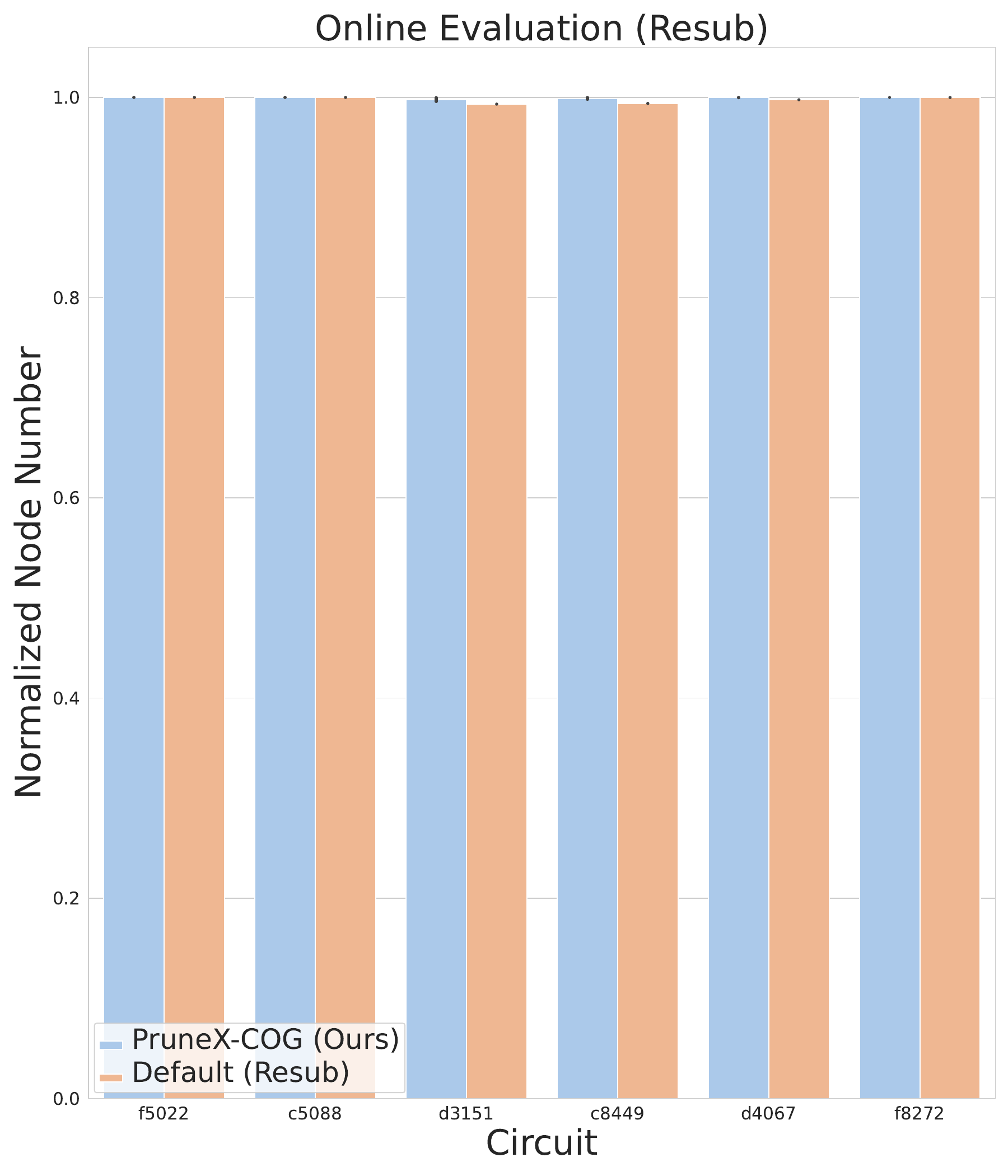}
        \vspace{-1.5mm}
        \caption{We evaluate our method on the industrial benchmark and Resub operator.}
        \label{fig:offline_evaluation_resub_haisi_acc_time_node}
    \end{subfigure}
    \begin{subfigure}{0.9\textwidth}
        \includegraphics[width=0.32\textwidth, height=0.255\textwidth]{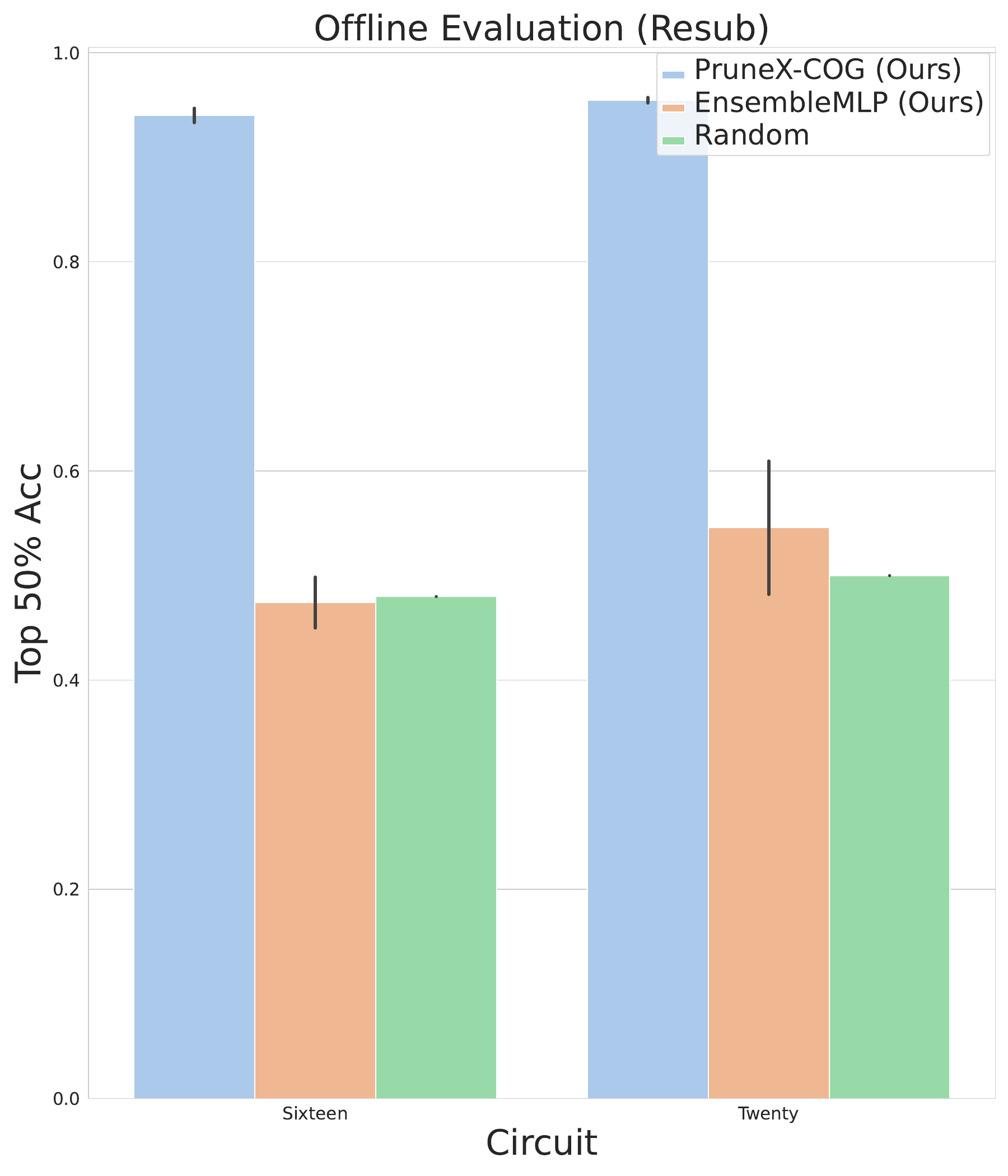}
        \includegraphics[width=0.32\textwidth, height=0.255\textwidth]{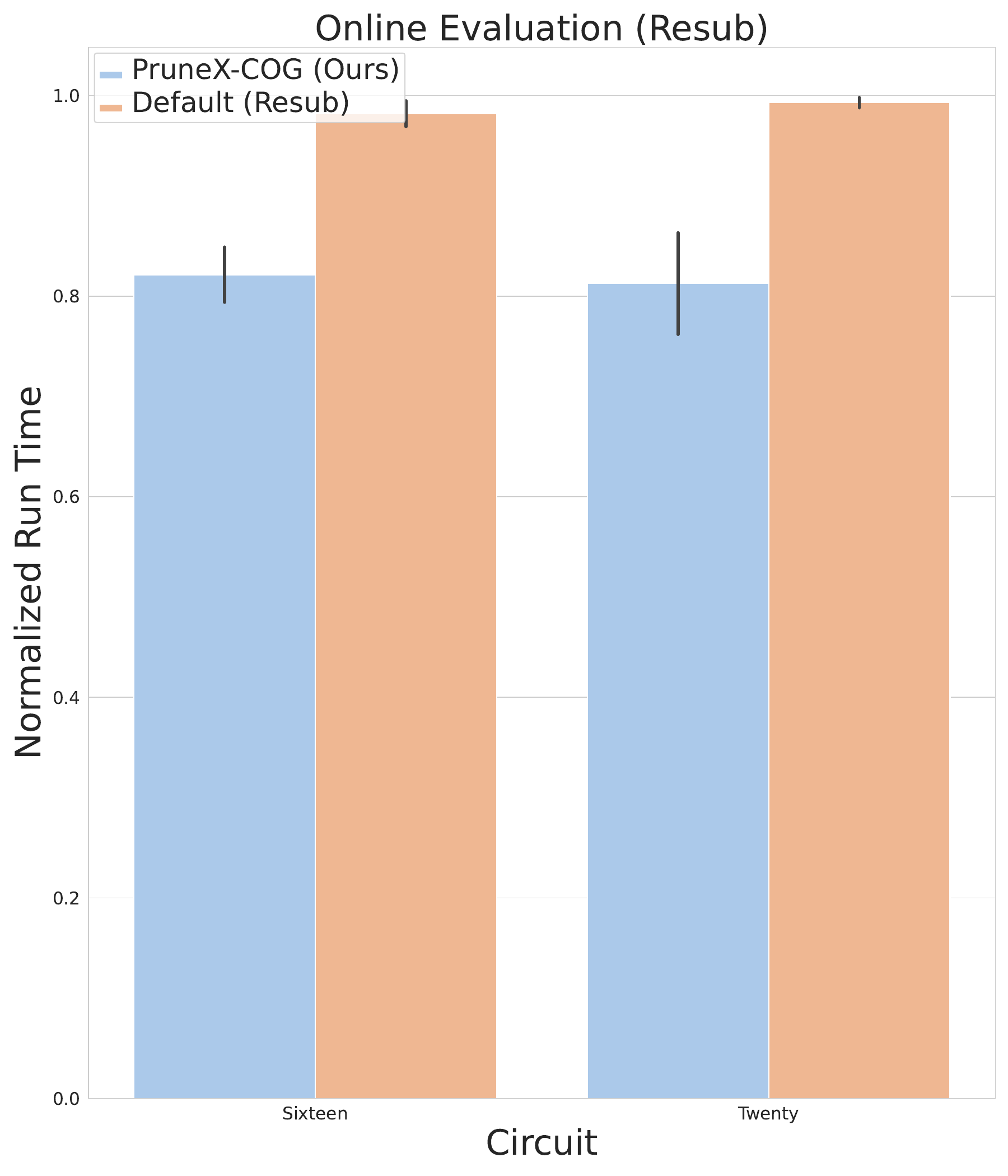}
        \includegraphics[width=0.32\textwidth, height=0.255\textwidth]{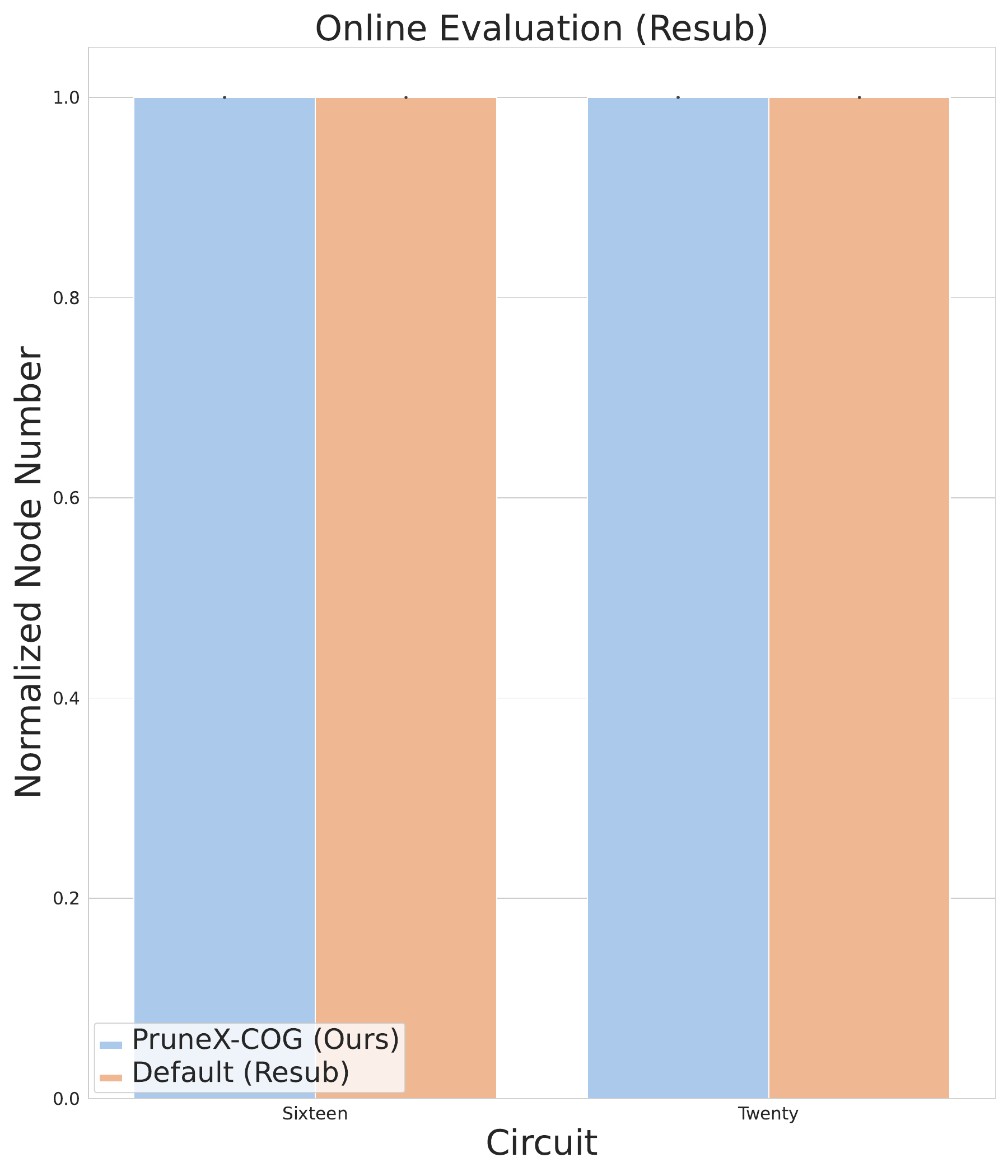}
        \vspace{-1.5mm}
        \caption{We evaluate our method on the very large-scale circuits and Resub operator.}
        \label{fig:offline_evaluation_resub_epfl_hard_acc_time_node}
    \end{subfigure}
    \vspace{-2.5mm} 
    \caption{The results demonstrate the strong ability of our method to promote efficient LS on industrial and very large-scale circuits. In particular, PruneX-COG reduces the runtime of applying the Mfs2 operator once by up to $10.9$ hours.}
    \label{fig:main_results2}
\end{figure*}

    \subsection{Evaluation on Industrial and Large-Scale Circuits}\label{results:real_industry}        
        In this subsection, we further demonstrate the effectiveness of our method by deploying it to industrial circuits from Huawei HiSilicon and very large-scale circuits from EPFL \cite{amaru2015epfl}. 
        \textbf{First}, the circuits from Huawei consist of $27$ industrial circuits, where the circuit sizes range from $2,775$ to $788,288$. We evaluate our method using Evaluation Strategy 2, with $21$ circuits for training and $6$ circuits for testing. \textbf{Second}, the very large-scale circuits from EPFL scale up to twenty million nodes, which entails a scalability challenge. Note that applying the Mfs2 operator to these large-scale circuits once can take $25.88$ hours, which is so long that it may significantly postpone the Time to Market. Please refer to Appendix \ref{appendix:datasets_industrial_vlsi} for more details of the datasets.  
        
        For the \textbf{offline evaluation} on industrial circuits, Figs. \ref{fig:offline_evaluation_mfs2_real_acc_time_node} and \ref{fig:offline_evaluation_resub_haisi_acc_time_node} show that PruneX-COG significantly improves the prediction recall compared to the Random baseline, achieving the recall of over $80\%$ on most testing circuits.
        In addition, although EnsembleMLP, i.e., our proposed simple baseline, achieves comparable prediction recall to PruneX-COG on some circuits, it struggles to consistently perform well on all testing circuits. The results highlight the effectiveness of our proposed OOD generalization framework. 

        For the \textbf{online evaluation} on industrial circuits, Figs. \ref{fig:offline_evaluation_mfs2_real_acc_time_node} and \ref{fig:offline_evaluation_resub_haisi_acc_time_node} show that PruneX-COG significantly reduces the runtime compared to the Default Mfs2 and Resub operators, respectively. 
        Moreover, the results demonstrate that our PruneX-COG achieves comparable sizes of optimized circuits with the Default operators.    
        Thus, the results highlight the strong ability to promote efficient LS and chip design with our proposed PruneX-COG on industrial circuits.

        For the evaluation on very large-scale circuits from EPFL, Figs. \ref{fig:offline_evaluation_mfs2_epfl_hard_acc_time_node} and \ref{fig:offline_evaluation_resub_epfl_hard_acc_time_node} show that PruneX-COG significantly improves prediction recall while reducing the runtime by up to $42\%$. In particular, PruneX-COG reduces the runtime by up to $10.9$ hours compared to the Default Mfs2.
        Moreover, PruneX-COG only degrades the size by $0.001\%$ on average. The results demonstrate the strong generalization ability and scalability of our PruneX-COG on very large-scale industrial circuits with up to twenty million nodes. We defer more detailed results to Appendix \ref{appendix:exps_real_circuits_vlsc}.
            
\begin{table*}[t]
\caption{We compare the Default Mfs2 operator with our 2PruneX-COG operator with the hyperparameter $k$ set as $30\%$ and $40\%$ on open-source and industrial circuits. Let Nd denote the node number (size) of circuits, and Lev denote the level (depth) of circuits. We define an Improvement metric by $\frac{M(\text{Default}) - M(\text{2PruneX-COG})}{M(\text{Default})}$, where $M(\cdot)$ denotes the Nd, Lev, or Time. We report the arithmetic mean (standard deviation) of the Nd, Lev, and Time.}
\label{table:mfs2_generalize_iwls_epfl_online_improving_ppa}
\centering
\resizebox{0.99\textwidth}{!}{
\begin{tabular}{@{}ccccccccc@{}}
\toprule
\toprule
 & \multicolumn{4}{c}{Multiplier} & \multicolumn{4}{c}{Square} \\ \midrule
\multirow{2}{*}{Method} & \multirow{2}{*}{Nd $\downarrow$} & \multirow{2}{*}{\begin{tabular}[c]{@{}c@{}}Improvement $\uparrow$\\      (Nd, \%)\end{tabular}} & \multirow{2}{*}{Time (s) $\downarrow$} & \multirow{2}{*}{\begin{tabular}[c]{@{}c@{}}Improvement $\uparrow$\\      (Time, \%)\end{tabular}} & \multirow{2}{*}{Nd $\downarrow$} & \multirow{2}{*}{\begin{tabular}[c]{@{}c@{}}Improvement $\uparrow$\\      (Nd, \%)\end{tabular}} & \multirow{2}{*}{Time (s) $\downarrow$} & \multirow{2}{*}{\begin{tabular}[c]{@{}c@{}}Improvement $\uparrow$ \\      (Time, \%)\end{tabular}} \\
 &  &  &  &  &  &  &  &  \\ \cmidrule(r){1-5} \cmidrule(l){6-9}
Default (Mfs2) & 7799.00 (0.0) & NA & 16.69 (0.07) & NA & 5701.00 (0.0) & NA & 21.41 (0.071) & NA \\
2PruneX-COG (0.3, Ours) & 7658.00 (4.32) & \textbf{1.81} & 8.86 (0.87) & \textbf{46.91} & 5550.33 (11.67) & \textbf{2.64} & 9.94 (0.55) & \textbf{53.57} \\
2PruneX-COG (0.4, Ours) & 7655.00 (4.54) & \textbf{1.85} & 14.09 (1.07) & \textbf{15.58} & 5523.33 (5.43) & \textbf{3.12} & 14.23 (0.66) & \textbf{33.54} \\ \midrule \midrule
 & \multicolumn{4}{c}{Hyp} & \multicolumn{4}{c}{Ethernet} \\ \midrule
\multirow{2}{*}{Method} & \multirow{2}{*}{Lev $\downarrow$} & \multirow{2}{*}{\begin{tabular}[c]{@{}c@{}}Improvement $\uparrow$\\      (Lev, \%)\end{tabular}} & \multirow{2}{*}{Time (s) $\downarrow$} & \multirow{2}{*}{\begin{tabular}[c]{@{}c@{}}Improvement $\uparrow$\\      (Time, \%)\end{tabular}} & \multirow{2}{*}{Nd $\downarrow$} & \multirow{2}{*}{\begin{tabular}[c]{@{}c@{}}Improvement $\uparrow$\\      (Nd, \%)\end{tabular}} & \multirow{2}{*}{Time (s) $\downarrow$} & \multirow{2}{*}{\begin{tabular}[c]{@{}c@{}}Improvement $\uparrow$\\      (Time, \%)\end{tabular}} \\
 &  &  &  &  &  &  &  &  \\ \cmidrule(r){1-5} \cmidrule(l){6-9}
Default (Mfs2) & 8259.00 (0.0) & NA & 274.13 (9.90) & NA & 13638.00 (0.0) & NA & 27.73 (0.45) & NA \\
2PruneX-COG (0.3, Ours) & 5762.00 (0.0) & \textbf{30.23} & 222.14 (27.98) & \textbf{18.97} & 13514.33 (4.71) & \textbf{0.91} & 8.87 (1.47) & \textbf{68.01} \\
2PruneX-COG (0.4, Ours) & 5762.00 (0.0) & \textbf{30.23} & 288.85 (34.96) & \textbf{-5.37} & 13511.00 (0.81) & \textbf{0.93} & 15.44 (0.76) & \textbf{44.32} \\ \midrule \midrule
 & \multicolumn{4}{c}{Wb conmax} & \multicolumn{4}{c}{Des perf} \\ \midrule 
\multirow{2}{*}{Method} & \multirow{2}{*}{Nd $\downarrow$} & \multirow{2}{*}{\begin{tabular}[c]{@{}c@{}}Improvement $\uparrow$\\      (Nd, \%)\end{tabular}} & \multirow{2}{*}{Time (s) $\downarrow$} & \multirow{2}{*}{\begin{tabular}[c]{@{}c@{}}Improvement $\uparrow$\\      (Time, \%)\end{tabular}} & \multirow{2}{*}{Nd $\downarrow$} & \multirow{2}{*}{\begin{tabular}[c]{@{}c@{}}Improvement $\uparrow$\\      (Nd, \%)\end{tabular}} & \multirow{2}{*}{Time (s) $\downarrow$} & \multirow{2}{*}{\begin{tabular}[c]{@{}c@{}}Improvement $\uparrow$\\      (Time, \%)\end{tabular}} \\
 &  &  &  &  &  &  &  &  \\ \cmidrule(r){1-5} \cmidrule(l){6-9}
Default (Mfs2) & 16509.00 (0.0) & NA & 21.24 (0.53) & NA & 30853.00 (0.0) & NA & 29.51 (0.28) & NA \\
2PruneX-COG (0.3, Ours) & 16111.00 (97.40) & \textbf{2.41} & 13.45 (0.73) & \textbf{36.68} & 29807.66 (13.69) & \textbf{3.39} & 24.79 (0.14) & \textbf{15.99} \\
2PruneX-COG (0.4, Ours) & 16006.66 (112.77) & \textbf{3.04} & 16.74 (0.72) & \textbf{21.19} & 29538.66 (11.61) & \textbf{4.26} & 31.67 (0.15) & \textbf{-7.32} \\
\end{tabular}
}
\newline
\newline
\resizebox{0.99\textwidth}{!}{
\begin{tabular}{@{}ccccccccc@{}}
\toprule
\toprule
 & \multicolumn{4}{c}{f5022} & \multicolumn{4}{c}{f8272} \\ \midrule
\multirow{2}{*}{Method} & \multirow{2}{*}{Lev $\downarrow$} & \multirow{2}{*}{\begin{tabular}[c]{@{}c@{}}Improvement $\uparrow$\\      (Lev, \%)\end{tabular}} & \multirow{2}{*}{Time (s) $\downarrow$} & \multirow{2}{*}{\begin{tabular}[c]{@{}c@{}}Improvement $\uparrow$\\      (Time, \%)\end{tabular}} & \multirow{2}{*}{Nd $\downarrow$} & \multirow{2}{*}{\begin{tabular}[c]{@{}c@{}}Improvement $\uparrow$\\      (Nd, \%)\end{tabular}} & \multirow{2}{*}{Time (s) $\downarrow$} & \multirow{2}{*}{\begin{tabular}[c]{@{}c@{}}Improvement $\uparrow$\\      (Time, \%)\end{tabular}} \\
 &  &  &  &  &  &  &  &  \\ \cmidrule(r){1-5} \cmidrule(l){6-9}
Default (Mfs2) & 47.00 (0.0) & NA & 177.93 (0.69) & NA & 99245.00 (0.0) & NA & 76.97 (0.12) & NA \\
2PruneX (COG, 0.3) & 45.00 (0.0) & \textbf{4.26} & 148.87 (19.79) & \textbf{16.33} & 95210.00 (0.0) & \textbf{4.07} & 59.03 (2.37) & \textbf{23.31} \\
2PruneX (COG, 0.4) & 45.00 (0.0) & \textbf{4.26} & 182.35 (25.45) & -2.48 & 95144.50 (65.5) & \textbf{4.13} & 57.51 (1.16) & \textbf{25.28} \\ \midrule \midrule
 & \multicolumn{4}{c}{c5088} & \multicolumn{4}{c}{d3151} \\ \midrule
\multirow{2}{*}{Method} & \multirow{2}{*}{Nd $\downarrow$} & \multirow{2}{*}{\begin{tabular}[c]{@{}c@{}}Improvement $\uparrow$\\      (Nd, \%)\end{tabular}} & \multirow{2}{*}{Time (s) $\downarrow$} & \multirow{2}{*}{\begin{tabular}[c]{@{}c@{}}Improvement $\uparrow$\\      (Time, \%)\end{tabular}} & \multirow{2}{*}{Nd $\downarrow$} & \multirow{2}{*}{\begin{tabular}[c]{@{}c@{}}Improvement $\uparrow$\\      (Nd, \%)\end{tabular}} & \multirow{2}{*}{Time (s) $\downarrow$} & \multirow{2}{*}{\begin{tabular}[c]{@{}c@{}}Improvement $\uparrow$\\      (Time, \%)\end{tabular}} \\
 &  &  &  &  &  &  &  &  \\ \cmidrule(r){1-5} \cmidrule(l){6-9}
Default (Mfs2) & 195665.00 (0.0) & NA & 487.51 (51.68) & NA & 6634.00 (0.0) & NA & 66.17 (5.25) & NA \\
2PruneX (COG, 0.3) & 193618.00 (0.81) & \textbf{1.05} & 264.32 (7.05) & \textbf{45.78} & 6538.33 (8.99) & \textbf{1.44} & 52.69 (0.26) & \textbf{20.37} \\
2PruneX (COG, 0.4) & 193618.00 (0.81) & \textbf{1.05} & 328.74 (4.89) & \textbf{32.57} & 6516.66 (10.84) & \textbf{1.77} & 85.80 (3.84) & -29.67 \\ \midrule \midrule
 & \multicolumn{4}{c}{c8449} & \multicolumn{4}{c}{d4067} \\ \midrule
\multirow{2}{*}{Method} & \multirow{2}{*}{Nd $\downarrow$} & \multirow{2}{*}{\begin{tabular}[c]{@{}c@{}}Improvement $\uparrow$\\      (Nd, \%)\end{tabular}} & \multirow{2}{*}{Time (s) $\downarrow$} & \multirow{2}{*}{\begin{tabular}[c]{@{}c@{}}Improvement $\uparrow$\\      (Time, \%)\end{tabular}} & \multirow{2}{*}{Nd $\downarrow$} & \multirow{2}{*}{\begin{tabular}[c]{@{}c@{}}Improvement $\uparrow$\\      (Nd, \%)\end{tabular}} & \multirow{2}{*}{Time (s) $\downarrow$} & \multirow{2}{*}{\begin{tabular}[c]{@{}c@{}}Improvement $\uparrow$\\      (Time, \%)\end{tabular}} \\
 &  &  &  &  &  &  &  &  \\ \cmidrule(r){1-5} \cmidrule(l){6-9}
Default (Mfs2) & 9538.00 (0.0) & NA & 234.58 (45.54) & NA & 215708.00 (0.0) & NA & 201.43 (12.17) & NA \\
2PruneX (COG, 0.3) & 9496.66 (16.99) & \textbf{0.43} & 184.02 (9.03) & \textbf{21.55} & 215447.00 (10.03) & \textbf{0.12} & 170.85 (10.19) & \textbf{15.18} \\
2PruneX (COG, 0.4) & 9452.66 (7.40) & \textbf{0.89} & 292.90 (14.26) & -24.86 & 215443.66 (8.80) & \textbf{0.12} & 308.35 (9.05) & -53.08 \\ \bottomrule
\end{tabular}
}
\end{table*}

    \subsection{Improving Quality of Results with PruneX-COG}\label{exps:improving_qor}
    In this subsection, we conduct experiments to demonstrate that efficient LS operators can not only reduce runtime but also improve Quality of Results (QoR), such as sizes and depths of optimized circuits. The size and depth are critical metrics in chip design, as they are proxy for the final area and delay of chips. Specifically, reducing the size of circuits could significantly save hardware resources to design a specific chip \cite{de2021fast}. Thus, a one percent improvement in the QoR may yield substantial economic value.   
    
    To improve QoR with PruneX-COG, we can sequentially apply PruneX-COG multiple times rather than once, as the runtime of PruneX-COG is significantly shorter than that of Default LS operators.
    Specifically, we focus on the Mfs2 operator here, and compare the runtime and QoR of 2PruneX-COG, i.e., applying the PruneX-COG operator twice, with the Default Mfs2 operator. We conduct extensive experiments on the open-source and industrial circuits. Moreover, we set the hyperparameter $k$ as $30\%$ and $40\%$ rather than $50\%$ to achieve faster runtime.  
    We provide more details and results in Appendix \ref{appendix:exps_improving_qor}. 
        
    Table \ref{table:mfs2_generalize_iwls_epfl_online_improving_ppa} shows that 2PruneX-COG significantly reduces the size and depth of optimized circuits while achieving faster runtime compared with the Default Mfs2 operator. Specifically, 2PruneX-COG with $k=40\%$ reduces the size/depth by $7.24\%$ on average while reducing the runtime by $17\%$ on the open-source circuits. Moreover, 2PruneX-COG reduces the size/depth by $2.04\%$ on average but increases the runtime by $8.7\%$ on the industrial circuits. Furthermore, suppose we want to achieve faster runtime in certain real-world scenarios, then we can set $k$ as a smaller value such as $30\%$. Table \ref{table:mfs2_generalize_iwls_epfl_online_improving_ppa} shows that 2PruneX-COG with $k=30\%$ reduces the size/depth by $6.9\%$ on average with $40.02\%$ runtime reduction on the open-source circuits, and reduces the size/depth by $1.9\%$ on average with $23.75\%$ runtime reduction on the industrial circuits.
    In particular, our method achieves a significant reduction over the depth on Hyp, improving the level by $30.23\%$. Overall, the results suggest that our efficient PruneX-COG can significantly improve the QoR while achieving faster runtime, yielding substantial economic value in chip design.

    \subsection{Ablation Study}
        In this section, we conduct an ablation study to understand the individual contribution of each component within our PruneX-COG. To this end, we compare our PruneX-COG with its variants, i.e., PruneX-COG without Domain-Aware Distributional Classifier (DADC) and PruneX-COG without DADC and Knowledge-Driven Subgraph Representation (KDSR), on open-source benchmarks under the Evaluation Strategy 2.
        Specifically, PruneX-COG without DADC aggregates all Circuit Datasets into a single domain, and uses our KDSR to learn node embeddings for classification. Then, PruneX-COG without DADC and KDSR further replaces our KDSR module with manually designed features (see Appendix \ref{appendix:node_features}). 
        Due to limited space, we defer additional details and results to Appendix \ref{exps:ablation_study}. 

\begin{table}[t]
\caption{We present the ablation study results.}
\label{table:mfs2_resub_eva_strategy2_ablation_study}
\centering
\resizebox{0.49\textwidth}{!}{
\begin{tabular}{@{}ccccc@{}}
\toprule
\toprule
Operator & \multicolumn{4}{c}{Mfs2} \\ \midrule
Method/Circuit & Log2 & Hyp & Multiplier & Sin \\ \midrule
 & top 50\% acc $\uparrow$ & top 50\% acc $\uparrow$ & top 50\% acc $\uparrow$ & top 50\% acc $\uparrow$ \\ \midrule
PruneX-COG & \textbf{0.88 (0.02)} & \textbf{0.85 (0.06)} & \textbf{0.87 (0.04)} & \textbf{0.79 (0.12)} \\
PruneX-COG without DADC & 0.87 (0.05) & \textbf{0.85 (0.01)} & 0.79 (0.02) & 0.75 (0.06) \\
PruneX-COG without DADC and KDSR & 0.08 (0.007) & 0.78 (0.01) & 0.33 (0.01) & 0.65 (0.08) \\ \midrule
Operator & \multicolumn{4}{c}{Resub} \\ \midrule
Method/Circuit & Log2 & Hyp & Multiplier & Sin \\ \midrule
 & top 50\% acc $\uparrow$ & top 50\% acc $\uparrow$ & top 50\% acc $\uparrow$ & top 50\% acc $\uparrow$ \\ \midrule
PruneX-COG & \textbf{0.80 (0.005)} & \textbf{0.87 (0.03)} & \textbf{0.87 (0.0)} & \textbf{0.81 (0.03)} \\
PruneX-COG without DADC & 0.45 (0.23) & 0.72 (0.17) & 0.41 (0.31) & 0.46 (0.23) \\
PruneX-COG without DADC and KDSR & 0.21 (0.02) & 0.67 (0.12) & 0.71 (0.06) & 0.28 (0.0) \\ \bottomrule
\end{tabular}
}
\end{table}

        The results in Table \ref{table:mfs2_resub_eva_strategy2_ablation_study} suggest the following two conclusions. First, PruneX-COG without DADC significantly outperforms PruneX-COG without DADC and KDSR in terms of offline prediction recall, demonstrating the importance of our KDSR module. That is, the results suggest that our representation module effectively learns domain-invariant representations for high generalization capability. 
        Second, PruneX-COG also significantly outperforms PruneX-COG without DADC in terms of offline prediction recall. This demonstrates that it is important to formulate the learning problem as a circuit domain generalization task and subsequently learn domain-aware classifiers for further improving the generalization capability of our method.

\section{Conclusion}
    In this paper, we found an important problem that leads to inefficient Logic Synthesis (LS), that is, a large number of node-level transformations in many LS operators are \textit{ineffective}, making applying these operators highly time-consuming.
    To address this challenge, we propose a novel data-driven LS operator paradigm (see Fig. \ref{fig:paradigm_operators}), namely PruneX, which promotes efficient LS by 
    significantly improving the efficiency of LS operators.
    The major challenge of developing PruneX is how to learn models that can well generalize to unseen circuits, that is, the out-of-distribution (OOD) generalization problem across circuits in LS operators.
    Thus, the major technical contribution of PruneX is the novel \textbf{c}ircuit d\textbf{o}main \textbf{g}eneralization (COG) framework (see Fig. \ref{fig:cog}), which is critical for the success of data-driven LS algorithms. 
    Extensive experiments demonstrate that PruneX significantly and consistently improves the efficiency of LS operators, achieving up to $3.1\times$ faster runtime while keeping comparable optimization performance. 


%

\ifCLASSOPTIONcompsoc


\ifCLASSOPTIONcaptionsoff
  \newpage
\fi



\bibliographystyle{IEEEtran}
\bibliography{IEEEtran}

\begin{thebibliography}{10}
\providecommand{\url}[1]{#1}
\csname url@samestyle\endcsname
\providecommand{\newblock}{\relax}
\providecommand{\bibinfo}[2]{#2}
\providecommand{\BIBentrySTDinterwordspacing}{\spaceskip=0pt\relax}
\providecommand{\BIBentryALTinterwordstretchfactor}{4}
\providecommand{\BIBentryALTinterwordspacing}{\spaceskip=\fontdimen2\font plus
\BIBentryALTinterwordstretchfactor\fontdimen3\font minus
  \fontdimen4\font\relax}
\providecommand{\BIBforeignlanguage}[2]{{%
\expandafter\ifx\csname l@#1\endcsname\relax
\typeout{** WARNING: IEEEtran.bst: No hyphenation pattern has been}%
\typeout{** loaded for the language `#1'. Using the pattern for}%
\typeout{** the default language instead.}%
\else
\language=\csname l@#1\endcsname
\fi
#2}}
\providecommand{\BIBdecl}{\relax}
\BIBdecl

\bibitem{review_ai4lo}
A.~A.~S. Berndt, M.~Foga{\c{c}}a, and C.~Meinhardt, ``A review of machine
  learning in logic synthesis,'' \emph{Journal of Integrated Circuits and
  Systems}, vol.~17, no.~3, pp. 1--12, 2022.

\bibitem{aisyn}
G.~Pasandi, S.~Pratty, and J.~Forsyth, ``Aisyn: Ai-driven reinforcement
  learning-based logic synthesis framework,'' \emph{arXiv preprint
  arXiv:2302.06415}, 2023.

\bibitem{de2021fast}
B.~A. De~Abreu, A.~Berndt, I.~S. Campos, C.~Meinhardt, J.~T. Carvalho,
  M.~Grellert, and S.~Bampi, ``Fast logic optimization using decision trees,''
  in \emph{2021 IEEE International Symposium on Circuits and Systems
  (ISCAS)}.\hskip 1em plus 0.5em minus 0.4em\relax IEEE, 2021, pp. 1--5.

\bibitem{rewrite}
V.~Bertacco and M.~Damiani, ``The disjunctive decomposition of logic
  functions,'' in \emph{iccad}, vol.~97, 1997, pp. 78--82.

\bibitem{syn_textbook}
G.~D. Micheli, \emph{Synthesis and optimization of digital circuits}.\hskip 1em
  plus 0.5em minus 0.4em\relax McGraw-Hill Higher Education, 1994.

\bibitem{farrahi1994complexity}
A.~H. Farrahi and M.~Sarrafzadeh, ``Complexity of the lookup-table minimization
  problem for fpga technology mapping,'' \emph{IEEE Transactions on
  Computer-Aided Design of Integrated Circuits and Systems}, vol.~13, no.~11,
  pp. 1319--1332, 1994.

\bibitem{abc}
R.~Brayton and A.~Mishchenko, ``Abc: An academic industrial-strength
  verification tool,'' in \emph{Computer Aided Verification: 22nd International
  Conference, CAV 2010, Edinburgh, UK, July 15-19, 2010. Proceedings 22}.\hskip
  1em plus 0.5em minus 0.4em\relax Springer, 2010, pp. 24--40.

\bibitem{mvsis}
J.-H. Jiang, Y.~Jiang, Y.~Li, A.~Mishchenko, S.~Sinha, T.~Villa, R.~Brayton,
  and R.~I. Parades, ``Mvsis v1. 1 manual.''

\bibitem{mfs2}
A.~Mishchenko, R.~Brayton, J.-H.~R. Jiang, and S.~Jang, ``Scalable
  don't-care-based logic optimization and resynthesis,'' \emph{ACM Transactions
  on Reconfigurable Technology and Systems (TRETS)}, vol.~4, no.~4, pp. 1--23,
  2011.

\bibitem{resub}
A.~M.~R. Brayton, ``Scalable logic synthesis using a simple circuit
  structure,'' vol.~6, pp. 15--22, 2006.

\bibitem{rewrite2}
A.~Mishchenko, S.~Chatterjee, and R.~Brayton, ``Dag-aware aig rewriting a fresh
  look at combinational logic synthesis,'' in \emph{Proceedings of the 43rd
  annual Design Automation Conference}, 12 2006, pp. 532--535.

\bibitem{refactor}
R.~K. Brayton, ``The decomposition and factorization of boolean expressions,''
  \emph{ISCA-82}, pp. 49--54, 1982.

\bibitem{neto2021read}
W.~L. Neto, M.~T. Moreira, L.~Amaru, C.~Yu, and P.-E. Gaillardon, ``Read your
  circuit: leveraging word embedding to guide logic optimization,'' in
  \emph{Proceedings of the 26th Asia and South Pacific Design Automation
  Conference}, 2021, pp. 530--535.

\bibitem{time_to_market_logic_synthesis}
S.~Sabbavarapu, K.~R. Basireddy, and A.~Acharyya, ``A new dynamic library based
  ic design automation methodology using functional symmetry with npn class
  representation approach to reduce nre costs and time-to-market,'' in
  \emph{2014 Fifth International Symposium on Electronic System Design}.\hskip
  1em plus 0.5em minus 0.4em\relax IEEE, 2014, pp. 115--119.

\bibitem{time_to_market_logic_synthesis2}
B.~K. Reddy, S.~Sabbavarapu, and A.~Acharyya, ``A new vlsi ic design automation
  methodology with reduced nre costs and time-to-market using the npn class
  representation and functional symmetry,'' in \emph{2014 IEEE International
  Symposium on Circuits and Systems (ISCAS)}.\hskip 1em plus 0.5em minus
  0.4em\relax IEEE, 2014, pp. 177--180.

\bibitem{technology_mapping}
A.~Mishchenko, S.~Cho, S.~Chatterjee, and R.~Brayton, ``Combinational and
  sequential mapping with priority cuts,'' in \emph{2007 IEEE/ACM International
  Conference on Computer-Aided Design}.\hskip 1em plus 0.5em minus 0.4em\relax
  IEEE, 2007, pp. 354--361.

\bibitem{ren2023machine}
H.~Ren and J.~Hu, \emph{Machine Learning Applications in Electronic Design
  Automation}.\hskip 1em plus 0.5em minus 0.4em\relax Springer Nature, 2023.

\bibitem{place_nature}
A.~Mirhoseini, A.~Goldie, M.~Yazgan, J.~W. Jiang, E.~Songhori, S.~Wang, Y.-J.
  Lee, E.~Johnson, O.~Pathak, A.~Nazi \emph{et~al.}, ``A graph placement
  methodology for fast chip design,'' \emph{Nature}, vol. 594, no. 7862, pp.
  207--212, 2021.

\bibitem{huang2021machine}
G.~Huang, J.~Hu, Y.~He, J.~Liu, M.~Ma, Z.~Shen, J.~Wu, Y.~Xu, H.~Zhang,
  K.~Zhong \emph{et~al.}, ``Machine learning for electronic design automation:
  A survey,'' \emph{ACM Transactions on Design Automation of Electronic Systems
  (TODAES)}, vol.~26, no.~5, pp. 1--46, 2021.

\bibitem{survey_gnn4eda}
\BIBentryALTinterwordspacing
D.~S\'{a}nchez, L.~Servadei, G.~N. Kiprit, R.~Wille, and W.~Ecker, ``A
  comprehensive survey on electronic design automation and graph neural
  networks: Theory and applications,'' \emph{ACM Trans. Des. Autom. Electron.
  Syst.}, vol.~28, no.~2, feb 2023. [Online]. Available:
  \url{https://doi.org/10.1145/3543853}
\BIBentrySTDinterwordspacing

\bibitem{lai2022maskplace}
\BIBentryALTinterwordspacing
Y.~Lai, Y.~Mu, and P.~Luo, ``Maskplace: Fast chip placement via reinforced
  visual representation learning,'' in \emph{Advances in Neural Information
  Processing Systems}, A.~H. Oh, A.~Agarwal, D.~Belgrave, and K.~Cho, Eds.,
  2022. [Online]. Available: \url{https://openreview.net/forum?id=T2DBbSh6_uY}
\BIBentrySTDinterwordspacing

\bibitem{lai2023chipformer}
\BIBentryALTinterwordspacing
Y.~Lai, J.~Liu, Z.~Tang, B.~Wang, H.~Jianye, and P.~Luo, ``Chipformer:
  Transferable chip placement via offline decision transformer,'' 2023.
  [Online]. Available: \url{https://openreview.net/pdf?id=j0miEWtw87}
\BIBentrySTDinterwordspacing

\bibitem{xppe}
\BIBentryALTinterwordspacing
H.~M. Makrani, H.~Sayadi, T.~Mohsenin, S.~rafatirad, A.~Sasan, and H.~Homayoun,
  ``Xppe: Cross-platform performance estimation of hardware accelerators using
  machine learning,'' in \emph{Proceedings of the 24th Asia and South Pacific
  Design Automation Conference}, ser. ASPDAC '19.\hskip 1em plus 0.5em minus
  0.4em\relax New York, NY, USA: Association for Computing Machinery, 2019, p.
  727–732. [Online]. Available: \url{https://doi.org/10.1145/3287624.3288756}
\BIBentrySTDinterwordspacing

\bibitem{hls_iccad18}
R.~G. Kim, J.~R. Doppa, and P.~P. Pande, ``Machine learning for design space
  exploration and optimization of manycore systems,'' in \emph{2018 IEEE/ACM
  International Conference on Computer-Aided Design (ICCAD)}, 2018, pp. 1--6.

\bibitem{hls_dac13}
H.-Y. Liu and L.~P. Carloni, ``On learning-based methods for design-space
  exploration with high-level synthesis,'' in \emph{2013 50th ACM/EDAC/IEEE
  Design Automation Conference (DAC)}, 2013, pp. 1--7.

\bibitem{lsoracle}
W.~L. Neto, M.~Austin, S.~Temple, L.~Amaru, X.~Tang, and P.-E. Gaillardon,
  ``Lsoracle: A logic synthesis framework driven by artificial intelligence,''
  in \emph{2019 IEEE/ACM International Conference on Computer-Aided Design
  (ICCAD)}.\hskip 1em plus 0.5em minus 0.4em\relax IEEE, 2019, pp. 1--6.

\bibitem{autodmp}
\BIBentryALTinterwordspacing
A.~Agnesina, P.~Rajvanshi, T.~Yang, G.~Pradipta, A.~Jiao, B.~Keller,
  B.~Khailany, and H.~Ren, ``Autodmp: Automated dreamplace-based macro
  placement,'' in \emph{Proceedings of the 2023 International Symposium on
  Physical Design}, ser. ISPD '23.\hskip 1em plus 0.5em minus 0.4em\relax New
  York, NY, USA: Association for Computing Machinery, 2023, p. 149–157.
  [Online]. Available: \url{https://doi.org/10.1145/3569052.3578923}
\BIBentrySTDinterwordspacing

\bibitem{cheng2022the}
\BIBentryALTinterwordspacing
R.~Cheng, X.~Lyu, Y.~Li, J.~Ye, J.~HAO, and J.~Yan, ``The policy-gradient
  placement and generative routing neural networks for chip design,'' in
  \emph{Advances in Neural Information Processing Systems}, A.~H. Oh,
  A.~Agarwal, D.~Belgrave, and K.~Cho, Eds., 2022. [Online]. Available:
  \url{https://openreview.net/forum?id=uNYqDfPEDD8}
\BIBentrySTDinterwordspacing

\bibitem{fawcett1994synthesis}
B.~Fawcett, ``Synthesis for fpgas: an overview,'' \emph{Proceedings of
  WESCON'94}, pp. 576--580, 1994.

\bibitem{dsoai}
\BIBentryALTinterwordspacing
Synopsys, ``Design space optimization ai,'' 2020. [Online]. Available:
  \url{https://www.synopsys.com/ai/chip-design/dso-ai.html}
\BIBentrySTDinterwordspacing

\bibitem{cadence_cerebrus}
\BIBentryALTinterwordspacing
Cadence, ``Cadence cerebrus,'' 2021. [Online]. Available:
  \url{https://www.cadence.com/en_US/home/tools/digital-design-and-signoff/soc-implementation-and-floorplanning/cerebrus-intelligent-chip-explorer.html}
\BIBentrySTDinterwordspacing

\bibitem{hosny2020drills}
A.~Hosny, S.~Hashemi, M.~Shalan, and S.~Reda, ``Drills: Deep reinforcement
  learning for logic synthesis,'' in \emph{2020 25th Asia and South Pacific
  Design Automation Conference (ASP-DAC)}.\hskip 1em plus 0.5em minus
  0.4em\relax IEEE, 2020, pp. 581--586.

\bibitem{grosnit2022boils}
A.~Grosnit, C.~Malherbe, R.~Tutunov, X.~Wan, J.~Wang, and H.~B. Ammar, ``Boils:
  Bayesian optimisation for logic synthesis,'' in \emph{2022 Design, Automation
  \& Test in Europe Conference \& Exhibition (DATE)}.\hskip 1em plus 0.5em
  minus 0.4em\relax IEEE, 2022, pp. 1193--1196.

\bibitem{kirby2019congestionnet}
R.~Kirby, S.~Godil, R.~Roy, and B.~Catanzaro, ``Congestionnet: Routing
  congestion prediction using deep graph neural networks,'' in \emph{2019
  IFIP/IEEE 27th International Conference on Very Large Scale Integration
  (VLSI-SoC)}.\hskip 1em plus 0.5em minus 0.4em\relax IEEE, 2019, pp. 217--222.

\bibitem{zhou2019primal}
Y.~Zhou, H.~Ren, Y.~Zhang, B.~Keller, B.~Khailany, and Z.~Zhang, ``Primal:
  Power inference using machine learning,'' in \emph{Proceedings of the 56th
  Annual Design Automation Conference 2019}, 2019, pp. 1--6.

\bibitem{neto2021slap}
W.~L. Neto, M.~T. Moreira, Y.~Li, L.~Amar{\`u}, C.~Yu, and P.-E. Gaillardon,
  ``Slap: a supervised learning approach for priority cuts technology
  mapping,'' in \emph{2021 58th ACM/IEEE Design Automation Conference
  (DAC)}.\hskip 1em plus 0.5em minus 0.4em\relax IEEE, 2021, pp. 859--864.

\bibitem{neto2019improving}
W.~L. Neto, X.~Tang, M.~Austin, L.~Amaru, and P.-E. Gaillardon, ``Improving
  logic optimization in sequential circuits using majority-inverter graphs,''
  in \emph{2019 IEEE Computer Society Annual Symposium on VLSI (ISVLSI)}.\hskip
  1em plus 0.5em minus 0.4em\relax IEEE, 2019, pp. 224--229.

\bibitem{congestionnet}
R.~Kirby, S.~Godil, R.~Roy, and B.~Catanzaro, ``Congestionnet: Routing
  congestion prediction using deep graph neural networks,'' in \emph{2019
  IFIP/IEEE 27th International Conference on Very Large Scale Integration
  (VLSI-SoC)}, 2019, pp. 217--222.

\bibitem{LHNN}
\BIBentryALTinterwordspacing
B.~Wang, G.~Shen, D.~Li, J.~Hao, W.~Liu, Y.~Huang, H.~Wu, Y.~Lin, G.~Chen, and
  P.~A. Heng, ``Lhnn: Lattice hypergraph neural network for vlsi congestion
  prediction,'' in \emph{Proceedings of the 59th ACM/IEEE Design Automation
  Conference}, ser. DAC '22.\hskip 1em plus 0.5em minus 0.4em\relax New York,
  NY, USA: Association for Computing Machinery, 2022, p. 1297–1302. [Online].
  Available: \url{https://doi.org/10.1145/3489517.3530675}
\BIBentrySTDinterwordspacing

\bibitem{cross_graph}
\BIBentryALTinterwordspacing
A.~Ghose, V.~Zhang, Y.~Zhang, D.~Li, W.~Liu, and M.~Coates, ``Generalizable
  cross-graph embedding for gnn-based congestion prediction,'' in \emph{2021
  IEEE/ACM International Conference On Computer Aided Design (ICCAD)}.\hskip
  1em plus 0.5em minus 0.4em\relax IEEE Press, 2021, p. 1–9. [Online].
  Available: \url{https://doi.org/10.1109/ICCAD51958.2021.9643446}
\BIBentrySTDinterwordspacing

\bibitem{yang2022versatile}
\BIBentryALTinterwordspacing
S.~Yang, Z.~Yang, D.~Li, Y.~Zhang, Z.~Zhang, G.~Song, and J.~HAO, ``Versatile
  multi-stage graph neural network for circuit representation,'' in
  \emph{Advances in Neural Information Processing Systems}, A.~H. Oh,
  A.~Agarwal, D.~Belgrave, and K.~Cho, Eds., 2022. [Online]. Available:
  \url{https://openreview.net/forum?id=nax3ATLrovW}
\BIBentrySTDinterwordspacing

\bibitem{khailany2020accelerating}
B.~Khailany, ``Accelerating chip design with machine learning,'' in
  \emph{Proceedings of the 2020 ACM/IEEE Workshop on Machine Learning for CAD},
  2020, pp. 33--33.

\bibitem{lopera2021survey}
D.~S. Lopera, L.~Servadei, G.~N. Kiprit, S.~Hazra, R.~Wille, and W.~Ecker, ``A
  survey of graph neural networks for electronic design automation,'' in
  \emph{2021 ACM/IEEE 3rd Workshop on Machine Learning for CAD (MLCAD)}.\hskip
  1em plus 0.5em minus 0.4em\relax IEEE, 2021, pp. 1--6.

\bibitem{drills}
A.~Hosny, S.~Hashemi, M.~Shalan, and S.~Reda, ``Drills: Deep reinforcement
  learning for logic synthesis,'' in \emph{2020 25th Asia and South Pacific
  Design Automation Conference (ASP-DAC)}.\hskip 1em plus 0.5em minus
  0.4em\relax IEEE, 2020, pp. 581--586.

\bibitem{braytontechnology}
A.~M. S. C.~R. Brayton and X.~W.~T. Kam, ``Technology mapping with boolean
  matching, supergates and choices.''

\bibitem{balance}
J.~Cortadella, ``Timing-driven logic bi-decomposition,'' \emph{IEEE
  Transactions on Computer-Aided Design of Integrated Circuits and Systems},
  vol.~22, no.~6, pp. 675--685, 2003.

\bibitem{amaru2015epfl}
L.~Amar{\'u}, P.-E. Gaillardon, and G.~De~Micheli, ``The epfl combinational
  benchmark suite,'' no. CONF, 2015.

\bibitem{focal_loss}
T.-Y. Lin, P.~Goyal, R.~Girshick, K.~He, and P.~Doll{\'a}r, ``Focal loss for
  dense object detection,'' in \emph{Proceedings of the IEEE international
  conference on computer vision}, 2017, pp. 2980--2988.

\bibitem{rota2017loss}
S.~Rota~Bulo, G.~Neuhold, and P.~Kontschieder, ``Loss max-pooling for semantic
  image segmentation,'' in \emph{Proceedings of the IEEE conference on computer
  vision and pattern recognition}, 2017, pp. 2126--2135.

\bibitem{goodfellow2016deep}
I.~Goodfellow, Y.~Bengio, and A.~Courville, \emph{Deep learning}.\hskip 1em
  plus 0.5em minus 0.4em\relax MIT press, 2016.

\bibitem{taud2018multilayer}
H.~Taud and J.~Mas, ``Multilayer perceptron (mlp),'' \emph{Geomatic approaches
  for modeling land change scenarios}, pp. 451--455, 2018.

\bibitem{zhou2021ensemble}
Z.-H. Zhou and Z.-H. Zhou, \emph{Ensemble learning}.\hskip 1em plus 0.5em minus
  0.4em\relax Springer, 2021.

\bibitem{mfs3}
A.~M.~R. Brayton, T.~B.~S. Govindarajan, H.~Arts, and P.~van Besouw,
  ``Versatile sat-based remapping for standard cells.''

\bibitem{resnet}
K.~He, X.~Zhang, S.~Ren, and J.~Sun, ``Deep residual learning for image
  recognition,'' in \emph{Proceedings of the IEEE conference on computer vision
  and pattern recognition}, 2016, pp. 770--778.

\bibitem{vit}
A.~Dosovitskiy, L.~Beyer, A.~Kolesnikov, D.~Weissenborn, X.~Zhai,
  T.~Unterthiner, M.~Dehghani, M.~Minderer, G.~Heigold, S.~Gelly \emph{et~al.},
  ``An image is worth 16x16 words: Transformers for image recognition at
  scale,'' in \emph{International Conference on Learning Representations}.

\bibitem{shen2021towards}
Z.~Shen, J.~Liu, Y.~He, X.~Zhang, R.~Xu, H.~Yu, and P.~Cui, ``Towards
  out-of-distribution generalization: A survey,'' \emph{arXiv preprint
  arXiv:2108.13624}, 2021.

\bibitem{wang2022generalizing}
J.~Wang, C.~Lan, C.~Liu, Y.~Ouyang, T.~Qin, W.~Lu, Y.~Chen, W.~Zeng, and P.~Yu,
  ``Generalizing to unseen domains: A survey on domain generalization,''
  \emph{IEEE Transactions on Knowledge and Data Engineering}, 2022.

\bibitem{wang2020generalizing}
Y.~Wang, Q.~Yao, J.~T. Kwok, and L.~M. Ni, ``Generalizing from a few examples:
  A survey on few-shot learning,'' \emph{ACM computing surveys (csur)},
  vol.~53, no.~3, pp. 1--34, 2020.

\bibitem{blanchard2011generalizing}
G.~Blanchard, G.~Lee, and C.~Scott, ``Generalizing from several related
  classification tasks to a new unlabeled sample,'' \emph{Advances in neural
  information processing systems}, vol.~24, 2011.

\bibitem{muandet2013domain}
K.~Muandet, D.~Balduzzi, and B.~Sch{\"o}lkopf, ``Domain generalization via
  invariant feature representation,'' in \emph{International conference on
  machine learning}.\hskip 1em plus 0.5em minus 0.4em\relax PMLR, 2013, pp.
  10--18.

\bibitem{albuquerque2019adversarial}
I.~Albuquerque, J.~Monteiro, T.~H. Falk, and I.~Mitliagkas, ``Adversarial
  target-invariant representation learning for domain generalization,''
  \emph{arXiv preprint arXiv:1911.00804}, vol.~8, 2019.

\bibitem{hamilton2020graph}
W.~L. Hamilton, ``Graph representation learning,'' \emph{Synthesis Lectures on
  Artifical Intelligence and Machine Learning}, vol.~14, no.~3, pp. 1--159,
  2020.

\bibitem{duvenaud2015convolutional}
D.~K. Duvenaud, D.~Maclaurin, J.~Iparraguirre, R.~Bombarell, T.~Hirzel,
  A.~Aspuru-Guzik, and R.~P. Adams, ``Convolutional networks on graphs for
  learning molecular fingerprints,'' \emph{Advances in neural information
  processing systems}, vol.~28, 2015.

\bibitem{kipfsemi}
T.~N. Kipf and M.~Welling, ``Semi-supervised classification with graph
  convolutional networks,'' in \emph{International Conference on Learning
  Representations}.

\bibitem{shi2023lmc}
Z.~Shi, X.~Liang, and J.~Wang, ``Lmc: Fast training of gnns via subgraph
  sampling with provable convergence,'' in \emph{The Eleventh International
  Conference on Learning Representations}, 2023.

\bibitem{gori2005new}
M.~Gori, G.~Monfardini, and F.~Scarselli, ``A new model for learning in graph
  domains,'' in \emph{Proceedings. 2005 IEEE International Joint Conference on
  Neural Networks, 2005.}, vol.~2.\hskip 1em plus 0.5em minus 0.4em\relax IEEE,
  2005, pp. 729--734.

\bibitem{gasse2019exact}
M.~Gasse, D.~Ch{\'e}telat, N.~Ferroni, L.~Charlin, and A.~Lodi, ``Exact
  combinatorial optimization with graph convolutional neural networks,''
  \emph{Advances in neural information processing systems}, vol.~32, 2019.

\bibitem{pets}
K.~Chua, R.~Calandra, R.~McAllister, and S.~Levine, ``Deep reinforcement
  learning in a handful of trials using probabilistic dynamics models,''
  \emph{Advances in neural information processing systems}, vol.~31, 2018.

\bibitem{tqc}
A.~Kuznetsov, P.~Shvechikov, A.~Grishin, and D.~Vetrov, ``Controlling
  overestimation bias with truncated mixture of continuous distributional
  quantile critics,'' in \emph{International Conference on Machine
  Learning}.\hskip 1em plus 0.5em minus 0.4em\relax PMLR, 2020, pp. 5556--5566.

\bibitem{cmbac}
Z.~Wang, J.~Wang, Q.~Zhou, B.~Li, and H.~Li, ``Sample-efficient reinforcement
  learning via conservative model-based actor-critic,'' in \emph{Proceedings of
  the AAAI Conference on Artificial Intelligence}, vol.~36, no.~8, 2022, pp.
  8612--8620.

\bibitem{albrecht2005iwls}
C.~Albrecht, ``Iwls 2005 benchmarks,'' 2005.

\bibitem{rai2021logic}
S.~Rai, W.~L. Neto, Y.~Miyasaka, X.~Zhang, M.~Yu, Q.~Yi, M.~Fujita, G.~B.
  Manske, M.~F. Pontes, L.~S. da~Rosa \emph{et~al.}, ``Logic synthesis meets
  machine learning: Trading exactness for generalization,'' in \emph{2021
  Design, Automation \& Test in Europe Conference \& Exhibition (DATE)}.\hskip
  1em plus 0.5em minus 0.4em\relax IEEE, 2021, pp. 1026--1031.

\bibitem{adam}
D.~P. Kingma and J.~Ba, ``Adam: A method for stochastic optimization,''
  \emph{arXiv preprint arXiv:1412.6980}, 2014.

\bibitem{torch}
A.~Paszke, S.~Gross, F.~Massa, A.~Lerer, J.~Bradbury, G.~Chanan, T.~Killeen,
  Z.~Lin, N.~Gimelshein, L.~Antiga \emph{et~al.}, ``Pytorch: An imperative
  style, high-performance deep learning library,'' \emph{Advances in neural
  information processing systems}, vol.~32, 2019.

\bibitem{grannite}
Y.~Zhang, H.~Ren, and B.~Khailany, ``Grannite: Graph neural network inference
  for transferable power estimation,'' in \emph{Proceedings of the 57th
  ACM/EDAC/IEEE Design Automation Conference}, ser. DAC '20.\hskip 1em plus
  0.5em minus 0.4em\relax IEEE Press, 2020.

\bibitem{tsne}
L.~Van~der Maaten and G.~Hinton, ``Visualizing data using t-sne.''
  \emph{Journal of machine learning research}, vol.~9, no.~11, 2008.

\bibitem{gulrajanisearch}
I.~Gulrajani and D.~Lopez-Paz, ``In search of lost domain generalization,'' in
  \emph{International Conference on Learning Representations}.

\end{thebibliography}
\clearpage
\newpage
\appendices
\section{Theoretical Analysis}
    \subsection{Proof of Theorem \ref{obj:bound_estimation}}\label{appendix:proof of error bound}
    The proof of Theorem \ref{obj:bound_estimation} draws inspiration from the methodology proposed by \cite{blanchard2011generalizing, muandet2013domain}.  
    Nevertheless, our problem setting is different from that of \cite{blanchard2011generalizing, muandet2013domain}. Specifically, they assume values of $M$ and $n_k\text{ for }k=1,2,\cdots,M$ are fixed as a prior. In contrast, our setting allows for flexibility in the values of $M$ and $n_k\text{ for } k = 1, 2, \dots, M$ as we propose a circuit aggregation mechanism $g$. Consequently, our conclusion holds a greater level of generality.

    To complete the proof, we utilize the kernel method and introduce the concept of reproducing kernel Hilbert space (RKHS)
    $\mathcal{H}_{\bar{K}}$ with kernel: 
    \begin{align}
    \bar{K}\left(\left(P_x^1, x_1\right), \left(P_x^2, x_2\right)\right) = K_\mathcal{P}\left(P_x^1, P_x^2\right)K_\mathcal{X}\left(x_1, x_2\right)\nonumber  
    \end{align}
    Here, $K_\mathcal{P}$ and $K_\mathcal{X}$ represent kernel functions on distributions and input space, respectively. Suppose the RKHS corresponding to $K$ ($K$ can be any kernel function) is $\mathcal{H}_K$. We consider the feature mapping $\Psi: \mathfrak{P}_\mathcal{X} \rightarrow \mathcal{H}_{K_\mathcal{X}}$:
    \begin{align}
        P_x \rightarrow \Psi(P_x) := \int_\mathcal{X} K_\mathcal{X}(x,\cdot)dP_X(x)\nonumber
    \end{align}
    a universal kernel $\kappa$ \cite{blanchard2011generalizing} on $\mathcal{H}_{K_\mathcal{X}}$ which satisfies:
    \begin{align}
        K_\mathcal{P}(P_x^1, P_x^2) = \kappa\left(\Psi(P_x^1),\Psi(P_x^2)\right)\nonumber
    \end{align}
    and the mapping $\Phi_\kappa : \mathcal{H}_{K_\mathcal{X}} \rightarrow \mathcal{H}_\kappa$ which satisfies:
    \begin{align}
        \kappa\left(\Psi(P_x^1),\Psi(P_x^2)\right)=
        \langle \Phi_\kappa(\Psi_{P_X^1}),\Phi_\kappa(\Psi_{P_X^2}) \rangle\nonumber
    \end{align}
    
    \textbf{Notations}
    Let $\mathcal{X}$ denote the input space and $\mathcal{Y} = \{0, 1\}$ the output space. Let $\mathfrak{P}_\mathcal{X}$ denote the set of probability distributions on $\mathcal{X}$. A decision function is a function $f: \mathfrak{P}_\mathcal{X} \times \mathcal{X} \rightarrow \mathcal{Y}$. The loss function has the form $l: \mathcal{Y} \times \mathcal{Y} \rightarrow \mathbb{R}_{+}$. We consider a scenario where iid training distributions $\mathbb{P}_{X}^k$ and test distribution $\mathbb{P}_{X}^t$ are drawn according to a hyper-distribution $\mathscr{P}$. $\hat{\mathbb{P}}_{X}^k$ denote the empirical estimation of $\mathbb{P}_{X}^k$.
    
    \textbf{Remark} 
    Recall that $M$ is the number of training domains, $n_k$ is the sample size of the $k$-th domain, and $n = \sum\limits_{k=1}^{M}n_k$ is the total number of samples. In our setting, $M$ and $n_k$ are variable and $n$ is fixed.   

    \textbf{Assumptions}
    (1) The loss function $l: \mathcal{Y} \times \mathcal{Y} \rightarrow \mathbb{R}_{+}$ is $\phi_{X}$-Lipschitz in its first variable and $\phi_{Y}$-Lipschitz in its second variable and bounded by $U_{l}$. (2) The kernel $K_\mathcal{X}$ and $\kappa$ are bounded by $U_{K}^2$ and $U_\kappa^2$, respectively. (3) The feature map $\Phi_\kappa: \mathcal{H}_{K_\mathcal{X}} \rightarrow \mathcal{H}_\kappa$ satisfies a specific $\text {H}\ddot{o}\text{lder}$ condition with constant $L_\kappa$ on $B_{K_\mathcal{X}}(U_K)$:
    \begin{align}
        \forall v,w \in B_{K_\mathcal{X}}(U_K): \lvert\lvert \Phi_\kappa(v) - \Phi_\kappa(w) \rvert\rvert \leq L_\kappa\lvert\lvert v-w \rvert\rvert\nonumber
    \end{align}
    Under the aforementioned assumptions, we present the following proof.
    \begin{proof}
    Based on the inequality $\lvert a_1+ \cdots +a_n \rvert^2 \leq n\lvert a_1 \rvert^2 + \cdots +n\lvert a_2 \rvert^2$ and $\sup(A+B) \leq \sup A +\sup B$, we decompose
    \begin{equation}
        \begin{aligned}[b]
        &\sup _{f \in B_{\bar{K}}\left(r\right)} \lvert \mathcal{R}(f)-\hat{\mathcal{R}}(f) \rvert^2\\
       =&\sup _{f \in B_{\bar{K}}\left(r\right)}\biggl\lvert\mathbb{E}_{\mathbb{P}_{XY}^t\sim \mathscr{P}} \mathbb{E}_{(\textbf{x},y)\sim \mathbb{P}_{XY}^t}[l(f(\mathbb{P}_{X}^t,\textbf{x}),y)] \\
       &-\frac{1}{M}\sum_{k=1}^{M} \frac{1}{n_k} \sum_{j=1}^{n_k} l(f(\hat{\mathbb{P}}_{X}^k,\textbf{x}_j^k),y_j^k)\biggr\rvert^2\\
        \leq & 3\sup _{f \in B_{\bar{K}}\left(r\right)} \biggl\lvert \mathbb{E}_{\mathbb{P}_{XY}^t\sim \mathscr{P}} \mathbb{E}_{(\textbf{x},y)\sim \mathbb{P}_{XY}^t}[l(f(\mathbb{P}_{X}^t,\textbf{x}),y)] \\
        &-\frac{1}{M}\sum_{k=1}^{M}\mathbb{E}_{(\textbf{x},y)\sim P_{XY}^k} [l(f(\mathbb{P}_{X}^k,\textbf{x}),y)]\biggr\rvert^2 \\
        +&3\sup _{f \in B_{\bar{K}}\left(r\right)}\biggl\lvert\frac{1}{M}\sum_{k=1}^{M}\mathbb{E}_{(\textbf{x},y)\sim \mathbb{P}_{XY}^k} [l(f(\mathbb{P}_{X}^k,\textbf{x}),y)] \\ 
        &-\frac{1}{M}\sum_{k=1}^{M} \frac{1}{n_k} \sum_{j=1}^{n_k} l(f(\mathbb{P}_{X}^k,\textbf{x}_j^k),y_j^k)\biggr\rvert^2\\
        +&3\sup _{f \in B_{\bar{K}}\left(r\right)}\biggl\lvert\frac{1}{M}\sum_{k=1}^{M} \frac{1}{n_k} \sum_{j=1}^{n_k} l(f(\mathbb{P}_{X}^k,\textbf{x}_j^k),y_j^k)\\
        &-\frac{1}{M}\sum_{k=1}^{M} \frac{1}{n_k} \sum_{j=1}^{n_k} l(f(\hat{\mathbb{P}}_{X}^k,\textbf{x}_j^k),y_j^k)\biggr\rvert^2\\
         =&(I)+(II)+(III)\nonumber
        \end{aligned}
    \end{equation}
    \textbf{Control of I}
    \begin{align}
        (I)=3\sup _{f \in B_{\bar{K}}\left(r\right)} \biggl\lvert \mathbb{E}_{\mathbb{P}_{XY}^t\sim \mathscr{P}} \mathbb{E}_{(\textbf{x},y)\sim \mathbb{P}_{XY}^t}[l(f(\mathbb{P}_{X}^t,\textbf{x}),y)]\nonumber\\
        -\frac{1}{M}\sum_{k=1}^{M}\mathbb{E}_{(\textbf{x},y)\sim P_{XY}^k} [l(f(\mathbb{P}_{X}^k,\textbf{x}),y)]\biggr\rvert^2\nonumber
    \end{align}
    Since the $(\mathbb{P}_{XY}^k)_{1\leq k \leq M}$ are iid, so we let
    \begin{align}
        \beta((\mathbb{P}_{XY}^k)_{1\leq k \leq M}) :=\sup _{f \in B_{\bar{K}}\left(r\right)} \biggl\lvert \mathbb{E}_{\mathbb{P}_{XY}^t\sim \mathscr{P}} \mathbb{E}_{(\textbf{x},y)\sim \mathbb{P}_{XY}^t}[l(f(\mathbb{P}_{X}^t,\textbf{x}),y)]\nonumber\\
        -\frac{1}{M}\sum_{k=1}^{M}\mathbb{E}_{(\textbf{x},y)\sim P_{XY}^k} [l(f(\mathbb{P}_{X}^k,\textbf{x}),y)]\biggr\rvert\nonumber
    \end{align}
    By McDiarmid inequality in Hilbert space, we have 
    \begin{align}
        \beta - \mathbb{E}[\beta] \leq U_l\sqrt{\frac{\log\delta^{-1}}{2M}}\nonumber
    \end{align}
    where $\mathbb{E}[\beta]$ is denoted as 
    \begin{align}
        \mathbb{E}[\beta]=\mathbb{E}_{(\mathbb{P}_{XY}^k)_{1\leq k \leq M}}\sup _{f \in B_{\bar{K}}\left(r\right)} \biggl\lvert \mathbb{E}_{\mathbb{P}_{XY}^t\sim \mathscr{P}} \mathbb{E}_{(\textbf{x},y)\sim \mathbb{P}_{XY}^t}[l(f(\mathbb{P}_{X}^t,\textbf{x}),y)]\nonumber\\
        -\frac{1}{M}\sum_{k=1}^{M}\mathbb{E}_{(\textbf{x},y)\sim P_{XY}^k} [l(f(\mathbb{P}_{X}^k,\textbf{x}),y)]\biggr\rvert\nonumber 
    \end{align}
    To bound $\mathbb{E}[\beta]$, we use Rademacher complexity analysis. We denote $(\textbf{x}_k, y_k)$ a single draw from distribution $\mathbb{P}_{XY}^k$ and these draws are independent. We also denote $(\sigma_k)_{1\leq k \leq M}$ $\{\pm 1\}$-valued iid Rademacher variables which are independent from everything else. We have:
    \begin{align}
        \mathbb{E}[\beta]
        \leq & \mathbb{E}_{(\mathbb{P}_{XY}^k)_{1\leq k \leq M}}\mathbb{E}_{\sigma_k} [  \nonumber \\
        & \sup _{f \in B_{\bar{K}}\left(r\right)} \biggl\lvert \frac{2}{M}\sum_{k=1}^{M}\sigma_{k}\mathbb{E}_{(\textbf{x}_k,y_k)\sim P_{XY}^k} [l(f(\mathbb{P}_{X}^k,\textbf{x}_k),y_k)]\biggr\rvert \nonumber ] \\
        \leq & \mathbb{E}_{(\mathbb{P}_{XY}^k)_{1\leq k \leq M}}\mathbb{E}_{(\textbf{x}_k,y_k)\sim P_{XY}^k}\mathbb{E}_{\sigma_k} [ \nonumber \\
        & \sup _{f \in B_{\bar{K}}\left(r\right)} \biggl\lvert \frac{2}{M}\sum_{k=1}^{M}\sigma_{k}l(f(\mathbb{P}_{X}^k,\textbf{x}_k),y_k)\biggr\rvert \nonumber ] \\
        \leq & \frac{2r\phi_{X}U_KU_\kappa}{\sqrt{M}}\nonumber
    \end{align}
    The first inequality is a standard symmetrization argument. The second inequality pulls the inner expectation on $(\textbf{x}_k, y_k)$ outwards. The last inequality is a standard bound for the Rademacher complexity of a Lipschitz loss function on the ball of radius $r$ of $\mathcal{H}_{\bar{K}}$, where the kernel $\bar{K}$ is bounded by $U_K^2U_\kappa^2$.
    Based on the above analysis, we can conclude that
    \begin{align}
        (I) &= 3\beta^2\nonumber\\
        & \leq 6(\mathbb{E}[\beta]^2+\frac{U_l^2\log\delta^{-1}}{2M})\nonumber\\
        & \leq \frac{24r^2\phi_{X}^2U_K^2U_\kappa^2}{M}+\frac{3U_l^2\log\delta^{-1}}{M}\nonumber
    \end{align}
    \textbf{Control of II}
    \begin{align}
        (II) = & 3\sup _{f \in B_{\bar{K}}\left(r\right)}\biggl\lvert\frac{1}{M}\sum_{k=1}^{M}\mathbb{E}_{(\textbf{x},y)\sim \mathbb{P}_{XY}^k} [l(f(\mathbb{P}_{X}^k,\textbf{x}),y)] \nonumber \\
        & -\frac{1}{M}\sum_{k=1}^{M} \frac{1}{n_k} \sum_{j=1}^{n_k} l(f(\mathbb{P}_{X}^k,\textbf{x}_j^k),y_j^k)\biggr\rvert^2 \nonumber \\
        = & 3\sup _{f \in B_{\bar{K}}\left(r\right)}\biggl\lvert \frac{1}{M}\sum_{k=1}^{M} \frac{1}{n_k} \sum_{j=1}^{n_k}\mathbb{E}_{(\textbf{x},y)\sim \mathbb{P}_{XY}^k} [l(f(\mathbb{P}_{X}^k,\textbf{x}),y) \nonumber \\
        & - l(f(\mathbb{P}_{X}^k,\textbf{x}_j^k),y_j^k)]\biggr\rvert^2\nonumber\\
        \leq & 3\sup _{f \in B_{\bar{K}}\left(r\right)}\frac{1}{M}\sum_{k=1}^{M} \frac{1}{n_k} \sum_{j=1}^{n_k}\mathbb{E}_{(\textbf{x},y)\sim \mathbb{P}_{XY}^k}\bigl\lvert l(f(\mathbb{P}_{X}^k,\textbf{x}),y) \nonumber \\
        & - l(f(\mathbb{P}_{X}^k,\textbf{x}_j^k),y_j^k)\bigr\rvert^2\nonumber
    \end{align}
    Based on the reproducing property of $\bar{K}$ and the condition that $l$ is $\phi_X$-Lipschitz in its first variable and $\phi_Y$-Lipschitz in its second variable, we have:
    \begin{align}
      &\bigl\lvert l(f(\mathbb{P}_{X}^k,\textbf{x}),y)-l(f(\mathbb{P}_{X}^k,\textbf{x}_j^k),y_j^k)\bigr\rvert^2\nonumber\\
      =&\bigl\lvert l(f(\mathbb{P}_{X}^k,\textbf{x}),y)-l(f(\mathbb{P}_{X}^k,\textbf{x}),y_j^k) \nonumber \\ 
      & + l(f(\mathbb{P}_{X}^k,\textbf{x}),y_j^k)-l(f(\mathbb{P}_{X}^k,\textbf{x}_j^k),y_j^k)\bigr\rvert^2\nonumber\\
      \leq & 2\bigl\lvert l(f(\mathbb{P}_{X}^k,\textbf{x}),y)-l(f(\mathbb{P}_{X}^k,\textbf{x}),y_j^k)\bigr\rvert^2 \nonumber \\
      & + 2\bigl\lvert l(f(\mathbb{P}_{X}^k,\textbf{x}),y_j^k)-l(f(\mathbb{P}_{X}^k,\textbf{x}_j^k),y_j^k)\bigr\rvert^2\nonumber\\
      \leq&2\phi_{Y}^2\bigl\lvert y-y_j^k\bigr\rvert^2+2\phi_{X}^2\bigl\lvert f(\mathbb{P}_{X}^k,\textbf{x})-f(\mathbb{P}_{X}^k,\textbf{x}_j^k)\bigr\rvert^2\nonumber\\
      \leq&2\phi_{Y}^2+4\phi_{X}^2\bigl\lvert \max_\textbf{x} f(\mathbb{P}_{X}^k,\textbf{x})\bigr\rvert^2\nonumber\\
      =&2\phi_{Y}^2+4\phi_{X}^2\bigl\lvert\max_\textbf{x} \langle \bar{K}((\mathbb{P}_{X}^k,\textbf{x}),\cdot),f \rangle \bigr\rvert^2\nonumber\\
      \leq&2\phi_{Y}^2+4\phi_{X}^2\lvert\lvert\max_\textbf{x} \bar{K}((\mathbb{P}_{X}^k,\textbf{x}),\cdot)\rvert\rvert^2\lvert\lvert f \rvert\rvert^2 \nonumber\\
      \leq&2\phi_{Y}^2+4\phi_{X}^2r^2U_K^2U_\kappa^2\nonumber
    \end{align}
    Therefore, we can conclude that
    \begin{align}
        (II) \leq 6\phi_{Y}^2+12\phi_{X}^2r^2U_K^2U_\kappa^2\nonumber
    \end{align}
    \textbf{Control of III}
    \begin{align}
        (III)=&3\sup _{f \in B_{\bar{K}}\left(r\right)}\biggl\lvert\frac{1}{M}\sum_{k=1}^{M} \frac{1}{n_k} \sum_{j=1}^{n_k} l(f(\mathbb{P}_{X}^k,\textbf{x}_j^k),y_j^k) \nonumber \\
        & -\frac{1}{M}\sum_{k=1}^{M} \frac{1}{n_k} \sum_{j=1}^{n_k} l(f(\hat{\mathbb{P}}_{X}^k,\textbf{x}_j^k),y_j^k)\biggr\rvert^2\nonumber\\
        =&3\sup _{f \in B_{\bar{K}}\left(r\right)}\biggl\lvert\frac{1}{M}\sum_{k=1}^{M} \frac{1}{n_k} \sum_{j=1}^{n_k} [l(f(\mathbb{P}_{X}^k,\textbf{x}_j^k),y_j^k) \nonumber \\ 
        & - l(f(\hat{\mathbb{P}}_{X}^k,\textbf{x}_j^k),y_j^k)]\biggr\rvert^2\nonumber\\
        \leq & 3\phi_X^2\sup _{f \in B_{\bar{K}}\left(r\right)}\biggl\lvert\frac{1}{M}\sum_{k=1}^{M} \frac{1}{n_k} \sum_{j=1}^{n_k} [f(\mathbb{P}_{X}^k,\textbf{x}_j^k)-f(\hat{\mathbb{P}}_{X}^k,\textbf{x}_j^k)]\biggr\rvert^2\nonumber\\
        \leq & 3\phi_X^2 \sup _{f \in B_{\bar{K}}\left(r\right)}  \frac {1}{M}\sum_{k=1}^{M}\lvert\lvert f(\mathbb{P}_X^k,\cdot) - f(\hat{\mathbb{P}}_X^k,\cdot)\rvert\rvert_\infty^2\nonumber
    \end{align}
    Here $f(P_X^k,\cdot)$ denotes a vector and $\lvert\lvert \cdot \rvert\rvert_\infty$ is the infinite norm of the vector. 
    Using the reproducing property of $\bar{K}$ and the condition that $\lvert\lvert\Psi(\mathbb{P}_X)\rvert\rvert \leq U_K$ , we have for any $X \in \mathcal{X}$ and $f \in B_{\bar{K}}\left(r\right)$
    \begin{align}
        &\lvert f(\mathbb{P}_X^k, X) - f(\hat{\mathbb{P}}_X^k,X) \rvert\nonumber \\
        =&\lvert \langle \bar{K}((\mathbb{P}_X^k,X),\cdot) - \bar{K}((\hat{\mathbb{P}}_X^k,X),\cdot), f \rangle \rvert\nonumber \\
        \leq& \lvert\lvert f \rvert\rvert \cdot \lvert\lvert \bar{K}((\mathbb{P}_X^k,X),\cdot) - \bar{K}((\hat{\mathbb{P}}_X^k,X),\cdot) \rvert\rvert \nonumber \\
        \leq& rK_\mathcal{X}(X,X)^{\frac1{2}} \cdot \nonumber \\
        & \left(\kappa(\Psi(\mathbb{P}_X^k),\Psi(\mathbb{P}_X^k)) + \kappa(\Psi(\hat{\mathbb{P}}_X^k),\Psi(\hat{\mathbb{P}}_X^k)) 
        - 2\kappa(\Psi(\mathbb{P}_X^k),\Psi(\hat{\mathbb{P}}_X^k)) \right)^{\frac 1{2}}\nonumber \\
        \leq&rU_K\lvert\lvert \Phi_\kappa(\Psi(\mathbb{P}_X^k)) - \Phi_\kappa(\Psi(\hat{\mathbb{P}}_X^k))\rvert\rvert \nonumber \\
        \leq& rU_KL_\kappa \lvert\lvert \Psi(\mathbb{P}_X^k) - \Psi(\hat{\mathbb{P}}_X^k) \rvert\rvert\nonumber
    \end{align}
    By Hoeffding's inequality in Hilbert space, we can conclude that with probability $1-\delta$
    \begin{align}
        \lvert\lvert \Psi(\hat{\mathbb{P}}_X^k) - \Psi(\mathbb{P}_X^k) \rvert\rvert & = \biggl\lvert\biggl\lvert \frac{1}{n_k}\sum_{j=1}^{n_k}K_\mathcal{X}(X_j^k,\cdot) - \mathbb{E}_{X \sim \mathbb{P}_X^k}[K_\mathcal{X}(X,\cdot)]\biggr\rvert\biggr\rvert \nonumber \\
        & \leq 3U_K\sqrt\frac{\log2\delta^{-1}}{n_k}\nonumber
    \end{align}
    Therefore, we have
    \begin{align}
        (III) &\leq 3\phi_{X}^2r^2U_K^2L_\kappa^2\frac{1}{M}\sum_{k=1}^{M}\lvert\lvert \Psi(\mathbb{P}_X^k) - \Psi(\hat{\mathbb{P}}_X^k) \rvert\rvert^2\nonumber \\
        &\leq 27\phi_{X}^2r^2U_K^4L_\kappa^2\frac{1}{M}\sum_{k=1}^{M}\frac{\log2\delta^{-1}}{n_k}\nonumber 
    \end{align}
    Combining all of the above  inequalities, we obtain the announced results of the theorem \ref{obj:bound_estimation} that with probability at least $1-\delta$
    \begin{align}
        \sup _{f \in B_{\bar{K}}\left(r\right)}\lvert \mathcal{R}(f)-\hat{\mathcal{R}}(f) \rvert^2 \leq c_1+c_2 \frac{\log 2\delta^{-1}}{M}\sum_{k=1}^M \frac{1}{n_k}+c_3 \frac{\log \delta^{-1}}{M}+\frac{c_4}{M}\nonumber
    \end{align}
    where $c_1 = 2\phi_{Y}^2+4\phi_{X}^2r^2U_K^2U_\kappa^2$, $c_2 = 27\phi_{X}^2r^2U_K^4L_\kappa^2$, $c_3 = 3U_l^2$, $c_4 = 24r^2\phi_{X}^2U_K^2U_\kappa^2$. 
    \end{proof}

    \subsection{Proof of Corollary \ref{proof:corollary3}}\label{appendix:corollary3_proof}
    \begin{proof}
    The generalization error bound in Theorem \ref{obj:bound_estimation} is
    \begin{align}
        c_1+c_2 \frac{\log 2\delta^{-1}}{M}\sum_{k=1}^M \frac{1}{n_k}+c_3 \frac{\log \delta^{-1}}{M}+\frac{c_4}{M}\nonumber
    \end{align}
    The condition to prove Corollary \ref{proof:corollary3} is that the number of domains $M$ and the total number of samples $n = \sum_{k=1}^{M} n_k$ are fixed. Besides, based on the Harmonic Arithmetic Mean Inequality 
    \begin{align}
        (n_1+n_2+\cdots+n_M)(\frac{1}{n_1}+\frac{1}{n_2}+\cdots+\frac{1}{n_M}) \geq M^2\nonumber
    \end{align}
    we have
    \begin{align}
        &c_1+c_2 \frac{\log 2\delta^{-1}}{M}\sum_{k=1}^M \frac{1}{n_k}+c_3 \frac{\log \delta^{-1}}{M}+\frac{c_4}{M}\nonumber\\ \geq  &c_1+c_2 \frac{\log 2\delta^{-1}}{M}\frac{M^2}{\sum_{k=1}^M n_k}+c_3 \frac{\log \delta^{-1}}{M}+\frac{c_4}{M}\nonumber\\
        = &c_1+c_2 \frac{M\log 2\delta^{-1}}{n} +c_3 \frac{\log \delta^{-1}}{M}+\frac{c_4}{M}\nonumber
    \end{align}
    The condition for the above inequality to be equal is that $n_k = \frac{n}{M} \text{ for } k = 1, 2, \cdots M$. Therefore, we can conclude that
    the error bound reaches its minimum when $n_k = \frac{n}{M} \text{ for } k = 1, 2, \cdots M$, which proves that Corollary \ref{proof:corollary3} holds.
    \end{proof}
    
    \subsection{Proof of Corollary \ref{proof:corollary1}}\label{appendix:corollary1_proof}
    \begin{proof}
    Under the condition
    \begin{align}\label{coro1_condition1}
        n_k = \frac{n}{M} \text{ for } k = 1, 2, \cdots, M
    \end{align}
    and
    \begin{align}\label{coro1_condition2}
          1 \leq M \leq \frac{c_3\log\delta^{-1}+c_4}{c_2\log2\delta^{-1}} \cdot n
    \end{align}
    we can conclude that using multi-domain dataset (i.e., $M>1$) will result in a smaller generalization error bound than just pooling them into one mixed dataset (i.e., $M$=1).
    
    Based on the condition \ref{coro1_condition1}, the generalization error bound in Theorem \ref{obj:bound_estimation} can be represented as a function on discrete variable $M$ 
    \begin{align}
        B(M) = c_1+c_2\frac{M\log{2\delta^{-1}}}{n}+c_3 \frac{\log \delta^{-1}}{M}+\frac{c_4}{M}\quad(M \geq 1)\nonumber
    \end{align} 
    where $n$ representing the total number of samples is a constant and $M$ denotes the number of domains. To prove Corollary \ref{proof:corollary1}, we just need to prove that $B(1) \geq B(M) \text{ for } M \geq 1$. Consequently, under the condition \ref{coro1_condition2}, we have:
    \begin{align}
        &1 \leq M \leq \frac{c_3\log\delta^{-1}+c_4}{c_2\log2\delta^{-1}} \cdot n\nonumber \\
        \Rightarrow &\frac{c_2\log{2\delta^{-1}}}{n}(1-M) - \frac{c_3\log\delta^{-1}+c_4}{M}(1-M) \geq 0\nonumber \\
        \Rightarrow&(c_1+c_2 \frac{\log 2\delta^{-1}}{n}+c_3 \log \delta^{-1}+c_4) \nonumber \\
        & - (c_1+c_2 \frac{M\log 2\delta^{-1}}{n}+c_3 \frac{\log \delta^{-1}}{M}+\frac{c_4}{M}) \geq 0\nonumber \\
        \Rightarrow&B(1) \geq B(M) \quad (M \geq 1)\nonumber
    \end{align}
    which proves that Corollary \ref{proof:corollary1} holds. 
   
    Finally, we explain the reasonableness of condition \ref{coro1_condition1} and \ref{coro1_condition2}. The condition \ref{coro1_condition1} is a result in Corollary \ref{proof:corollary3} and can be realized easily. Besides, in practical applications, $n$ is often significantly larger than $M$ and the constants $c_1, c_2, c_3$ are typically fixed values as demonstrated in \ref{appendix:proof of error bound}. Therefore, the condition \ref{coro1_condition1} and \ref{coro1_condition2} are both considered reasonable.
    \end{proof}

\section{Additional Results}\label{appendix:exps}
    \subsection{Additional Results of Ablation Study}\label{exps:ablation_study}
        In this subsection, we perform ablation study to understand the contribution of each component in COG. We report the offline prediction accuracy of COG, COG without Distributional Classifier (COG without DC), and EnsembleMLP in Tables \ref{table:mfs2_epfl_iwls_ablation}, \ref{table:resub_epfl_iwls_ablation}, and \ref{table:mfs2_resub_iwls_to_epfl_ablation}.  COG without DC aggregates the datasets from all circuits into a mixed domain, and uses our knowledge-driven subgraph representation to learn node embeddings for classification. Compared to COG without DC, EnsembleMLP further replaces our knowledge-driven subgraph representation module with manually designed node features (see Section \ref{appendix:node_features}). 

        From the results in Tables \ref{table:mfs2_epfl_iwls_ablation}, \ref{table:resub_epfl_iwls_ablation}, and \ref{table:mfs2_resub_iwls_to_epfl_ablation}, we can draw two conclusions. First, COG without DC significantly outperforms EnsembleMLP in terms of the prediction accuracy, demonstrating the importance of our knowledge-driven subgraph representation module. Moreover, the results suggest that our representation module effectively learns domain-invariant representations for high generalization capability. 
        Second, COG further improves COG without DC in terms of the prediction accuracy on most circuits. This demonstrates that formulating the circuits as multiple domains and learning domain-aware classifiers is important for further improving the generalization capability.

\begin{table*}[t]
\caption{The ablation study of COG 
 on the Mfs2 operator under Evaluation Strategy 1. The best performance is marked in bold.}
\label{table:mfs2_epfl_iwls_ablation}
\centering
\resizebox{0.99\textwidth}{!}{
\begin{tabular}{@{}cccccccccc@{}}
\toprule
\toprule
 & \multicolumn{3}{c}{Log2} & \multicolumn{3}{c}{Hyp} & \multicolumn{3}{c}{Multiplier} \\ \midrule
Method & top 40\% acc & top 50\% acc & top 60\% acc & top 40\% acc & top 50\% acc & top 60\% acc & top 40\% acc & top 50\% acc & top 60\% acc \\ \cmidrule(r){1-4} \cmidrule(lr){5-7} \cmidrule(l){8-10}
COG & \textbf{0.89 (0.04)} & \textbf{0.97 (0.01)} & \textbf{0.99 (0.01)} & \textbf{0.79 (0.04)} & \textbf{0.90 (0.03)} & \textbf{0.95 (0.008} & \textbf{0.91 (0.03)} & \textbf{0.96 (0.03)} & \textbf{0.98 (0.01)} \\
COG without DC & 0.74 (0.12) & 0.86 (0.10) & 0.92 (0.04) & 0.78 (0.06) & 0.85 (0.03) & 0.92 (0.02) & 0.52 (0.12) & 0.75 (0.15) & 0.90 (0.03) \\
EnsembleMLP & 0.61 (0.0) & 0.65 (0.007) & 0.85 (0.03) & 0.62 (0.005) & 0.73 (0.004) & 0.82 (0.003) & 0.63 (0.0) & 0.71 (0.0) & 0.75 (0.0) \\ \midrule
 & \multicolumn{3}{c}{Sin} & \multicolumn{3}{c}{Square} & \multicolumn{3}{c}{Vga lcd} \\ \midrule
Method & top 40\% acc & top 50\% acc & top 60\% acc & top 40\% acc & top 50\% acc & top 60\% acc & top 40\% acc & top 50\% acc & top 60\% acc \\ \cmidrule(r){1-4} \cmidrule(lr){5-7} \cmidrule(l){8-10}
COG & 0.71 (0.06) & 0.87 (0.04) & \textbf{0.95 (0.02)} & \textbf{0.90 (0.01)} & \textbf{0.94 (0.002)} & \textbf{0.95 (0.0)} & \textbf{0.93 (0.007)} & \textbf{0.95 (0.01)} & \textbf{0.98 (0.007)} \\
COG without DC & \textbf{0.72 (0.06)} & \textbf{0.91 (0.013)} & 0.94 (0.02) & 0.40 (0.009) & 0.54 (0.04) & 0.70 (0.02) & 0.88 (0.08) & 0.93 (0.07) & 0.98 (0.0) \\
EnsembleMLP & 0.56 (0.0) & 0.58 (0.0) & 0.67 (0.0) & 0.85 (0.02) & 0.91 (0.004) & 0.92 (0.0) & 0.19 (0.04) & 0.29 (0.02) & 0.53 (0.02) \\ \midrule
 & \multicolumn{3}{c}{Ethernet} & \multicolumn{3}{c}{Wb conmax} & \multicolumn{3}{c}{Des perf} \\ \midrule
Method & top 40\% acc & top 50\% acc & top 60\% acc & top 40\% acc & top 50\% acc & top 60\% acc & top 40\% acc & top 50\% acc & top 60\% acc \\ \cmidrule(r){1-4} \cmidrule(lr){5-7} \cmidrule(l){8-10}
COG & \textbf{0.91 (0.09)} & \textbf{0.96 (0.04)} & \textbf{0.97 (0.03)} & \textbf{0.69 (0.04)} & \textbf{0.81 (0.05)} & \textbf{0.89 (0.04)} & \textbf{0.53 (0.0006)} & \textbf{0.68 (0.0007)} & \textbf{0.82 (0.0004)} \\
COG without DC & \textbf{0.97 (0.004)} & \textbf{0.98 (0.01)} & \textbf{0.98 (0.01)} & \textbf{0.71 (0.05)} & \textbf{0.84 (0.05)} & \textbf{0.92 (0.04)} & \textbf{0.53 (0.005)} & \textbf{0.68 (0.003)} & \textbf{0.82 (0.003)} \\
EnsembleMLP & 0.53 (0.004) & 0.62 (0.01) & 0.63 (0.03) & 0.43 (0.07) & 0.57 (0.05) & 0.68 (0.03) & 0.44 (0.07) & 0.49 (0.06) & 0.58 (0.03) \\ \bottomrule
\end{tabular}
}
\end{table*}

\begin{table*}[t]
\caption{The ablation study of COG 
 on the Resub operator under Evaluation Strategy 1. The best performance is marked in bold.}
\label{table:resub_epfl_iwls_ablation}
\centering
\resizebox{0.99\textwidth}{!}{
\begin{tabular}{@{}cccccccccc@{}}
\toprule
\toprule
 & \multicolumn{3}{c}{Log2} & \multicolumn{3}{c}{Hyp} & \multicolumn{3}{c}{Multiplier} \\ \midrule
Method & top 40\% acc & top 50\% acc & top 60\% acc & top 40\% acc & top 50\% acc & top 60\% acc & top 40\% acc & top 50\% acc & top 60\% acc \\ \cmidrule(r){1-4} \cmidrule(lr){5-7} \cmidrule(l){8-10} 
COG & \textbf{0.66 (0.02)} & \textbf{0.88 (0.02)} & \textbf{0.97 (0.004)} & \textbf{0.88 (0.008)} & \textbf{0.91 (0.02)} & 0.94 (0.03) & 0.74 (0.17) & \textbf{0.97 (0.02)} & \textbf{0.99 (0.003)} \\
COG without DC & 0.55 (0.22) & 0.67 (0.22) & 0.69 (0.21) & 0.86 (0.03) & 0.91 (0.007) & \textbf{0.98 (0.01)} & 0.85 (0.10) & 0.95 (0.03) & 0.99 (0.004) \\
EnsembleMLP & 0.39 (0.1) & 0.57 (0.2) & 0.62 (0.20) & 0.72 (0.02) & 0.86 (0.005) & 0.97 (0.01) & \textbf{0.88 (0.12)} & 0.93 (0.08) & 0.99 (0.01) \\ \midrule
 & \multicolumn{3}{c}{Sin} & \multicolumn{3}{c}{Square} & \multicolumn{3}{c}{Vga lcd} \\ \midrule
Method & top 40\% acc & top 50\% acc & top 60\% acc & top 40\% acc & top 50\% acc & top 60\% acc & top 40\% acc & top 50\% acc & top 60\% acc \\ \cmidrule(r){1-4} \cmidrule(lr){5-7} \cmidrule(l){8-10}
COG & 0.46 (0.17) & \textbf{0.69 (0.07)} & 0.76 (0.07) & \textbf{0.60 (0.11)} & \textbf{0.73 (0.08)} & \textbf{0.81 (0.07)} & 0.47 (0.34) & \textbf{0.77 (0.04)} & \textbf{0.94 (0.02)} \\
COG without DC & 0.37 (0.11) & 0.52 (0.05) & \textbf{0.78 (0.04)} & 0.56 (0.12) & 0.68 (0.11) & 0.76 (0.08) & 0.48 (0.33) & 0.56 (0.39) & 0.62 (0.43) \\
EnsembleMLP & \textbf{0.48 (0.03)} & 0.63 (0.1) & 0.76 (0.07) & 0.39 (0.09) & 0.52 (0.12) & 0.63 (0.08) & \textbf{0.68 (0.03)} & 0.71 (0.01) & 0.87 (0.01) \\ \midrule
 & \multicolumn{3}{c}{Ethernet} & \multicolumn{3}{c}{Wb conmax} & \multicolumn{3}{c}{Des perf} \\ \midrule
Method & top 40\% acc & top 50\% acc & top 60\% acc & top 40\% acc & top 50\% acc & top 60\% acc & top 40\% acc & top 50\% acc & top 60\% acc \\ \cmidrule(r){1-4} \cmidrule(lr){5-7} \cmidrule(l){8-10} 
COG & 0.82 (0.02) & \textbf{0.98 (0.009)} & \textbf{0.99 (0.0)} & \textbf{0.66 (0.07)} & \textbf{0.88 (0.08)} & \textbf{0.96 (0.01)} & \textbf{0.55 (0.07)} & \textbf{0.72 (0.05)} & \textbf{0.89 (0.009)} \\
COG without DC & \textbf{0.85 (0.08)} & 0.97 (0.009) & 0.98 (0.01) & 0.49 (0.39) & 0.56 (0.40) & 0.69 (0.37) & 0.40 (0.004) & 0.56 (0.02) & 0.71 (0.04) \\
EnsembleMLP & 0.61 (0.01) & 0.82 (0.04) & 0.87 (0.05) & 0.18 (0.26) & 0.18 (0.26) & 0.40 (0.11) & 0.36 (0.04) & 0.46 (0.03) & 0.56 (0.03) \\ \bottomrule
\end{tabular}
}
\end{table*}

\begin{table*}[t]
\caption{The ablation study of COG 
 on the Mfs2 and Resub operators under Evaluation Strategy 2. The best performance is marked in bold.}
\label{table:mfs2_resub_iwls_to_epfl_ablation}
\centering
\resizebox{0.99\textwidth}{!}{
\begin{tabular}{@{}ccccccc@{}}
\toprule
\toprule
Operator/Circuit &  & Log2 & Hyp & Multiplier & Sin & Square \\ \midrule
 & Method & top 50\% acc & top 50\% acc & top 50\% acc & top 50\% acc & top 50\% acc \\ \midrule
Mfs2 & COG & \textbf{0.88 (0.02)} & \textbf{0.85 (0.06)} & \textbf{0.87 (0.04)} & \textbf{0.79 (0.12)} & \textbf{0.58 (0.05)} \\
 & COG without DC & 0.87 (0.05) & 0.85 (0.01) & 0.79 (0.02) & 0.75 (0.06) & 0.55 (0.02) \\
 & EnsembleMLP & 0.08 (0.007) & 0.78 (0.01) & 0.33 (0.01) & 0.65 (0.08) & 0.45 (0.007) \\ \midrule
Resub & COG & \textbf{0.80 (0.005)} & \textbf{0.87 (0.03)} & \textbf{0.87 (0.0)} & \textbf{0.81 (0.03)} & \textbf{0.89 (0.01)} \\
 & COG without DC & 0.45 (0.23) & 0.72 (0.17) & 0.41 (0.31) & 0.46 (0.23) & 0.75 (0.03) \\
 & EnsembleMLP & 0.21 (0.02) & 0.67 (0.12) & 0.71 (0.06) & 0.28 (0.0) & \textbf{0.92 (0.008)} \\ \bottomrule
\end{tabular}
}
\end{table*}

\begin{table*}[t]
\caption{The results show that the proportion of effective nodes is very low, with an average of $6.19\%$.}
\label{table:mfs2_resub_int}
\centering
\resizebox{0.99\textwidth}{!}{
\begin{tabular}{@{}cccccccccc@{}}
\toprule
\toprule
 & \multicolumn{8}{c}{Mfs2} & \textbf{Avg} \\ \midrule
Stats/Circuit & hyp & log2 & multiplier & square & sin & adder & sqrt & div &  \\ \midrule
Num (effective nodes) & 664.00 & 67.00 & 93.00 & 182.00 & 36.00 & 0.00 & 27.00 & 16.00 &  \\
Num (total nodes) & 64245.00 & 10648.00 & 7821.00 & 5709.00 & 2000.00 & 339.00 & 5673.00 & 9318.00 &  \\
Percent (\%) & 1.03 & 0.63 & 1.19 & 3.19 & 1.80 & 0.00 & 0.48 & 0.17 & \textbf{1.06} \\ \midrule
Stats/Circuit & mem ctrl & priority & router & int2float & cavlc & voter & ctrl & i2c &  \\ \midrule
Num (effective nodes) & 4016.00 & 58.00 & 32.00 & 12.00 & 20.00 & 205.00 & 4.00 & 18.00 &  \\
Num (total nodes) & 17430.00 & 261.00 & 111.00 & 91.00 & 286.00 & 2454.00 & 53.00 & 464.00 &  \\
Percent (\%) & 23.04 & 22.22 & 28.83 & 13.19 & 6.99 & 8.35 & 7.55 & 3.88 & \textbf{14.26} \\ \midrule
 & \multicolumn{8}{c}{Resub} &  \\ \midrule
Stats/Circuit & hyp & log2 & multiplier & square & sin & adder & sqrt & div &  \\ \midrule
Num (effective nodes) & 6745.00 & 106.00 & 120.00 & 763.00 & 18.00 & 126.00 & 121.00 & 30.00 &  \\
Num (total nodes) & 211330.00 & 29370.00 & 24556.00 & 16623.00 & 5039.00 & 1019.00 & 19437.00 & 40772.00 &  \\
Percent (\%) & 3.19 & 0.36 & 0.49 & 4.59 & 0.36 & 12.37 & 0.62 & 0.07 & \textbf{2.76} \\ \midrule
Stats/Circuit & mem ctrl & priority & router & int2float & cavlc & voter & ctrl & i2c &  \\ \midrule
Num (effective nodes) & 999.00 & 93.00 & 2.00 & 4.00 & 24.00 & 974.00 & 17.00 & 62.00 &  \\
Num (total nodes) & 45614.00 & 676.00 & 177.00 & 214.00 & 662.00 & 9756.00 & 108.00 & 1162.00 &  \\
Percent (\%) & 2.19 & 13.76 & 1.13 & 1.87 & 3.63 & 9.98 & 15.74 & 5.34 & \textbf{6.70} \\ \midrule
\textbf{Total} &  &  &  &  &  &  &  &  & \textbf{6.19} \\ \bottomrule
\end{tabular}
}
\end{table*}

\begin{table*}[t]
\caption{The results show that the proportion of effective nodes is low on the Rewrite and Refactor operators, with an average of $11.29\%$ and $7.72\%$, respectively.}
\label{table:mfs2_rewrite_refactor}
\centering
\resizebox{0.99\textwidth}{!}{
\begin{tabular}{@{}ccccccccccc@{}}
\toprule
\toprule
 & \multicolumn{9}{c}{Rewrite} & Avg \\ \midrule
Stats/Circuit & hyp & log2 & multiplier & square & sin & des perf & ethernet & vga lcd & wb conmax &  \\ \midrule
Num (effective nodes) & 83.00 (0.0) & 1574.00 (0.0) & 1716.00 (0.0) & 605.00 (0.0) & 198.00 (0.0) & 6648.00 (0.0) & 16636.00 (0.0) & 34163.00 (0.0) & 2250.00 (0.0) &  \\
Num (total nodes) & 214252.00 (0.0) & 30486.00 (0.0) & 25346.00 (0.0) & 17879.00 (0.0) & 5218.00 (0.0) & 71651.00 (0.0) & 53048.00 (0.0) & 92548.00 (0.0) & 45603.00 (0.0) &  \\
Percent (\%) & 0.04 & 5.16 & 6.77 & 3.38 & 3.79 & 9.28 & 31.36 & 36.91 & 4.93 & \textbf{11.29} \\ \midrule
 & \multicolumn{9}{c}{Refactor} &  \\ \midrule
Stats/Circuit & hyp & log2 & multiplier & square & sin & des perf & ethernet & vga lcd & wb conmax &  \\ \midrule
Num (effective nodes) & 1886.00 (0.0) & 568.00 (0.0) & 268.00 (0.0) & 191.00 (0.0) & 101.00 (0.0) & 3339.00 (0.0) & 12884.00 (0.0) & 25350.00 (0.0) & 4617.00 (0.0) &  \\
Num (total nodes) & 212449.00 (0.0) & 31492.00 (0.0) & 26794.00 (0.0) & 18293.00 (0.0) & 5315.00 (0.0) & 74960.00 (0.0) & 56800.00 (0.0) & 101361.00 (0.0) & 43236.00 (0.0) &  \\
Percent (\%) & 0.89 & 1.80 & 1.00 & 1.04 & 1.90 & 4.45 & 22.68 & 25.01 & 10.68 & \textbf{7.72} \\ \bottomrule
\end{tabular}
}
\end{table*}

\begin{table*}[t]
\caption{We analyze the runtime of commonly used LS operators on open-source circuits. Ratio denotes the ratio of the runtime to that of the Rewrite operator.}
\label{table:operators_time_analysis}
\centering
\resizebox{0.99\textwidth}{!}{ 
\begin{tabular}{@{}ccccccc@{}}
\toprule
\toprule
 & \multicolumn{2}{c}{Log2} & \multicolumn{2}{c}{Hyp} & \multicolumn{2}{c}{Multiplier} \\ \midrule
Operators & RunTime (s) & Ratio & RunTime (s) & Ratio & RunTime (s) & Ratio \\ \cmidrule(r){1-3} \cmidrule(lr){4-5} \cmidrule(l){6-7}
Rewrite & 1.73 (0.09) & 1.00 & 29.97 (0.33) & 1.00 & 0.74 (0.014) & 1.00 \\
Balance & 0.04 (0.0) & 0.02 & 0.36 (0.035) & 0.01 & 0.03 (0.0) & 0.04 \\
Refactor & 2.12 (0.04) & 1.23 & 24.01 (0.07) & 0.80 & 1.72 (0.02) & 2.32 \\
Resub (K=12) & 4.33 (0.042) & 2.50 & 9.17 (0.04) & 0.31 & 1.81 (0.02) & 2.45 \\
Resub (K=16) & 76.44 (0.55) & 44.18 & 116.14 (0.10) & 3.88 & 23.33 (0.05) & 31.53 \\
Mfs2 & 128.25 (0.33) & 74.13 & 259.07 (0.33) & 8.64 & 13.70 (0.07) & 18.51 \\ \midrule
 & \multicolumn{2}{c}{Sin} & \multicolumn{2}{c}{Square} & \multicolumn{2}{c}{Vga lcd} \\ \midrule
Operators & RunTime (s) & Ratio & RunTime (s) & Ratio & RunTime (s) & Ratio \\ \cmidrule(r){1-3} \cmidrule(lr){4-5} \cmidrule(l){6-7}
Rewrite & 0.23 (0.0) & 1.00 & 0.50 (0.04) & 1.00 & 6.99 (0.22) & 1.00 \\
Balance & 0.01 (0.0) & 0.04 & 0.02 (0.0) & 0.04 & 0.38 (0.022) & 0.05 \\
Refactor & 0.21 (0.016) & 0.91 & 1.19 (0.02) & 2.38 & 7.83 (0.12) & 1.12 \\
Resub (K=12) & 0.90 (0.0) & 3.91 & 1.03 (0.005) & 2.06 & 97.32 (0.06) & 13.92 \\
Resub (K=16) & 17.05 (0.016) & 74.13 & 11.87 (0.08) & 23.74 & 649.19 (1.16) & 92.87 \\
Mfs2 & 10.15 (0.06) & 44.13 & 25.33 (0.41) & 50.66 & 128.55 (1.08) & 18.39 \\ \midrule
 & \multicolumn{2}{c}{Ethernet} & \multicolumn{2}{c}{Wb conmax} & \multicolumn{2}{c}{Des perf} \\ \midrule
Operators & RunTime (s) & Ratio & RunTime (s) & Ratio & RunTime (s) & Ratio \\ \cmidrule(r){1-3} \cmidrule(lr){4-5} \cmidrule(l){6-7}
Rewrite & 1.51 (0.03) & 1.00 & 0.86 (0.03) & 1.00 & 1.86 (0.04) & 1.00 \\
Balance & 0.13 (0.005) & 0.09 & 0.07 (0.009) & 0.08 & 0.16 (0.016) & 0.09 \\
Refactor & 1.06 (0.009) & 0.70 & 0.59 (0.016) & 0.69 & 1.37 (0.09) & 0.74 \\
Resub (K=12) & 10.72 (0.02) & 7.10 & 7.96 (0.04) & 9.26 & 23.37 (0.27) & 12.56 \\
Resub (K=16) & 145.77 (0.19) & 96.54 & 149.80 (0.38) & 174.19 & 223.09 (1.44) & 119.94 \\
Mfs2 & 27.53 (0.15) & 18.23 & 25.40 (0.04) & 29.53 & 30.11 (0.065) & 16.19 \\ \midrule
\multicolumn{7}{c}{Avg Time Ratio to Rewrite} \\ \midrule
Operators & Rewrite & Balance & Refactor & Resub (K=12) & Resub (K=16) & Mfs2 \\ \midrule
Time Ratio & 1.00 & 0.05 & 1.21 & \textbf{6.01} & \textbf{73.44} & \textbf{30.94} \\ \bottomrule
\end{tabular}
}
\end{table*}

\begin{table*}[t]
\caption{We analyze the runtime of commonly used LS operators on industrial circuits. Ratio denotes the ratio of the runtime to that of the Rewrite operator.}
\label{table:operators_time_analysis_industrial}
\centering
\resizebox{0.99\textwidth}{!}{
\begin{tabular}{@{}ccccccc@{}}
\toprule
\toprule
 & \multicolumn{2}{c}{f5022} & \multicolumn{2}{c}{f8272} & \multicolumn{2}{c}{c5088} \\ \midrule
Operators & RunTime (s) & Ratio & RunTime (s) & Ratio & RunTime (s) & Ratio \\ \cmidrule(r){1-3} \cmidrule(lr){4-5} \cmidrule(l){6-7}
Rewrite & 36.06 (0.40) & 1.00 & 29.41 (0.15) & 1.00 & 25.05 (0.10) & 1.00 \\
Balance & 0.75 (0.005) & 0.02 & 0.74 (0.04) & 0.03 & 0.76 (0.005) & 0.03 \\
Refactor & 42.24 (0.10) & 1.17 & 24.65 (0.08) & 0.84 & 31.51 (0.33) & 1.26 \\
Resub (K=12) & 228.92 (2.11) & 6.35 & 517.74 (27.07) & 17.60 & 336.13 (22.22) & 13.42 \\
Resub (K=16) & 1144.56 (10.57) & 31.74 & 743.51 (4.53) & 25.28 & 1660.70 (7.03) & 66.30 \\
Mfs2 & 177.93 (0.69) & 4.93 & 76.97 (0.13) & 2.62 & 487.51 (51.68) & 19.46 \\ \midrule
 & \multicolumn{2}{c}{d3151} & \multicolumn{2}{c}{c8449} & \multicolumn{2}{c}{d4067} \\ \midrule
Operators & RunTime (s) & Ratio & RunTime (s) & Ratio & RunTime (s) & Ratio \\ \cmidrule(r){1-3} \cmidrule(lr){4-5} \cmidrule(l){6-7}
Rewrite & 0.94 (0.012) & 1.00 & 1.44 (0.02) & 1.00 & 33.52 (0.15) & 1.00 \\
Balance & 0.03 (0.0) & 0.03 & 0.04 (0.0) & 0.03 & 1.16 (0.008) & 0.03 \\
Refactor & 0.80 (0.008) & 0.85 & 1.43 (0.008) & 0.99 & 44.53 (0.12) & 1.33 \\
Resub (K=12) & 6.13 (0.12) & 6.52 & 8.03 (0.49) & 5.58 & 1061.05 (6.96) & 31.65 \\
Resub (K=16) & 61.22 (0.03) & 65.13 & 72.80 (0.008) & 50.56 & 5300.23 (20.44) & 158.12 \\
Mfs2 & 66.17 (5.25) & 70.39 & 234.58 (45.54) & 162.90 & 201.43 (12.17) & 6.01 \\ \midrule
\multicolumn{7}{c}{Avg Time Ratio to Rewrite} \\ \midrule
Operators & Rewrite & Balance & Refactor & Resub (K=12) & Resub (K=16) & Mfs2 \\ \midrule
Time Ratio & 1.00 & 0.03 & 1.07 & \textbf{13.52} & \textbf{66.19} & \textbf{44.39} \\ \bottomrule
\end{tabular}
}
\end{table*}

\begin{table*}[t]
\caption{We analyze the runtime of the Rewrite and Refactor operators on very large-scale circuits.}
\label{table:rewrite_refactor_operators_time_analysis_epfl_hard}
\centering
\resizebox{0.99\textwidth}{!}{
\begin{tabular}{@{}cccc@{}}
\toprule
\toprule
\multicolumn{2}{c}{Sixteen} & \multicolumn{2}{c}{Twenty} \\ \midrule
Operators & RunTime (s) & Operators & RunTime (s) \\ \midrule
Rewrite & 8502.77 (297.2) & Rewrite & 12715.52 (52.83) \\
Refactor & 7486.48 (212.83) & Refactor & 10297.4 (42.0) \\ \bottomrule
\end{tabular}
}
\end{table*}

\begin{table*}[t]
\caption{We analyze the runtime percentage of the Mfs2 and Resub operators over common LS optimization sequences.}
\label{table:mfs2_resub_runtime_percent}
\centering
\resizebox{0.85\textwidth}{!}{
\begin{tabular}{@{}cccc@{}}
\toprule
\toprule
 & \multicolumn{3}{c}{Mfs2 Operator} \\ \midrule
Circuit & RunTime of Mfs2 & Total RunTime & Percent (\%) \\ \midrule
Log2 & 155.33 (1.31) & 175.46 (2.26) & 88.53 \\
Hyp & 263.91 (0.36) & 272.52 (0.38) & 96.84 \\
Multiplier & 16.69 (0.07) & 23.27 (0.03) & 71.72 \\
Sin & 12.64 (0.27) & 15.34 (0.29) & 82.40 \\
Square & 21.41 (0.07) & 26.6 (0.11) & 80.49 \\
Des perf & 29.51 (0.28) & 51.54 (0.3) & 57.26 \\
Ethernet & 27.73 (0.46) & 49.12 (0.85) & 56.45 \\
Wb conmax & 21.24 (0.54) & 31.32 (0.57) & 67.82 \\
Vga lcd & 189.1 (3.37) & 261.58 (4.42) & 72.29 \\
Average & - & - & \textbf{74.87} \\ \midrule
 & \multicolumn{3}{c}{Resub Operator} \\ \midrule
Circuit & RunTime of Resub & Total RunTime & Percent (\%) \\ \midrule
Log2 & 65.87 (1.44) & 73.82 (2.05) & 89.23 \\
Hyp & 115.86 (0.17) & 235.3 (2.49) & 49.24 \\
Multiplier & 19.91 (0.06) & 25.25 (0.02) & 78.85 \\
Sin & 12.74 (0.02) & 13.73 (0.01) & 92.79 \\
Square & 9.37 (0.13) & 13.93 (0.2) & 67.26 \\
Des perf & 228.12 (13.89) & 237.71 (14.09) & 95.97 \\
Ethernet & 83.54 (0.27) & 88.45 (0.37) & 94.45 \\
Wb conmax & 102.31 (0.25) & 106.14 (0.47) & 96.39 \\
Vga lcd & 403.3 (1.87) & 415.16 (2.4) & 97.14 \\
Average & - & - & \textbf{84.59} \\ \bottomrule
\end{tabular}
}
\end{table*}

\begin{table*}[t]
\caption{We report the recall and optimization performance of the Resub operator incorporated with Random models. And Reduction denotes the reduced number of nodes, i.e., optimization performance. Percent denotes the hyperparameter $k$, i.e., the percent of nodes to apply transformations. Normalized AR denotes the ratio of the AR to that of the default operator.}
\label{table:accuracy_optimization_performance}
\centering
\resizebox{0.9\textwidth}{!}{
\begin{tabular}{@{}cccccccc@{}}
\toprule
\toprule
\multicolumn{4}{c}{Log2} & \multicolumn{4}{c}{Hyp} \\ \midrule
Percent & Recall & And Reduction (AR) & Normalized AR & Percent & Recall & And Reduction (AR) & Normalized AR \\ \cmidrule(r){1-4} \cmidrule(l){5-8}
0.10 & 0.06 & 8.00 (2.0) & 0.07 & 0.10 & 0.11 & 702.00 (16.31) & 0.10 \\
0.20 & 0.15 & 16.00 (0.0) & 0.15 & 0.20 & 0.20 & 1362.67 (21.23) & 0.20 \\
0.30 & 0.25 & 37.50 (0.5) & 0.34 & 0.30 & 0.30 & 2027.00 (8.52) & 0.30 \\
0.40 & 0.31 & 35.00 (2.0) & 0.32 & 0.40 & 0.39 & 2667.33 (30.65) & 0.39 \\
0.50 & 0.48 & 52.00 (1.0) & 0.47 & 0.50 & 0.50 & 3407.33 (38.76) & 0.50 \\
0.60 & 0.60 & 64.50 (2.5) & 0.59 & 0.60 & 0.59 & 4053.33 (25.72) & 0.60 \\
0.70 & 0.72 & 80.50 (2.5) & 0.73 & 0.70 & 0.69 & 4733.33 (53.32) & 0.70 \\
0.80 & 0.81 & 85.00 (1.0) & 0.77 & 0.80 & 0.81 & 5423.33 (21.31) & 0.80 \\
0.90 & 0.87 & 99.50 (2.5) & 0.90 & 0.90 & 0.91 & 6108.33 (20.00) & 0.90 \\
1.00 & 1.00 & 110.00 (0.0) & 1.00 & 1.00 & 1.00 & 6797.00 (0.0) & 1.00 \\ \midrule
\multicolumn{4}{c}{Multiplier} & \multicolumn{4}{c}{Square} \\ \midrule
Percent & Recall & And Reduction (AR) & Normalized AR & Percent & Recall & And Reduction (AR) & Normalized AR \\ \cmidrule(r){1-4} \cmidrule(l){5-8}
0.10 & 0.13 & 8.00 (0.81) & 0.07 & 0.10 & 0.12 & 84.33 (6.23) & 0.11 \\
0.20 & 0.28 & 22.00 (1.41) & 0.18 & 0.20 & 0.15 & 154.66 (9.39) & 0.20 \\
0.30 & 0.40 & 36.00 (4.32) & 0.30 & 0.30 & 0.25 & 239.00 (15.12) & 0.31 \\
0.40 & 0.47 & 45.00 (2.44) & 0.38 & 0.40 & 0.37 & 311.00 (8.28) & 0.40 \\
0.50 & 0.58 & 58.33 (1.24) & 0.49 & 0.50 & 0.47 & 404.33 (14.88) & 0.52 \\
0.60 & 0.69 & 72.66 (2.35) & 0.61 & 0.60 & 0.55 & 463.00 (4.96) & 0.60 \\
0.70 & 0.78 & 81.66 (0.94) & 0.68 & 0.70 & 0.66 & 550.00 (11.04) & 0.71 \\
0.80 & 0.84 & 94.33 (2.86) & 0.79 & 0.80 & 0.76 & 629.00 (1.63) & 0.81 \\
0.90 & 0.97 & 107.66 (2.49) & 0.90 & 0.90 & 0.86 & 696.33 (5.31) & 0.90 \\
1.00 & 1.00 & 120.00 (0.0) & 1.00 & 1.00 & 1.00 & 778.00 (0.0) & 1.00 \\ \midrule
\multicolumn{4}{c}{Des perf} & \multicolumn{4}{c}{Vga lcd} \\ \midrule
Percent & Recall & And Reduction (AR) & Normalized AR & Percent & Recall & And Reduction (AR) & Normalized AR \\ \cmidrule(r){1-4} \cmidrule(l){5-8}
0.10 & 0.10 & 292.00 (8.98) & 0.13 & 0.10 & 0.10 & 7.66 (3.68) & 0.09 \\
0.20 & 0.21 & 505.00 (3.56) & 0.22 & 0.20 & 0.15 & 22.33 (2.49) & 0.26 \\
0.30 & 0.30 & 775.00 (6.53) & 0.34 & 0.30 & 0.20 & 32.00 (3.74) & 0.38 \\
0.40 & 0.41 & 1031.33 (27.34) & 0.45 & 0.40 & 0.34 & 37.00 (6.97) & 0.44 \\
0.50 & 0.51 & 1281.00 (18.02) & 0.56 & 0.50 & 0.46 & 50.66 (5.18) & 0.60 \\
0.60 & 0.60 & 1467.67 (22.23) & 0.64 & 0.60 & 0.54 & 60.33 (3.09) & 0.71 \\
0.70 & 0.71 & 1738.67 (22.16) & 0.75 & 0.70 & 0.61 & 69.00 (4.54) & 0.81 \\
0.80 & 0.81 & 1913.33 (12.92) & 0.83 & 0.80 & 0.80 & 77.00 (3.74) & 0.91 \\
0.90 & 0.90 & 2124.00 (4.55) & 0.92 & 0.90 & 0.90 & 79.33 (3.85) & 0.93 \\
1.00 & 1.00 & 2307.00 (0.0) & 1.00 & 1.00 & 1.00 & 85.00 (0.0) & 1.00 \\ \midrule
\multicolumn{4}{c}{Ethernet} & \multicolumn{4}{c}{Wb conmax} \\ \midrule
Percent & Recall & And Reduction (AR) & Normalized AR & Percent & Recall & And Reduction (AR) & Normalized AR \\ \cmidrule(r){1-4} \cmidrule(l){5-8}
0.10 & 0.13 & 26.33 (1.69) & 0.13 & 0.10 & 0.11 & 325.00 (17.8) & 0.16 \\
0.20 & 0.26 & 52.66 (9.80) & 0.26 & 0.20 & 0.20 & 596.67 (28.52) & 0.29 \\
0.30 & 0.34 & 70.33 (16.54) & 0.34 & 0.30 & 0.30 & 879.67 (27.01) & 0.43 \\
0.40 & 0.47 & 110.66 (15.54) & 0.54 & 0.40 & 0.39 & 1042.33 (21.36) & 0.51 \\
0.50 & 0.58 & 125.00 (9.09) & 0.61 & 0.50 & 0.49 & 1310.33 (20.89) & 0.64 \\
0.60 & 0.67 & 138.33 (2.05) & 0.68 & 0.60 & 0.60 & 1502.33 (3.09) & 0.74 \\
0.70 & 0.77 & 161.00 (9.89) & 0.79 & 0.70 & 0.70 & 1660.00 (53.39) & 0.81 \\
0.80 & 0.82 & 180.33 (3.39) & 0.88 & 0.80 & 0.79 & 1789.33 (8.38) & 0.88 \\
0.90 & 0.92 & 191.00 (8.64) & 0.94 & 0.90 & 0.90 & 1946.33 (8.26) & 0.96 \\
1.00 & 1.00 & 204.00 (0.0) & 1.00 & 1.00 & 1.00 & 2037.00 (0.0) & 1.00 \\ \bottomrule
\end{tabular}
}
\end{table*}

\begin{table*}[t]
\caption{We report the number of applied transformations and runtime of the Resub operator incorporated with a Random model. Percent denotes the hyperparameter $k$, i.e., the percent of nodes to apply transformations. Normalized RT denotes the ratio of the runtime to that of the default operator. 
}
\label{table:percent_runtime}
\centering
\resizebox{0.7\textwidth}{!}{
\begin{tabular}{@{}cccccc@{}}
\toprule
\toprule
\multicolumn{3}{c}{Log2} & \multicolumn{3}{c}{Hyp} \\ \midrule
Percent & Run Time (RT, s) & Normalized RT & Percent & Run Time (RT, s) & Normalized RT \\ \cmidrule(r){1-3} \cmidrule(l){4-6}
0.10 & 6.99 (0.24) & 0.11 & 0.10 & 14.64 (0.03) & 0.13 \\
0.20 & 13.56 (0.07) & 0.21 & 0.20 & 26.73 (0.07) & 0.23 \\
0.30 & 19.24 (0.36) & 0.30 & 0.30 & 37.93 (0.30) & 0.33 \\
0.40 & 26.06 (0.17) & 0.41 & 0.40 & 49.30 (0.21) & 0.43 \\
0.50 & 31.40 (0.13) & 0.50 & 0.50 & 60.50 (0.11) & 0.53 \\
0.60 & 37.80 (0.15) & 0.60 & 0.60 & 71.82 (0.19) & 0.63 \\
0.70 & 44.42 (0.42) & 0.70 & 0.70 & 82.42 (0.43) & 0.72 \\
0.80 & 50.84 (0.42) & 0.81 & 0.80 & 92.93 (0.81) & 0.82 \\
0.90 & 57.32 (0.62) & 0.91 & 0.90 & 103.25 (0.65) & 0.91 \\
1.00 & 63.13 (0.5) & 1.00 & 1.00 & 113.98 (0.71) & 1.00 \\ \midrule 
\multicolumn{3}{c}{Multiplier} & \multicolumn{3}{c}{Square} \\ \midrule
Percent & Run Time (RT, s) & Normalized RT & Percent & Run Time (RT, s) & Normalized RT \\ \cmidrule(r){1-3} \cmidrule(l){4-6}
0.10 & 2.36 (0.04) & 0.12 & 0.10 & 1.15 (0.071) & 0.13 \\
0.20 & 4.32 (0.04) & 0.22 & 0.20 & 2.15 (0.06) & 0.23 \\
0.30 & 6.35 (0.06) & 0.32 & 0.30 & 2.86 (0.03) & 0.31 \\
0.40 & 8.06 (0.10) & 0.41 & 0.40 & 3.89 (0.11) & 0.43 \\
0.50 & 10.41 (0.23) & 0.53 & 0.50 & 4.74 (0.05) & 0.52 \\
0.60 & 12.12 (0.15) & 0.61 & 0.60 & 5.58 (0.05) & 0.61 \\
0.70 & 13.99 (0.04) & 0.71 & 0.70 & 6.50 (0.04) & 0.71 \\
0.80 & 15.92 (0.07) & 0.80 & 0.80 & 7.39 (0.04) & 0.81 \\
0.90 & 17.84 (0.13) & 0.90 & 0.90 & 8.25 (0.05) & 0.90 \\
1.00 & 19.79 (0.02) & 1.00 & 1.00 & 9.15 (0.005) & 1.00 \\ \midrule
\multicolumn{3}{c}{Des perf} & \multicolumn{3}{c}{Vga lcd} \\ \midrule
Percent & Run Time (RT, s) & Normalized RT & Percent & Run Time (RT, s) & Normalized RT \\ \cmidrule(r){1-3} \cmidrule(l){4-6}
0.10 & 37.01 (0.41) & 0.17 & 0.10 & 67.04 (1.38) & 0.17 \\
0.20 & 57.51 (0.77) & 0.27 & 0.20 & 103.03 (1.72) & 0.26 \\
0.30 & 76.87 (0.64) & 0.36 & 0.30 & 138.10 (0.51) & 0.35 \\
0.40 & 97.43 (0.15) & 0.46 & 0.40 & 173.13 (0.89) & 0.44 \\
0.50 & 116.11 (0.31) & 0.55 & 0.50 & 211.32 (1.82) & 0.54 \\
0.60 & 135.97 (0.88) & 0.64 & 0.60 & 246.12 (3.85) & 0.63 \\
0.70 & 154.60 (1.53) & 0.73 & 0.70 & 283.42 (2.39) & 0.73 \\
0.80 & 174.10 (1.04) & 0.82 & 0.80 & 318.32 (2.08) & 0.82 \\
0.90 & 193.21 (1.37) & 0.91 & 0.90 & 352.70 (1.53) & 0.91 \\
1.00 & 211.95 (1.39) & 1.00 & 1.00 & 389.20 (2.91) & 1.00 \\ \midrule
\multicolumn{3}{c}{Ethernet} & \multicolumn{3}{c}{Wb conmax} \\ \midrule
Percent & Run Time (RT, s) & Normalized RT & Percent & Run Time (RT, s) & Normalized RT \\ \cmidrule(r){1-3} \cmidrule(l){4-6}
0.10 & 16.72 (0.40) & 0.21 & 0.10 & 12.03 (0.4) & 0.12 \\
0.20 & 24.12 (0.36) & 0.30 & 0.20 & 22.30 (0.4) & 0.22 \\
0.30 & 31.21 (0.53) & 0.38 & 0.30 & 33.03 (0.26) & 0.33 \\
0.40 & 39.16 (1.04) & 0.48 & 0.40 & 42.97 (0.62) & 0.42 \\
0.50 & 45.29 (1.09) & 0.56 & 0.50 & 53.14 (0.4) & 0.52 \\
0.60 & 52.99 (0.45) & 0.65 & 0.60 & 62.52 (0.35) & 0.62 \\
0.70 & 59.35 (0.19) & 0.73 & 0.70 & 72.15 (0.13) & 0.71 \\
0.80 & 66.56 (0.97) & 0.82 & 0.80 & 82.14 (0.21) & 0.81 \\
0.90 & 73.85 (0.35) & 0.91 & 0.90 & 91.82 (0.25) & 0.91 \\
1.00 & 81.34 (0.78) & 1.00 & 1.00 & 101.43 (0.16) & 1.00 \\ \bottomrule
\end{tabular}
}
\end{table*}

\begin{table*}[t]
\caption{The results show that only applying transformations on effective nodes (i.e., Orcale) significantly improves the runtime while keeping the same size and depth of the Log2 circuit. Note that Nd denotes the number of nodes on the circuit, i.e., the size of the circuit. Lev denotes the depth of the circuit.}
\label{oracle_prediction_results}
\centering
\resizebox{0.99\textwidth}{!}{
\begin{tabular}{@{}cccccc@{}}
\toprule
\toprule
 & \multicolumn{5}{c}{Log2} \\ \midrule
\multirow{2}{*}{Method} & \multirow{2}{*}{Nd} & \multirow{2}{*}{\begin{tabular}[c]{@{}c@{}}Decrease\\      (Nd, \%)\end{tabular}} & \multirow{2}{*}{Lev} & \multirow{2}{*}{Time (s)} & \multirow{2}{*}{\begin{tabular}[c]{@{}c@{}}Improvement\\      (Time, \%)\end{tabular}} \\
 &  &  &  &  &  \\ \midrule
Oracle & 10621.00 & \textbf{0.00} & 104.00 & 5.40 & \textbf{96.54} \\
Default (Mfs2) & 10621.00 (0.0) & NA & 104.00 (0.0) & 156.19 (1.66) & NA \\ \bottomrule
\end{tabular}
}
\end{table*}

    \subsection{More Motivating Results}\label{appendix:motivating_results}        
        Here we present more motivating results. 

        \subsubsection{Ineffective Node-Level Transformations problem}\label{appendix:motivation_int_problem}
        As shown in Table \ref{table:mfs2_resub_int}, the Mfs2 and Resub operators apply a large number of ineffective node-level transformations, with average of $93.81\%$. Moreover, the results in Table \ref{table:mfs2_rewrite_refactor} demonstrate that the Rewrite and Refactor operators apply a large number of ineffective node-level transformations as well.  

        \subsubsection{Analysis on the Runtime of Commonly Used LS Operators}\label{appendix:motivation_time_analysis} 

        \textbf{Runtime Percentage of the Mfs2 and Resub Operators over Common LS Optimization Processes} To evaluate the runtime percentage of applying the Mfs2 and Resub operators over common LS process (i.e., common LS optimization sequences of operators), we apply the following optimization sequence flows. (1) For the Resub, we apply the flow \textit{strash; resyn2; resub -K 16 -N 3 -z}, which is commonly used in industrial LS process \cite{resub}. Note that \textit{resyn2} denotes a fixed sequence of LS operators, i.e., \textit{balance; rewrite; refactor; balance; rewrite; rewrite -z; balance; refactor -z; rewrite -z; balance}.
        (2) For the Mfs2, we apply the flow \textit{strash; dch; if -C 12; mfs2 -W 4 -M 5000 -l}, which is commonly used in industrial LS process as well \cite{mfs2}.

        The results in Table \ref{table:mfs2_resub_runtime_percent} show that the runtime percentage of applying the Mfs2 and Resub operators is about 79\% of the total runtime of common LS optimization sequences. Thus, the results demonstrate that the runtime of applying the two operators acts as a bottleneck to the efficiency of LS.
        
        \textbf{Runtime of Operators} To evaluate the runtime of applying the commonly used operators, we apply the following optimization sequnece flows. (1) Given a logic optimization operator X, we apply the flow \textit{strash; X}. Specifically, we apply the flow for the Rewrite, Refactor, Resub, and Balance operators. (2) Given a post-mapping optimization operator X, we apply the flow \textit{strash; if -C 12; X}. Speicifically, we apply the flow for the Mfs2 operator.  
        
        We provide detailed results on the runtime analysis of these operators in the industrial setting in Tables \ref{table:operators_time_analysis} and \ref{table:operators_time_analysis_industrial}. The results demonstrate that applying the Resub and Mfs2 operators take the longest runtime among the commonly used LS operators. For the Resub operator, $K$ is an important hyperparameter, and represents the number of primary input nodes of subgraphs at each node when applying the operator. 
        Moreover, the results in Tables \ref{table:operators_time_analysis} and \ref{table:operators_time_analysis_industrial} demonstrate that applying the Rewrite and Refactor operators are much faster than the Resub and Mfs2 operators. Nevertheless, we found that applying the Rewrite and Refactor operators are highly time-consuming on very large-scale circuits as shown in Table \ref{table:rewrite_refactor_operators_time_analysis_epfl_hard}. Therefore, it is also valuable to improve the efficiency of the Rewrite and Refactor operators. Fortunately, the Rewrite/Refactor operators follow the same paradigm as the Resub/Mfs2 operators as shown in Fig. \ref{method:prunex} in the main text. Moreover, the results in Table \ref{table:mfs2_rewrite_refactor} demonstrate that the Rewrite and Refactor operators apply a large number of ineffective node-level transformations as well. Therefore, our proposed PruneX is applicable to the Rewrite and Refactor operators as well to improve their efficiency, especially on very large-scale circuits. We provide more discussion on how to apply PruneX to the two operators in Appendix \ref{appendix:discuss_apply_to_rewrite}.

        \subsubsection{More Details of the Out-of-Distribution (OOD) Generalization Problem in LS}\label{appendix:motivation_ood} 
        In terms of the visualization experiments in Fig. \ref{fig:generalization_motivation_tsne_visu_sample_batch} in the main text, we present more implementation details. We use the t-distributed stochastic neighbor embedding (t-SNE) \cite{tsne} algorithm to reduce the node features to two-dimensional space. That is, each point illustrates a reduced node feature. Moreover, the visualized points of the training data points can be too dense, as the training data points are much more than the testing data points. Thus, for visual clarity, we sample the same number of training data points as the testing set for visualization. In addition, we visualize data points from different circuits as well in Fig. \ref{fig:more_visualization}. The results show that the data distributions from different circuits are similar but different, which demonstrates the reasonableness of our circuit domain formulation as well.  

        
        \subsubsection{The Importance of the Prediction Recall on Optimization Performance}\label{appendix:recall_qor}
        To analyze the relationship between the prediction recall of effective nodes and the optimization performance of operators, we evaluate the optimization performance of the Random method with different values of the hyperparameter $k$. Note that Random is a baseline that randomly predicts a score between $\left[0,1\right]$ for each node, and selects the top $k$ nodes to apply node-level transformations. Specifically, we report the recall and optimization performance (i.e., And Reduction) of Random with different values of $k$ in Table \ref{table:accuracy_optimization_performance}. 
        The results show that the value of $k$ is approximately linearly positively correlated with the recall, and the recall is approximately linearly positively correlated with the optimization performance as well. Therefore, in order not to degrade the optimization performance of operators, the prediction recall of our model should be as high as possible. 
        Thus, it is critical to tackle the out-of-distribution generalization problem in LS to improve the prediction recall on unseen circuits. 
        

    \subsection{Oracle Prediction Results}\label{exps:oracle_results}
        Here we present the oracle prediction results on the Mfs2 operator and Log2 circuit. 
        To evaluate whether only applying transformations on effective nodes can achieve similar optimization performance to that of the default operator, we conducted the following oracle prediction experiment. Specifically, we apply the Mfs2 operator to the Log2 circuit and recorded the set of node ids for all effective nodes. Based on this set, we implemented an Oracle version of Mfs2 that only apply transformations to nodes in the set.
        The results in Table \ref{oracle_prediction_results} show that Oracle significantly improves the runtime while achieving the same size (i.e., the number of nodes) and depth (i.e., level) compared to Default (i.e., the default Mfs2 operator). 

    \subsection{The Relationship between the Number of Nodes to Apply Transformations and Runtime}\label{appendix:num_trans_runtime} 

    To evaluate whether reducing node-level transformations can reduce the runtime of operators, we test the Random method with different values of $k$ on open-source circuits. Note that Random is a baseline that randomly predicts a score between $\left[0,1\right]$ for each node, and selects the top $k$ nodes to apply node-level transformations. 
    The results in Table \ref{table:percent_runtime} demonstrate that the runtime of applying the resub operator significantly increases with the number of applied node-level transformations. Therefore, our method can significantly reduce the runtime of applying operators by reducing a large number of ineffective node-level transformations.

\begin{table*}[t]
\caption{Detailed offline evaluation results using Evaluation Strategy 1 on the Mfs2 operator and open-source circuits.}
\label{table:mfs2_generalize_one_benchmarks}
\centering
\resizebox{0.99\textwidth}{!}{
\begin{tabular}{@{}ccccccccccl@{}}
\toprule
\toprule
 & \multicolumn{3}{c}{Log2} & \multicolumn{3}{c}{Hyp} & \multicolumn{3}{c}{Multiplier} &  \\ \midrule
Method & top 40\% acc & top 50\% acc & top 60\% acc & top 40\% acc & top 50\% acc & top 60\% acc & top 40\% acc & top 50\% acc & top 60\% acc &  \\ \cmidrule(r){1-4} \cmidrule(lr){5-7} \cmidrule(l){8-10}
COG & \textbf{0.89 (0.04)} & \textbf{0.97 (0.01)} & \textbf{0.99 (0.01)} & \textbf{0.79 (0.04)} & \textbf{0.90 (0.03)} & \textbf{0.95 (0.008} & \textbf{0.91 (0.03)} & \textbf{0.96 (0.03)} & \textbf{0.98 (0.01)} &  \\
EnsembleMLP & 0.61 (0.0) & 0.65 (0.007) & 0.85 (0.03) & 0.62 (0.005) & 0.73 (0.004) & 0.82 (0.003) & 0.63 (0.0) & 0.71 (0.0) & 0.75 (0.0) &  \\
Random & 0.31 & 0.48 & 0.60 & 0.39 & 0.50 & 0.59 & 0.47 & 0.58 & 0.69 &  \\ \midrule
 & \multicolumn{3}{c}{Sin} & \multicolumn{3}{c}{Square} & \multicolumn{3}{c}{Vga lcd} &  \\ \midrule
Method & top 40\% acc & top 50\% acc & top 60\% acc & top 40\% acc & top 50\% acc & top 60\% acc & top 40\% acc & top 50\% acc & top 60\% acc &  \\ \cmidrule(r){1-4} \cmidrule(lr){5-7} \cmidrule(l){8-10}
COG & \textbf{0.71 (0.06)} & \textbf{0.87 (0.04)} & \textbf{0.95 (0.02)} & \textbf{0.90 (0.01)} & \textbf{0.94 (0.002)} & \textbf{0.95 (0.0)} & \textbf{0.93 (0.007)} & \textbf{0.95 (0.01)} & \textbf{0.98 (0.007)} &  \\
EnsembleMLP & 0.56 (0.0) & 0.58 (0.0) & 0.67 (0.0) & 0.85 (0.02) & 0.91 (0.004) & 0.92 (0.0) & 0.19 (0.04) & 0.29 (0.02) & 0.53 (0.02) &  \\
Random & 0.47 & 0.64 & 0.72 & 0.37 & 0.47 & 0.55 & 0.34 & 0.46 & 0.54 &  \\ \midrule
 & \multicolumn{3}{c}{Ethernet} & \multicolumn{3}{c}{Wb conmax} & \multicolumn{3}{c}{Des perf} &  \\ \midrule
Method & top 40\% acc & top 50\% acc & top 60\% acc & top 40\% acc & top 50\% acc & top 60\% acc & top 40\% acc & top 50\% acc & top 60\% acc &  \\ \cmidrule(r){1-4} \cmidrule(lr){5-7} \cmidrule(l){8-10}
COG & \textbf{0.91 (0.09)} & \textbf{0.96 (0.04)} & \textbf{0.97 (0.03)} & \textbf{0.69 (0.04)} & \textbf{0.81 (0.05)} & \textbf{0.89 (0.04)} & \textbf{0.53 (0.0006)} & \textbf{0.68 (0.0007)} & \textbf{0.82 (0.0004)} &  \\
EnsembleMLP & 0.53 (0.004) & 0.62 (0.01) & 0.63 (0.03) & 0.43 (0.07) & 0.57 (0.05) & 0.68 (0.03) & 0.44 (0.07) & 0.49 (0.06) & 0.58 (0.03) &  \\
Random & 0.47 & 0.58 & 0.67 & 0.39 & 0.49 & 0.60 & 0.41 & 0.51 & 0.60 &  \\ \bottomrule
\end{tabular}

}
\end{table*}

\begin{table*}[t]
\caption{Detailed offline evaluation results using Evaluation Strategy 1 on the Resub operator and open-source circuits.}
\label{table:resub_generalize_one_benchmarks}
\centering
\resizebox{0.99\textwidth}{!}{
\begin{tabular}{@{}ccccccccccc@{}}
\toprule
\toprule
 & \multicolumn{3}{c}{Log2} & \multicolumn{3}{c}{Hyp} & \multicolumn{3}{c}{Multiplier} &  \\ \midrule
Method & top 40\% acc & top 50\% acc & top 60\% acc & top 40\% acc & top 50\% acc & top 60\% acc & top 40\% acc & top 50\% acc & top 60\% acc &  \\ \cmidrule(r){1-4} \cmidrule(lr){5-7} \cmidrule(l){8-10}
COG & \textbf{0.66 (0.02)} & \textbf{0.88 (0.02)} & \textbf{0.97 (0.004)} & \textbf{0.88 (0.008)} & \textbf{0.91 (0.02)} & 0.94 (0.03) & 0.74 (0.17) & \textbf{0.97 (0.02)} & \textbf{0.99 (0.003)} &  \\
EnsembleMLP & 0.39 (0.1) & 0.57 (0.2) & 0.62 (0.20) & 0.72 (0.02) & 0.86 (0.005) & \textbf{0.97 (0.01)} & \textbf{0.88 (0.12)} & 0.93 (0.08) & 0.99 (0.01) &  \\
Random & 0.39 & 0.48 & 0.63 & 0.40 & 0.50 & 0.60 & 0.34 & 0.43 & 0.58 &  \\ \midrule
 & \multicolumn{3}{c}{Sin} & \multicolumn{3}{c}{Square} & \multicolumn{3}{c}{Vga lcd} &  \\ \midrule
Method & top 40\% acc & top 50\% acc & top 60\% acc & top 40\% acc & top 50\% acc & top 60\% acc & top 40\% acc & top 50\% acc & top 60\% acc &  \\ \cmidrule(r){1-4} \cmidrule(lr){5-7} \cmidrule(l){8-10}
COG & 0.46 (0.17) & \textbf{0.69 (0.07)} & \textbf{0.76 (0.07)} & \textbf{0.60 (0.11)} & \textbf{0.73 (0.08)} & \textbf{0.81 (0.07)} & 0.47 (0.34) & \textbf{0.77 (0.04)} & \textbf{0.94 (0.02)} &  \\
EnsembleMLP & \textbf{0.48 (0.03)} & 0.63 (0.1) & 0.76 (0.07) & 0.39 (0.09) & 0.52 (0.12) & 0.63 (0.08) & \textbf{0.68 (0.03)} & 0.71 (0.01) & 0.87 (0.01) &  \\
Random & 0.42 & 0.54 & 0.57 & 0.38 & 0.48 & 0.59 & 0.34 & 0.46 & 0.54 &  \\ \midrule
 & \multicolumn{3}{c}{Ethernet} & \multicolumn{3}{c}{Wb conmax} & \multicolumn{3}{c}{Des perf} &  \\ \midrule
Method & top 40\% acc & top 50\% acc & top 60\% acc & top 40\% acc & top 50\% acc & top 60\% acc & top 40\% acc & top 50\% acc & top 60\% acc &  \\ \cmidrule(r){1-4} \cmidrule(lr){5-7} \cmidrule(l){8-10}
COG & \textbf{0.82 (0.02)} & \textbf{0.98 (0.009)} & \textbf{0.99 (0.0)} & \textbf{0.66 (0.07)} & \textbf{0.88 (0.08)} & \textbf{0.96 (0.01)} & \textbf{0.55 (0.07)} & \textbf{0.72 (0.05)} & \textbf{0.89 (0.009)} &  \\
EnsembleMLP & 0.61 (0.01) & 0.82 (0.04) & 0.87 (0.05) & 0.18 (0.26) & 0.18 (0.26) & 0.40 (0.11) & 0.36 (0.04) & 0.46 (0.03) & 0.56 (0.03) &  \\
Random & 0.47 & 0.58 & 0.67 & 0.39 & 0.49 & 0.60 & 0.41 & 0.51 & 0.60 &  \\ \bottomrule
\end{tabular}
}
\end{table*}

\begin{table*}[t]
\caption{Detailed offline evaluation results using Evaluation Strategy 2 on the Mfs2 and Resub operators and open-source circuits.}
\label{table:mfs2_resub_generalize_from_iwls_to_epfl}
\centering
\resizebox{0.99\textwidth}{!}{
\begin{tabular}{@{}ccccccc@{}}
\toprule
\toprule
 &  & Log2 & Hyp & Multiplier & Sin & Square \\ \midrule
Operator & Method & top 50\% acc & top 50\% acc & top 50\% acc & top 50\% acc & top 50\% acc \\ \midrule
mfs2 & COG & \textbf{0.88 (0.02)} & \textbf{0.85 (0.06)} & \textbf{0.87 (0.04)} & \textbf{0.79 (0.12)} & \textbf{0.58 (0.05)} \\
 & EnsembleMLP & 0.08 (0.007) & 0.78 (0.01) & 0.33 (0.01) & 0.65 (0.08) & 0.45 (0.007) \\
 & Random & 0.48 & 0.50 & 0.58 & 0.54 & 0.47 \\ \midrule
resub & COG & \textbf{0.80 (0.005)} & \textbf{0.87 (0.03)} & \textbf{0.87 (0.0)} & \textbf{0.81 (0.03)} & 0.89 (0.01) \\
 & EnsembleMLP & 0.21 (0.02) & 0.67 (0.12) & 0.71 (0.06) & 0.28 (0.0) & \textbf{0.92 (0.008)} \\
 & Random & 0.48 & 0.50 & 0.43 & 0.54 & 0.48 \\ \bottomrule
\end{tabular}
}
\end{table*}

\begin{table*}[t]
\caption{Detailed online evaluation results using Evaluation Strategy 1 on the Mfs2 operator and open-source circuits. Improvement denotes the improvement of our PruneX compared to the default operator.}
\label{table:mfs2_generalize_in_iwls_epfl_online}
\centering
\resizebox{0.99\textwidth}{!}{
\begin{tabular}{@{}cccccccccccccccc@{}}
\toprule
\toprule
 & \multicolumn{5}{c}{Log2} & \multicolumn{5}{c}{Hyp} & \multicolumn{5}{c}{Multiplier} \\ \midrule
\multirow{2}{*}{Method} & \multirow{2}{*}{Nd} & \multirow{2}{*}{\begin{tabular}[c]{@{}c@{}}Decrease\\      (Nd, \%)\end{tabular}} & \multirow{2}{*}{Lev} & \multirow{2}{*}{Time (s)} & \multirow{2}{*}{\begin{tabular}[c]{@{}c@{}}Improvement\\      (Time, \%)\end{tabular}} & \multirow{2}{*}{Nd} & \multirow{2}{*}{\begin{tabular}[c]{@{}c@{}}Decrease\\      (Nd, \%)\end{tabular}} & \multirow{2}{*}{Lev} & \multirow{2}{*}{Time (s)} & \multirow{2}{*}{\begin{tabular}[c]{@{}c@{}}Improvement\\      (Time, \%)\end{tabular}} & \multirow{2}{*}{Nd} & \multirow{2}{*}{\begin{tabular}[c]{@{}c@{}}Decrease\\      (Nd, \%)\end{tabular}} & \multirow{2}{*}{Lev} & \multirow{2}{*}{Time (s)} & \multirow{2}{*}{\begin{tabular}[c]{@{}c@{}}Improvement\\      (Time, \%)\end{tabular}} \\
 &  &  &  &  &  &  &  &  &  &  &  &  &  &  &  \\ \cmidrule(r){1-6} \cmidrule(lr){7-11} \cmidrule(l){12-16}
PruneX (COG) & 10621.00 (0.0) & 0.00 & 104.00 (0.0) & 137.87 (3.96) & \textbf{11.73} & 63819.00 (58.65) & -0.37 & 8259.00 (0.0) & 149.97 (19.89) & \textbf{45.30} & 7800.33 (1.24) & -0.02 & 87.00 (0.0) & 9.04 (0.37) & \textbf{45.24} \\
Default (Mfs2) & 10621.00 (0.0) & NA & 104.00 (0.0) & 156.19 (1.66) & NA & 63581.00 (0.0) & NA & 8259.00 (0.0) & 274.13 (9.90) & NA & 7799.00 (0.0) & NA & 87.00 (0.0) & 16.69 (0.07) & NA \\ \midrule
 & \multicolumn{5}{c}{Sin} & \multicolumn{5}{c}{Square} & \multicolumn{5}{c}{Vga lcd} \\
\multirow{2}{*}{Method} & \multirow{2}{*}{Nd} & \multirow{2}{*}{\begin{tabular}[c]{@{}c@{}}Decrease\\      (Nd, \%)\end{tabular}} & \multirow{2}{*}{Lev} & \multirow{2}{*}{Time (s)} & \multirow{2}{*}{\begin{tabular}[c]{@{}c@{}}Improvement\\      (Time, \%)\end{tabular}} & \multirow{2}{*}{Nd} & \multirow{2}{*}{\begin{tabular}[c]{@{}c@{}}Decrease\\      (Nd, \%)\end{tabular}} & \multirow{2}{*}{Lev} & \multirow{2}{*}{Time (s)} & \multirow{2}{*}{\begin{tabular}[c]{@{}c@{}}Improvement\\      (Time, \%)\end{tabular}} & \multirow{2}{*}{Nd} & \multirow{2}{*}{\begin{tabular}[c]{@{}c@{}}Decrease\\      (Nd, \%)\end{tabular}} & \multirow{2}{*}{Lev} & \multirow{2}{*}{Time (s)} & \multirow{2}{*}{\begin{tabular}[c]{@{}c@{}}Improvement\\      (Time, \%)\end{tabular}} \\
 &  &  &  &  &  &  &  &  &  &  &  &  &  &  &  \\ \cmidrule(r){1-6} \cmidrule(lr){7-11} \cmidrule(l){12-16}
PruneX (COG) & 1991.00 (0.81) & -0.05 & 53.00 (0.0) & 10.52 (0.41) & \textbf{15.11} & 5701.67 (0.47) & -0.01 & 83.00 (0.0) & 9.30 (0.43) & \textbf{56.61} & 38328.00 (0.0) & -0.01 & 7.00 (0.0) & 96.16 (3.02) & \textbf{49.15} \\
Default (Mfs2) & 1990.00 (0.0) & NA & 53.00 (0.0) & 12.38 (0.016) & NA & 5701.00 (0.0) & NA & 83.00 (0.0) & 21.41 (0.071) & NA & 38326.00 (0.0) & NA & 7.00 (0.0) & 189.10 (3.37) & NA \\ \midrule
 & \multicolumn{5}{c}{Ethernet} & \multicolumn{5}{c}{Wb conmax} & \multicolumn{5}{c}{Des perf} \\
\multirow{2}{*}{Method} & \multirow{2}{*}{Nd} & \multirow{2}{*}{\begin{tabular}[c]{@{}c@{}}Decrease\\      (Nd, \%)\end{tabular}} & \multirow{2}{*}{Lev} & \multirow{2}{*}{Time (s)} & \multirow{2}{*}{\begin{tabular}[c]{@{}c@{}}Improvement\\      (Time, \%)\end{tabular}} & \multirow{2}{*}{Nd} & \multirow{2}{*}{\begin{tabular}[c]{@{}c@{}}Decrease\\      (Nd, \%)\end{tabular}} & \multirow{2}{*}{Lev} & \multirow{2}{*}{Time (s)} & \multirow{2}{*}{\begin{tabular}[c]{@{}c@{}}Improvement\\      (Time, \%)\end{tabular}} & \multirow{2}{*}{Nd} & \multirow{2}{*}{\begin{tabular}[c]{@{}c@{}}Decrease\\      (Nd, \%)\end{tabular}} & \multirow{2}{*}{Lev} & \multirow{2}{*}{Time (s)} & \multirow{2}{*}{\begin{tabular}[c]{@{}c@{}}Improvement\\      (Time, \%)\end{tabular}} \\
 &  &  &  &  &  &  &  &  &  &  &  &  &  &  &  \\ \cmidrule(r){1-6} \cmidrule(lr){7-11} \cmidrule(l){12-16}
PruneX (COG) & 13639.33 (0.47) & -0.01 & 9.00 (0.0) & 11.42 (0.48) & \textbf{58.82} & 16813.33 (87.64) & -1.84 & 9.00 (0.0) & 10.41 (0.25) & \textbf{51.04} & 31139.00 (7.78) & -0.93 & 6.00 (0.0) & 18.98 (0.25) & \textbf{35.68} \\
Default (Mfs2) & 13638.00 (0.0) & NA & 9.00 (0.0) & 27.73 (0.45) & NA & 16509.00 (0.0) & NA & 9.00 (0.0) & 21.24 (0.53) & NA & 30853.00 (0.0) & NA & 6.00 (0.0) & 29.51 (0.28) & NA \\ \bottomrule
\end{tabular}
}
\end{table*}

\begin{table*}[t]
\caption{Detailed online evaluation results using Evaluation Strategy 1 on the Resub operator and open-source circuits. Improvement denotes the improvement of our PruneX compared to the default operator.}
\label{table:resub_generalize_in_iwls_epfl_online}
\centering
\resizebox{0.99\textwidth}{!}{
\begin{tabular}{@{}cccccccccccccccc@{}}
\toprule
\toprule
 & \multicolumn{5}{c}{Log2} & \multicolumn{5}{c}{Hyp} & \multicolumn{5}{c}{Multiplier} \\ \midrule
\multirow{2}{*}{Method} & \multirow{2}{*}{And} & \multirow{2}{*}{\begin{tabular}[c]{@{}c@{}}Decrease\\      (And, \%)\end{tabular}} & \multirow{2}{*}{Lev} & \multirow{2}{*}{Time (s)} & \multirow{2}{*}{\begin{tabular}[c]{@{}c@{}}Improvement\\      (Time, \%)\end{tabular}} & \multirow{2}{*}{And} & \multirow{2}{*}{\begin{tabular}[c]{@{}c@{}}Decrease\\      (And, \%)\end{tabular}} & \multirow{2}{*}{Lev} & \multirow{2}{*}{Time (s)} & \multirow{2}{*}{\begin{tabular}[c]{@{}c@{}}Improvement\\      (Time, \%)\end{tabular}} & \multirow{2}{*}{And} & \multirow{2}{*}{\begin{tabular}[c]{@{}c@{}}Decrease\\      (And, \%)\end{tabular}} & \multirow{2}{*}{Lev} & \multirow{2}{*}{Time (s)} & \multirow{2}{*}{\begin{tabular}[c]{@{}c@{}}Improvement\\      (Time, \%)\end{tabular}} \\
 &  &  &  &  &  &  &  &  &  &  &  &  &  &  &  \\ \cmidrule(r){1-6} \cmidrule(lr){7-11} \cmidrule(l){12-16}
PruneX (COG) & 29271.00 (3.74) & -0.04 & 376.00 (0.0) & 35.73 (0.72) & \textbf{44.92} & 205183.33 (177.48) & -0.32 & 24794.00 (0.0) & 37.34 (0.79) & \textbf{67.77} & 24440.00 (3.26) & -0.02 & 262.00 (0.0) & 11.07 (0.68) & \textbf{44.90} \\
Default (Resub) & 29260.00 (0.0) & NA & 376.00 (0.0) & 64.87 (0.86) & NA & 204533.00 (0.0) & NA & 24792.00 (0.0) & 115.86 (0.16) & NA & 24436.00 (0.0) & NA & 262.00 (0.0) & 20.09 (0.10) & NA \\ \midrule
 & \multicolumn{5}{c}{Sin} & \multicolumn{5}{c}{Square} & \multicolumn{5}{c}{Vga lcd} \\
\multirow{2}{*}{Method} & \multirow{2}{*}{And} & \multirow{2}{*}{\begin{tabular}[c]{@{}c@{}}Decrease\\      (And, \%)\end{tabular}} & \multirow{2}{*}{Lev} & \multirow{2}{*}{Time (s)} & \multirow{2}{*}{\begin{tabular}[c]{@{}c@{}}Improvement\\      (Time, \%)\end{tabular}} & \multirow{2}{*}{And} & \multirow{2}{*}{\begin{tabular}[c]{@{}c@{}}Decrease\\      (And, \%)\end{tabular}} & \multirow{2}{*}{Lev} & \multirow{2}{*}{Time (s)} & \multirow{2}{*}{\begin{tabular}[c]{@{}c@{}}Improvement\\      (Time, \%)\end{tabular}} & \multirow{2}{*}{And} & \multirow{2}{*}{\begin{tabular}[c]{@{}c@{}}Decrease\\      (And, \%)\end{tabular}} & \multirow{2}{*}{Lev} & \multirow{2}{*}{Time (s)} & \multirow{2}{*}{\begin{tabular}[c]{@{}c@{}}Improvement\\      (Time, \%)\end{tabular}} \\
 &  &  &  &  &  &  &  &  &  &  &  &  &  &  &  \\ \cmidrule(r){1-6} \cmidrule(lr){7-11} \cmidrule(l){12-16}
PruneX (COG) & 5025.67 (1.69) & -0.15 & 177.00 (0.0) & 8.15 (0.67) & \textbf{37.88} & 16053.67 (73.05) & -1.32 & 248.00 (0.0) & 4.92 (0.22) & \textbf{47.49} & 90870.67 (11.78) & -0.08 & 19.00 (0.0) & 242.76 (83.81) & \textbf{39.81} \\
Default (Resub) & 5018.00 (0.0) & NA & 177.00 (0.0) & 13.12 (0.12) & NA & 15845.00 (0.0) & NA & 248.00 (0.0) & 9.37 (0.13) & NA & 90795.00 (0.0) & NA & 19.00 (0.0) & 403.30 (1.86) & NA \\ \midrule
 & \multicolumn{5}{c}{Ethernet} & \multicolumn{5}{c}{Wb conmax} & \multicolumn{5}{c}{Des perf} \\
\multirow{2}{*}{Method} & \multirow{2}{*}{And} & \multirow{2}{*}{\begin{tabular}[c]{@{}c@{}}Decrease\\      (And, \%)\end{tabular}} & \multirow{2}{*}{Lev} & \multirow{2}{*}{Time (s)} & \multirow{2}{*}{\begin{tabular}[c]{@{}c@{}}Improvement\\      (Time, \%)\end{tabular}} & \multirow{2}{*}{And} & \multirow{2}{*}{\begin{tabular}[c]{@{}c@{}}Decrease\\      (And, \%)\end{tabular}} & \multirow{2}{*}{Lev} & \multirow{2}{*}{Time (s)} & \multirow{2}{*}{\begin{tabular}[c]{@{}c@{}}Improvement\\      (Time, \%)\end{tabular}} & \multirow{2}{*}{And} & \multirow{2}{*}{\begin{tabular}[c]{@{}c@{}}Decrease\\      (And, \%)\end{tabular}} & \multirow{2}{*}{Lev} & \multirow{2}{*}{Time (s)} & \multirow{2}{*}{\begin{tabular}[c]{@{}c@{}}Improvement\\      (Time, \%)\end{tabular}} \\
 &  &  &  &  &  &  &  &  &  &  &  &  &  &  &  \\ \cmidrule(r){1-6} \cmidrule(lr){7-11} \cmidrule(l){12-16}
PruneX (COG) & 43415.00 (64.65) & -0.16 & 26.00 (0.0) & 55.27 (10.02) & \textbf{33.84} & 39277.00 (120.14) & -0.39 & 18.00 (0.0) & 85.87 (1.12) & \textbf{16.07} & 68458.00 (114.77) & -1.88 & 16.00 (0.0) & 110.89 (10.31) & \textbf{51.39} \\
Default (Resub) & 43345.00 (0.0) & NA & 26.00 (0.0) & 83.54 (0.27) & NA & 39126.00 (0.0) & NA & 18.00 (0.0) & 102.31 (0.24) & NA & 67193.00 (0.0) & NA & 16.00 (0.0) & 228.12 (13.89) & NA \\ \bottomrule
\end{tabular}
}
\end{table*}

\begin{table*}[t]
\caption{Detailed online evaluation results using Evaluation Strategy 2 on the Mfs2 and Resub operators and open-source circuits. Improvement denotes the improvement of our PruneX compared to the default operator.}
\label{table:mfs2_resub_generalize_from_iwls_to_epfl_online}
\centering
\resizebox{0.99\textwidth}{!}{
\begin{tabular}{@{}cccccccccccc@{}}
\toprule
\toprule
 & \multicolumn{11}{c}{Log2} \\ \midrule
\multirow{2}{*}{Method} & \multirow{2}{*}{Nd} & \multirow{2}{*}{\begin{tabular}[c]{@{}c@{}}Decrease\\      (Nd, \%)\end{tabular}} & \multirow{2}{*}{Lev} & \multirow{2}{*}{Time (s)} & \multirow{2}{*}{\begin{tabular}[c]{@{}c@{}}Improvement\\      (Time, \%)\end{tabular}} & \multirow{2}{*}{Method} & \multirow{2}{*}{And} & \multirow{2}{*}{\begin{tabular}[c]{@{}c@{}}Decrease\\      (And, \%)\end{tabular}} & \multirow{2}{*}{Lev} & \multirow{2}{*}{Time (s)} & \multirow{2}{*}{\begin{tabular}[c]{@{}c@{}}Improvement\\      (Time, \%)\end{tabular}} \\
 &  &  &  &  &  &  &  &  &  &  &  \\ \cmidrule(r){1-6} \cmidrule(l){7-12}
PruneX (COG) & 10621.67 (0.47) & -0.01 & 104.00 (0.0) & 88.70 (17.85) & \textbf{42.90} & PruneX (COG) & 29284.00 (2.0) & -0.08 & 376.00 (0.0) & 40.03 (1.57) & \textbf{39.23} \\
Default (Mfs2) & 10621.00 (0.0) & NA & 104.00 (0.0) & 155.33 (1.31) & NA & Default (Resub) & 29260.00 (0.0) & NA & 376.00 (0.0) & 65.87 (1.43) & NA \\ \midrule
 & \multicolumn{11}{c}{Hyp} \\ \midrule
\multirow{2}{*}{Method} & \multirow{2}{*}{Nd} & \multirow{2}{*}{\begin{tabular}[c]{@{}c@{}}Decrease\\      (Nd, \%)\end{tabular}} & \multirow{2}{*}{Lev} & \multirow{2}{*}{Time (s)} & \multirow{2}{*}{\begin{tabular}[c]{@{}c@{}}Improvement\\      (Time, \%)\end{tabular}} & \multirow{2}{*}{Method} & \multirow{2}{*}{And} & \multirow{2}{*}{\begin{tabular}[c]{@{}c@{}}Decrease\\      (And, \%)\end{tabular}} & \multirow{2}{*}{Lev} & \multirow{2}{*}{Time (s)} & \multirow{2}{*}{\begin{tabular}[c]{@{}c@{}}Improvement\\      (Time, \%)\end{tabular}} \\
 &  &  &  &  &  &  &  &  &  &  &  \\ \cmidrule(r){1-6} \cmidrule(l){7-12}
PruneX (COG) & 63682.00 (18.23) & -0.16 & 8259.00 (0.0) & 210.52 (7.99) & \textbf{20.23} & PruneX (COG) & 205436.50 (216.5) & -0.44 & 24793.00 (0.0) & 43.01 (5.01) & \textbf{62.61} \\
Default (Mfs2) & 63581.00 (0.0) & NA & 8259.00 (0.0) & 263.91 (0.36) & NA & Default (Resub) & 204533.00 (0.0) & NA & 24792.00 (0.0) & 115.07 (0.009) & NA \\ \midrule
 & \multicolumn{11}{c}{Multiplier} \\ \midrule
\multirow{2}{*}{Method} & \multirow{2}{*}{Nd} & \multirow{2}{*}{\begin{tabular}[c]{@{}c@{}}Decrease\\      (Nd, \%)\end{tabular}} & \multirow{2}{*}{Lev} & \multirow{2}{*}{Time (s)} & \multirow{2}{*}{\begin{tabular}[c]{@{}c@{}}Improvement\\      (Time, \%)\end{tabular}} & \multirow{2}{*}{Method} & \multirow{2}{*}{And} & \multirow{2}{*}{\begin{tabular}[c]{@{}c@{}}Decrease\\      (And, \%)\end{tabular}} & \multirow{2}{*}{Lev} & \multirow{2}{*}{Time (s)} & \multirow{2}{*}{\begin{tabular}[c]{@{}c@{}}Improvement\\      (Time, \%)\end{tabular}} \\
 &  &  &  &  &  &  &  &  &  &  &  \\ \cmidrule(r){1-6} \cmidrule(l){7-12}
PruneX (COG) & 7799.00 (0.0) & 0.00 & 87.00 (0.0) & 14.67 (0.43) & \textbf{16.41} & PruneX (COG) & 24452.50 (0.5) & -0.07 & 262.00 (0.0) & 12.35 (0.14) & \textbf{37.97} \\
Default (Mfs2) & 7799.00 (0.0) & NA & 87.00 (0.0) & 17.55 (0.16) & NA & Default (Resub) & 24436.00 (0.0) & NA & 262.00 (0.0) & 19.91 (0.06) & NA \\ \midrule
 & \multicolumn{11}{c}{Sin} \\ \midrule
\multirow{2}{*}{Method} & \multirow{2}{*}{Nd} & \multirow{2}{*}{\begin{tabular}[c]{@{}c@{}}Decrease\\      (Nd, \%)\end{tabular}} & \multirow{2}{*}{Lev} & \multirow{2}{*}{Time (s)} & \multirow{2}{*}{\begin{tabular}[c]{@{}c@{}}Improvement\\      (Time, \%)\end{tabular}} & \multirow{2}{*}{Method} & \multirow{2}{*}{And} & \multirow{2}{*}{\begin{tabular}[c]{@{}c@{}}Decrease\\      (And, \%)\end{tabular}} & \multirow{2}{*}{Lev} & \multirow{2}{*}{Time (s)} & \multirow{2}{*}{\begin{tabular}[c]{@{}c@{}}Improvement\\      (Time, \%)\end{tabular}} \\
 &  &  &  &  &  &  &  &  &  &  &  \\ \cmidrule(r){1-6} \cmidrule(l){7-12}
PruneX (COG) & 1992.33 (0.94) & -0.12 & 53.00 (0.0) & 9.66 (1.75) & \textbf{23.58} & PruneX (COG) & 5019.50 (0.5) & -0.03 & 177.00 (0.0) & 7.85 (0.64) & \textbf{38.38} \\
Default (Mfs2) & 1990.00 (0.0) & NA & 53.00 (0.0) & 12.64 (0.27) & NA & Default (Resub) & 5018.00 (0.0) & NA & 177.00 (0.0) & 12.74 (0.024) & NA \\ \midrule
 & \multicolumn{11}{c}{Square} \\ \midrule
\multirow{2}{*}{Method} & \multirow{2}{*}{Nd} & \multirow{2}{*}{\begin{tabular}[c]{@{}c@{}}Decrease\\      (Nd, \%)\end{tabular}} & \multirow{2}{*}{Lev} & \multirow{2}{*}{Time (s)} & \multirow{2}{*}{\begin{tabular}[c]{@{}c@{}}Improvement\\      (Time, \%)\end{tabular}} & \multirow{2}{*}{Method} & \multirow{2}{*}{And} & \multirow{2}{*}{\begin{tabular}[c]{@{}c@{}}Decrease\\      (And, \%)\end{tabular}} & \multirow{2}{*}{Lev} & \multirow{2}{*}{Time (s)} & \multirow{2}{*}{\begin{tabular}[c]{@{}c@{}}Improvement\\      (Time, \%)\end{tabular}} \\
 &  &  &  &  &  &  &  &  &  &  &  \\ \cmidrule(r){1-6} \cmidrule(l){7-12}
PruneX (COG) & 5703.33 (1.24) & -0.04 & 83.00 (0.0) & 15.15 (0.96) & \textbf{30.18} & PruneX (COG) & 15918.50 (12.5) & -0.46 & 248.00 (0.0) & 5.14 (0.1) & \textbf{43.89} \\
Default (Mfs2) & 5701.00 (0.0) & NA & 83.00 (0.0) & 21.70 (0.32) & NA & Default (Resub) & 15845.00 (0.0) & NA & 248.00 (0.0) & 9.16 (0.009) & NA \\ \bottomrule
\end{tabular}
}
\end{table*}

\begin{table*}[t]
\caption{Detailed offline evaluation results on the Mfs2 operator and industrial circuits.}
\label{table:mfs2_haisi_offline}
\centering
\resizebox{0.99\textwidth}{!}{
\begin{tabular}{@{}ccccccc@{}}
\toprule
\toprule
 & \multicolumn{3}{c}{d3151} & \multicolumn{3}{c}{f5022} \\ \midrule
Method & top 40\% acc & top 50\% acc & top 60\% acc & top 40\% acc & top 50\% acc & top 60\% acc \\ \cmidrule(r){1-4} \cmidrule(l){5-7}
COG (Ours) & \textbf{0.78 (0.01)} & \textbf{0.85 (0.006)} & \textbf{0.90 (0.006)} & \textbf{1.00 (0.0001)} & \textbf{1.00 (0.0)} & \textbf{1.00 (0.0)} \\
EnsembleMLP (Ours) & 0.43 (0.002) & 0.54 (0.008) & 0.69 (0.008) & 1.00 (0.00006) & 1.00 (0.00006) & 1.00 (0.00006) \\
Random & 0.40 & 0.50 & 0.60 & 0.40 & 0.50 & 0.60 \\ \midrule
 & \multicolumn{3}{c}{c8449} & \multicolumn{3}{c}{f8272} \\ \midrule
Method & top 40\% acc & top 50\% acc & top 60\% acc & top 40\% acc & top 50\% acc & top 60\% acc \\ \cmidrule(r){1-4} \cmidrule(l){5-7}
COG (Ours) & \textbf{0.78 (0.01)} & \textbf{0.85 (0.01)} & \textbf{0.90 (0.02)} & \textbf{0.71 (0.003)} & \textbf{0.84 (0.02)} & \textbf{0.91 (0.02)} \\
EnsembleMLP (Ours) & 0.50 (0.003) & 0.61 (0.009) & 0.68 (0.004) & 0.48 (0.00003) & 0.66 (0.0004) & 0.75 (0.00003) \\
Random & 0.42 & 0.52 & 0.61 & 0.40 & 0.50 & 0.60 \\ \midrule
 & \multicolumn{3}{c}{d4067} & \multicolumn{3}{c}{c5088} \\ \midrule
Method & top 40\% acc & top 50\% acc & top 60\% acc & top 40\% acc & top 50\% acc & top 60\% acc \\ \cmidrule(r){1-4} \cmidrule(l){5-7}
COG (Ours) & \textbf{0.82 (0.003)} & \textbf{0.83 (0.002)} & \textbf{0.84 (0.01)} & \textbf{0.89 (0.008)} & \textbf{0.89 (0.01)} & \textbf{0.92 (0.02)} \\
EnsembleMLP (Ours) & \textbf{0.86 (0.007)} & \textbf{0.94 (0.0004)} & \textbf{0.95 (0.0008)} & \textbf{0.98 (0.0)} & \textbf{0.98 (0.0)} & \textbf{0.99 (0.004)} \\
Random & 0.36 & 0.48 & 0.58 & 0.33 & 0.47 & 0.61 \\ \bottomrule
\end{tabular}
}
\end{table*}

\begin{table*}[t]
\caption{Detailed offline evaluation results on the Resub operator and industrial circuits.}
\label{table:resub_haisi_offline}
\centering
\resizebox{0.99\textwidth}{!}{
\begin{tabular}{@{}ccccccc@{}}
\toprule
\toprule
 & \multicolumn{3}{c}{d3151} & \multicolumn{3}{c}{f5022} \\ \midrule
Method & top 40\% acc & top 50\% acc & top 60\% acc & top 40\% acc & top 50\% acc & top 60\% acc \\ \cmidrule(r){1-4} \cmidrule(l){5-7}
COG (Ours) & \textbf{0.91 (0.02)} & \textbf{0.94 (0.008)} & \textbf{0.96 (0.01)} & \textbf{0.89 (0.14)} & \textbf{1.00 (0.0)} & \textbf{1.00 (0.0)} \\
EnsembleMLP (Ours) & 0.39 (0.03) & 0.43 (0.04) & 0.53 (0.07) & 0.31 (0.22) & 0.67 (0.47) & 0.68 (0.45) \\
Random & 0.40 & 0.50 & 0.60 & 0.40 & 0.50 & 0.60 \\ \midrule
 & \multicolumn{3}{c}{c8449} & \multicolumn{3}{c}{f8272} \\ \midrule
Method & top 40\% acc & top 50\% acc & top 60\% acc & top 40\% acc & top 50\% acc & top 60\% acc \\ \cmidrule(r){1-4} \cmidrule(l){5-7}
COG (Ours) & \textbf{0.63 (0.01)} & \textbf{0.76 (0.02)} & \textbf{0.79 (0.002)} & \textbf{0.84 (0.02)} & \textbf{0.85 (0.03)} & \textbf{0.88 (0.02)} \\
EnsembleMLP (Ours) & 0.41 (0.02) & 0.50 (0.02) & 0.59 (0.013) & 0.25 (0.01) & 0.33 (0.02) & 0.45 (0.016) \\
Random & 0.42 & 0.52 & 0.61 & 0.40 & 0.50 & 0.60 \\ \midrule
 & \multicolumn{3}{c}{d4067} & \multicolumn{3}{c}{c5088} \\ \midrule
Method & top 40\% acc & top 50\% acc & top 60\% acc & top 40\% acc & top 50\% acc & top 60\% acc \\ \cmidrule(r){1-4} \cmidrule(l){5-7}
COG (Ours) & \textbf{0.99 (0.001)} & \textbf{1.00 (0.001)} & \textbf{1.00 (0.0005)} & \textbf{0.98 (0.0004)} & \textbf{0.98 (0.0008)} & \textbf{0.98 (0.001)} \\
EnsembleMLP (Ours) & 0.56 (0.22) & 0.61 (0.19) & 0.63 (0.18) & 0.17 (0.09) & 0.20 (0.11) & 0.33 (0.20) \\
Random & 0.36 & 0.48 & 0.58 & 0.33 & 0.47 & 0.61 \\ \bottomrule
\end{tabular}
}
\end{table*}

\begin{table*}[t]
\caption{Detailed online evaluation results on the Mfs2 operator and industrial circuits. Improvement denotes the improvement of our PruneX compared to the default operator.} 
\label{table:mfs2_haisi_online}
\centering
\resizebox{0.99\textwidth}{!}{
\begin{tabular}{@{}ccccccccccc@{}}
\toprule
\toprule
 & \multicolumn{5}{c}{f5022} & \multicolumn{5}{c}{f8272} \\ \midrule
\multirow{2}{*}{Method} & \multirow{2}{*}{Nd} & \multirow{2}{*}{\begin{tabular}[c]{@{}c@{}}Decrease\\      (Nd, \%)\end{tabular}} & \multirow{2}{*}{Lev} & \multirow{2}{*}{Time (s)} & \multirow{2}{*}{\begin{tabular}[c]{@{}c@{}}Improvement\\      (Time, \%)\end{tabular}} & \multirow{2}{*}{Nd} & \multirow{2}{*}{\begin{tabular}[c]{@{}c@{}}Decrease\\      (Nd, \%)\end{tabular}} & \multirow{2}{*}{Lev} & \multirow{2}{*}{Time (s)} & \multirow{2}{*}{\begin{tabular}[c]{@{}c@{}}Improvement\\      (Time, \%)\end{tabular}} \\
 &  &  &  &  &  &  &  &  &  &  \\ \cmidrule(r){1-6} \cmidrule(l){7-11}
PruneX-COG (Ours) & 165538.00 (0.0) & 0.00 & 47.00 (0.0) & 93.58 (0.49) & \textbf{47.41} & 99258.00 (0.0) & -0.01 & 12.00 (0.0) & 31.71 (1.17) & \textbf{58.80} \\
Default (Mfs2) & 165538.00 (0.0) & NA & 47.00 (0.0) & 177.93 (0.69) & NA & 99245.00 (0.0) & NA & 12.00 (0.0) & 76.97 (0.13) & NA \\ \midrule
 & \multicolumn{5}{c}{c5088} & \multicolumn{5}{c}{d3151} \\ \midrule
\multirow{2}{*}{Method} & \multirow{2}{*}{Nd} & \multirow{2}{*}{\begin{tabular}[c]{@{}c@{}}Decrease\\      (Nd, \%)\end{tabular}} & \multirow{2}{*}{Lev} & \multirow{2}{*}{Time (s)} & \multirow{2}{*}{\begin{tabular}[c]{@{}c@{}}Improvement\\      (Time, \%)\end{tabular}} & \multirow{2}{*}{Nd} & \multirow{2}{*}{\begin{tabular}[c]{@{}c@{}}Decrease\\      (Nd, \%)\end{tabular}} & \multirow{2}{*}{Lev} & \multirow{2}{*}{Time (s)} & \multirow{2}{*}{\begin{tabular}[c]{@{}c@{}}Improvement\\      (Time, \%)\end{tabular}} \\
 &  &  &  &  &  &  &  &  &  &  \\ \cmidrule(r){1-6} \cmidrule(l){7-11}
PruneX-COG (Ours) & 195670.33 (1.24) & -0.003 & 35.00 (0.0) & 212.10 (5.13) & \textbf{56.49} & 6648.33 (2.49) & -0.22 & 15.00 (0.0) & 51.58 (2.55) & \textbf{22.05} \\
Default (Mfs2) & 195665.00 (0.0) & NA & 35.00 (0.0) & 487.51 (51.68) & NA & 6634.00 (0.0) & NA & 15.00 (0.0) & 66.17 (5.25) & NA \\ \midrule
 & \multicolumn{5}{c}{c8449} & \multicolumn{5}{c}{d4067} \\ \midrule
\multirow{2}{*}{Method} & \multirow{2}{*}{Nd} & \multirow{2}{*}{\begin{tabular}[c]{@{}c@{}}Decrease\\      (Nd, \%)\end{tabular}} & \multirow{2}{*}{Lev} & \multirow{2}{*}{Time (s)} & \multirow{2}{*}{\begin{tabular}[c]{@{}c@{}}Improvement\\      (Time, \%)\end{tabular}} & \multirow{2}{*}{Nd} & \multirow{2}{*}{\begin{tabular}[c]{@{}c@{}}Decrease\\      (Nd, \%)\end{tabular}} & \multirow{2}{*}{Lev} & \multirow{2}{*}{Time (s)} & \multirow{2}{*}{\begin{tabular}[c]{@{}c@{}}Improvement\\      (Time, \%)\end{tabular}} \\
 &  &  &  &  &  &  &  &  &  &  \\ \cmidrule(r){1-6} \cmidrule(l){7-11}
PruneX-COG (Ours) & 9612.66 (17.55) & -0.78 & 17.00 (0.0) & 169.65 (40.92) & \textbf{27.68} & 215714.00 (2.82) & -0.003 & 26.00 (0.0) & 162.59 (16.74) & \textbf{19.28} \\
Default (Mfs2) & 9538.00 (0.0) & NA & 17.00 (0.0) & 234.58 (45.54) & NA & 215708.00 (0.0) & NA & 26.00 (0.0) & 201.43 (12.17) & NA \\ \bottomrule
\end{tabular}
}
\end{table*}

\begin{table*}[t]
\caption{Detailed online evaluation results on the Resub operator and industrial circuits. Improvement denotes the improvement of our PruneX compared to the default operator.}
\label{table:resub_haisi_online}
\centering
\resizebox{0.99\textwidth}{!}{
\begin{tabular}{@{}ccccccccccc@{}}
\toprule
\toprule
 & \multicolumn{5}{c}{f5022} & \multicolumn{5}{c}{f8272} \\ \midrule
\multirow{2}{*}{Method} & \multirow{2}{*}{And} & \multirow{2}{*}{\begin{tabular}[c]{@{}c@{}}Decrease\\      (And, \%)\end{tabular}} & \multirow{2}{*}{Lev} & \multirow{2}{*}{Time (s)} & \multirow{2}{*}{\begin{tabular}[c]{@{}c@{}}Improvement\\      (Time, \%)\end{tabular}} & \multirow{2}{*}{And} & \multirow{2}{*}{\begin{tabular}[c]{@{}c@{}}Decrease\\      (And, \%)\end{tabular}} & \multirow{2}{*}{Lev} & \multirow{2}{*}{Time (s)} & \multirow{2}{*}{\begin{tabular}[c]{@{}c@{}}Improvement\\      (Time, \%)\end{tabular}} \\
 &  &  &  &  &  &  &  &  &  &  \\ \cmidrule(r){1-6} \cmidrule(l){7-11}
PruneX-COG (Ours) & 474604.66 (6.5) & -0.001 & 141.0 (0.0) & 793.10 (110.67) & \textbf{30.71} & 278680.33 (1.88) & -0.015 & 31.0 (0.0) & 370.89 (4.00) & \textbf{50.12} \\
Default (Resub) & 474600.0 (0.0) & NA & 141.0 (0.0) & 1144.56 (10.57) & NA & 278637.0 (0.0) & NA & 31.0 (0.0) & 743.51 (4.53) & NA \\ \midrule
 & \multicolumn{5}{c}{c5088} & \multicolumn{5}{c}{d3151} \\ \midrule
\multirow{2}{*}{Method} & \multirow{2}{*}{And} & \multirow{2}{*}{\begin{tabular}[c]{@{}c@{}}Decrease\\      (And, \%)\end{tabular}} & \multirow{2}{*}{Lev} & \multirow{2}{*}{Time (s)} & \multirow{2}{*}{\begin{tabular}[c]{@{}c@{}}Improvement\\      (Time, \%)\end{tabular}} & \multirow{2}{*}{And} & \multirow{2}{*}{\begin{tabular}[c]{@{}c@{}}Decrease\\      (And, \%)\end{tabular}} & \multirow{2}{*}{Lev} & \multirow{2}{*}{Time (s)} & \multirow{2}{*}{\begin{tabular}[c]{@{}c@{}}Improvement\\      (Time, \%)\end{tabular}} \\
 &  &  &  &  &  &  &  &  &  &  \\ \cmidrule(r){1-6} \cmidrule(l){7-11}
PruneX-COG (Ours) & 506986.66 (15.36) & -0.003 & 173.0 (0.0) & 1908.63 (692.66) & \textbf{24.09} & 18531.33 (29.16) & -0.46 & 57.0 (0.0) & 13.51 (0.33) & \textbf{46.81} \\
Default (Resub) & 506972.0 (0.0) & NA & 173.0 (0.0) & 2514.46 (377.94) & NA & 18447.0 (0.0) & NA & 57.0 (0.0) & 25.40 (2.94) & NA \\ \midrule
 & \multicolumn{5}{c}{c8449} & \multicolumn{5}{c}{d4067} \\ \midrule
\multirow{2}{*}{Method} & \multirow{2}{*}{And} & \multirow{2}{*}{\begin{tabular}[c]{@{}c@{}}Decrease\\      (And, \%)\end{tabular}} & \multirow{2}{*}{Lev} & \multirow{2}{*}{Time (s)} & \multirow{2}{*}{\begin{tabular}[c]{@{}c@{}}Improvement\\      (Time, \%)\end{tabular}} & \multirow{2}{*}{And} & \multirow{2}{*}{\begin{tabular}[c]{@{}c@{}}Decrease\\      (And, \%)\end{tabular}} & \multirow{2}{*}{Lev} & \multirow{2}{*}{Time (s)} & \multirow{2}{*}{\begin{tabular}[c]{@{}c@{}}Improvement\\      (Time, \%)\end{tabular}} \\
 &  &  &  &  &  &  &  &  &  &  \\ \cmidrule(r){1-6} \cmidrule(l){7-11}
PruneX-COG (Ours) & 27834.33 (20.53) & -0.50 & 65.0 (0.0) & 17.33 (2.84) & \textbf{50.74} & 643664.66 (76.85) & -0.22 & 100.0 (0.0) & 1638.94 (255.20) & \textbf{67.60} \\
Default (Resub) & 27695.0 (0.0) & NA & 65.0 (0.0) & 35.18 (0.15) & NA & 642243.0 (0.0) & NA & 95.0 (0.0) & 5058.70 (51.38) & NA \\ \bottomrule
\end{tabular}
}
\end{table*}

\begin{table*}[t]
\caption{Detailed offline evaluation results on the Mfs2 operator and very large-scale circuits.}
\label{table:mfs2_epfl_hard_offline}
\centering
\resizebox{0.8\textwidth}{!}{
\begin{tabular}{@{}cccc@{}}
\toprule
\toprule
 & \multicolumn{3}{c}{Sixteen} \\ \midrule
Method & top 40\% acc & top 50\% acc & top 60\% acc \\ \midrule
COG & \textbf{1.00 (9e-05)} & \textbf{1.00 (4e-05)} & \textbf{1.00 (8e-05)} \\
EnsembleMLP & 0.34 (0.03) & 0.44 (0.03) & 0.53 (0.02) \\
Random & 0.31 & 0.48 & 0.60 \\ \midrule
 & \multicolumn{3}{c}{Twenty} \\ \midrule
Method & top 40\% acc & top 50\% acc & top 60\% acc \\ \midrule
COG & \textbf{1.00 (4e-05)} & \textbf{1.00 (0.0001)} & \textbf{1.00 (0.0001)} \\
EnsembleMLP & 0.54 (0.17) & 0.61 (0.15) & 0.68 (0.11) \\
Random & 0.40 & 0.50 & 0.60 \\ \bottomrule
\end{tabular}
}
\end{table*}

\begin{table*}[t]
\caption{Detailed offline evaluation results on the Resub operator and very large-scale circuits.}
\label{table:resub_epfl_hard_offline}
\centering
\resizebox{0.8\textwidth}{!}{
\begin{tabular}{@{}cccc@{}}
\toprule
\toprule
 & \multicolumn{3}{c}{Sixteen} \\ \midrule
Method & top 40\% acc & top 50\% acc & top 60\% acc \\ \midrule
COG & \textbf{0.89 (0.008)} & \textbf{0.94 (0.007)} & \textbf{0.97 (0.004)} \\
EnsembleMLP & 0.42 (0.06) & 0.54 (0.09) & 0.67 (0.13) \\
Random & 0.31 & 0.48 & 0.60 \\ \midrule
 & \multicolumn{3}{c}{Twenty} \\ \midrule
Method & top 40\% acc & top 50\% acc & top 60\% acc \\ \midrule
COG & \textbf{0.91 (0.006)} & \textbf{0.95 (0.003)} & \textbf{0.98 (0.001)} \\
EnsembleMLP & 0.41 (0.0) & 0.52 (0.0) & 0.61 (0.0) \\
Random & 0.42 & 0.48 & 0.62 \\ \bottomrule
\end{tabular}
}
\end{table*}

\begin{table*}[t]
\caption{Detailed online evaluation results on the Mfs2 operator and very large-scale circuits. Improvement denotes the improvement of our PruneX compared to the default operator.}
\label{table:mfs2_epfl_hard_online}
\centering
\resizebox{0.8\textwidth}{!}{
\begin{tabular}{@{}cccccc@{}}
\toprule
\toprule
 & \multicolumn{5}{c}{Sixteen} \\ \midrule
\multirow{2}{*}{Method} & \multirow{2}{*}{Nd} & \multirow{2}{*}{\begin{tabular}[c]{@{}c@{}}Decrease\\      (Nd, \%)\end{tabular}} & \multirow{2}{*}{Lev} & \multirow{2}{*}{Time (s)} & \multirow{2}{*}{\begin{tabular}[c]{@{}c@{}}Improvement\\      (Time, \%)\end{tabular}} \\
 &  &  &  &  &  \\ \midrule
PruneX (COG) & 6017637.0 (0.81) & -9.9E-05 & 48.0 (0.0) & 38581.40 (933.59) & \textbf{27.43} \\
Default (Mfs2) & 6017631.0 (0.0) & NA & 48.0 (0.0) & 53162.36 (337.79) & NA \\ \midrule
 & \multicolumn{5}{c}{Twenty} \\ \midrule
\multirow{2}{*}{Method} & \multirow{2}{*}{Nd} & \multirow{2}{*}{\begin{tabular}[c]{@{}c@{}}Decrease\\      (Nd, \%)\end{tabular}} & \multirow{2}{*}{Lev} & \multirow{2}{*}{Time (s)} & \multirow{2}{*}{\begin{tabular}[c]{@{}c@{}}Improvement\\      (Time, \%)\end{tabular}} \\
 &  &  &  &  &  \\ \midrule
PruneX (COG) & 7693099.0 (0.0) & -0.0001 & 54.0 (0.0) & 54046.50 (851.80) & \textbf{42.00} \\
Default (Mfs2) & 7693089.0 (0.0) & NA & 54.0 (0.0) & 93189.12 (8806.55) & NA \\ \bottomrule
\end{tabular}
}
\end{table*}

\begin{table*}[t]
\caption{Detailed online evaluation results on the Resub operator and very large-scale circuits. Improvement denotes the improvement of our PruneX compared to the default operator.}
\label{table:resub_epfl_hard_online}
\centering
\resizebox{0.8\textwidth}{!}{
\begin{tabular}{@{}cccccc@{}}
\toprule
\toprule
 & \multicolumn{5}{c}{Sixteen} \\ \midrule
\multirow{2}{*}{Method} & \multirow{2}{*}{And} & \multirow{2}{*}{\begin{tabular}[c]{@{}c@{}}Decrease\\      (And, \%)\end{tabular}} & \multirow{2}{*}{Lev} & \multirow{2}{*}{Time (s)} & \multirow{2}{*}{\begin{tabular}[c]{@{}c@{}}Improvement\\      (Time, \%)\end{tabular}} \\
 &  &  &  &  &  \\ \midrule
PruneX-COG (Ours) & 11972490.66 (267.26) & -0.005 & 99.0 (0.0) & 11166.83 (374.72) & \textbf{16.35} \\
Default (Resub) & 11971930.0 (0.0) & NA & 99.0 (0.0) & 13349.62 (175.76) & NA \\ \midrule
 & \multicolumn{5}{c}{Twenty} \\ \midrule
\multirow{2}{*}{Method} & \multirow{2}{*}{And} & \multirow{2}{*}{\begin{tabular}[c]{@{}c@{}}Decrease\\      (And, \%)\end{tabular}} & \multirow{2}{*}{Lev} & \multirow{2}{*}{Time (s)} & \multirow{2}{*}{\begin{tabular}[c]{@{}c@{}}Improvement\\      (Time, \%)\end{tabular}} \\
 &  &  &  &  &  \\
PruneX-COG (Ours) & 15310801.33 (214.67) & -0.0001 & 86.0 (0.0) & 14114.51 (881.11) & \textbf{18.17} \\
Default (Resub) & 15310242.0 (0.0) & NA & 86.0 (0.0) & 17249.11 (90.69) & NA \\ \bottomrule
\end{tabular}
}
\end{table*}

\begin{table*}[t]
\caption{We provide more online evaluation results of improving QoR with our proposed PruneX-COG. Improvement denotes the improvement of our PruneX compared to the default operator.}
\label{table:mfs2_epfl_iwls_online_improving_ppa}
\centering
\resizebox{0.99\textwidth}{!}{
\begin{tabular}{@{}ccccccccc@{}}
\toprule
\toprule
 & \multicolumn{4}{c}{Log2} & \multicolumn{4}{c}{Sin} \\ \midrule
\multirow{2}{*}{Method} & \multirow{2}{*}{Nd} & \multirow{2}{*}{\begin{tabular}[c]{@{}c@{}}Improvement\\      (Nd, \%)\end{tabular}} & \multirow{2}{*}{Time (s)} & \multirow{2}{*}{\begin{tabular}[c]{@{}c@{}}Improvement\\      (Time, \%)\end{tabular}} & \multirow{2}{*}{Nd} & \multirow{2}{*}{\begin{tabular}[c]{@{}c@{}}Improvement\\      (Nd, \%)\end{tabular}} & \multirow{2}{*}{Time (s)} & \multirow{2}{*}{\begin{tabular}[c]{@{}c@{}}Improvement\\      (Time, \%)\end{tabular}} \\
 &  &  &  &  &  &  &  &  \\ \cmidrule(r){1-5} \cmidrule(l){6-9}
Default (Mfs2) & 10621.00 (0.0) & NA & 156.19 (1.66) & NA & 1990.00 (0.0) & NA & 12.38 (0.016) & NA \\
2PruneX (COG, 0.3) & 10556.00 (3.74) & 0.61 & 179.04 (6.16) & -14.63 & 1985.00 (0.0) & 0.25 & 11.65 (0.60) & 5.90 \\
2PruneX (COG, 0.4) & 10548.00 (2.16) & 0.69 & 237.79 (8.28) & -52.24 & 1980.66 (0.94) & 0.47 & 16.23 (0.06) & -31.10 \\ \bottomrule
\end{tabular}
}
\end{table*}
    
    \subsection{More Results on the Open-Source Benchmarks}\label{appendix:exps_open_source}
        We first provide details of the results in Section \ref{results:open_source} in the main text. Specifically, we use three metrics, i.e., top $50\%$ acc, normalized runtime, and normalized node number. The top $50\%$ acc denotes the top $50\%$ accuracy metric. The normalized runtime denotes the ratio of the runtime to that of the default operator. The normalized node number denotes the ratio of the node number to that of the circuits optimized by the default operator.
    
        Then, in terms of experiments on the open-source benchmarks, we provide detailed \textbf{offline} evaluation results of our PruneX on both the Mfs2 and Resub operators in Tables \ref{table:mfs2_generalize_one_benchmarks}, \ref{table:resub_generalize_one_benchmarks}, and \ref{table:mfs2_resub_generalize_from_iwls_to_epfl}. Moreover, we provide detailed \textbf{online} evaluation results of our PruneX on both the Mfs2 and Resub operators in Tables \ref{table:mfs2_generalize_in_iwls_epfl_online}, \ref{table:resub_generalize_in_iwls_epfl_online}, and \ref{table:mfs2_resub_generalize_from_iwls_to_epfl_online}. 
        Overall, the results demonstrate that our proposed PruneX significantly improves the efficiency of the Resub and Mfs2 operators, while keeping comparable optimization performance.
            
        

        
    \subsection{More Results on Industrial Circuits and Very Large-Scale Circuits}\label{appendix:exps_real_circuits_vlsc}
        We first provide details of the results in Section \ref{results:real_industry} in the main text. Specifically, we use three metrics, i.e., top $50\%$ acc, normalized runtime, and normalized node number. The top $50\%$ acc denotes the top $50\%$ accuracy metric. The normalized runtime denotes the ratio of the runtime to that of the default operator. The normalized node number denotes the ratio of the node number to that of the circuits optimized by the default operator.

        Then, in terms of experiments on the industrial circuits and very large-scale circuits,
        we provide detailed \textbf{offline} evaluation results of our PruneX on both the Mfs2 and Resub operators in Tables \ref{table:mfs2_haisi_offline}, \ref{table:resub_haisi_offline},  \ref{table:mfs2_epfl_hard_offline}, and \ref{table:resub_epfl_hard_offline}. 
        Moreover, we provide detailed \textbf{online} evaluation results of our PruneX on both the Mfs2 and Resub operators in Tables \ref{table:mfs2_haisi_online}, \ref{table:resub_haisi_online}, \ref{table:mfs2_epfl_hard_online}, and \ref{table:resub_epfl_hard_online}. 
        Overall, the results demonstrate the strong performance of our proposed PruneX on the industrial circuits and very large-scale circuits.

        

        
    \subsection{More Results on Improving QoR with PruneX-COG}\label{appendix:exps_improving_qor}
        \textbf{Optimization Sequence Flows for 2PruneX-COG} To apply our PruneX-COG twice, we apply the sequence of operators, i.e., \textit{strash; dch; if -C 12; mfs2 -W 4 -M 5000; strash; if -C 12; mfs2 -W 4 -M 5000}, to evaluate the performance of 2PruneX-COG. Note that the mfs2 operator is a post-mapping optimization operator, whose input DAG is a k-input look up table graph (K-LUTs). Moreover, the strash operator transforms the current circuit representation into an And-Inverter Graph (AIG) by one-level structural hashing. Then, the if \cite{technology_mapping} operator maps an AIG into a K-LUTs. Finally, the Mfs2 operator optimizes the input K-LUTs. Note that the strash and if operators are much faster than the Mfs2 operator.
    
        We provide additional results of improving QoR with 2PruneX-COG on the Log2 and Sin circuit in Table \ref{appendix:exps_improving_qor}. The results demonstrate that our PruneX can not only reduce the runtime of operators, but also improve its optimization performance in terms of the size and depth. 
         
        

\begin{figure*}[t]
    \centering
    \begin{subfigure}{0.24\textwidth}
        \includegraphics[width=\textwidth, height=0.8\textwidth]{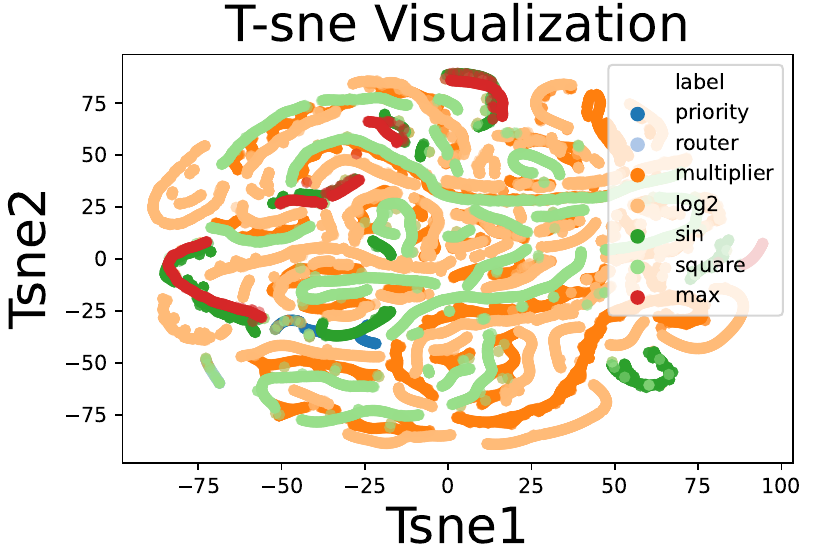}
        \caption{Visualization}
        \label{fig:more_visualization}
    \end{subfigure}
    \begin{subfigure}{0.72\textwidth}
        \includegraphics[width=0.3\textwidth, height=0.26\textwidth]{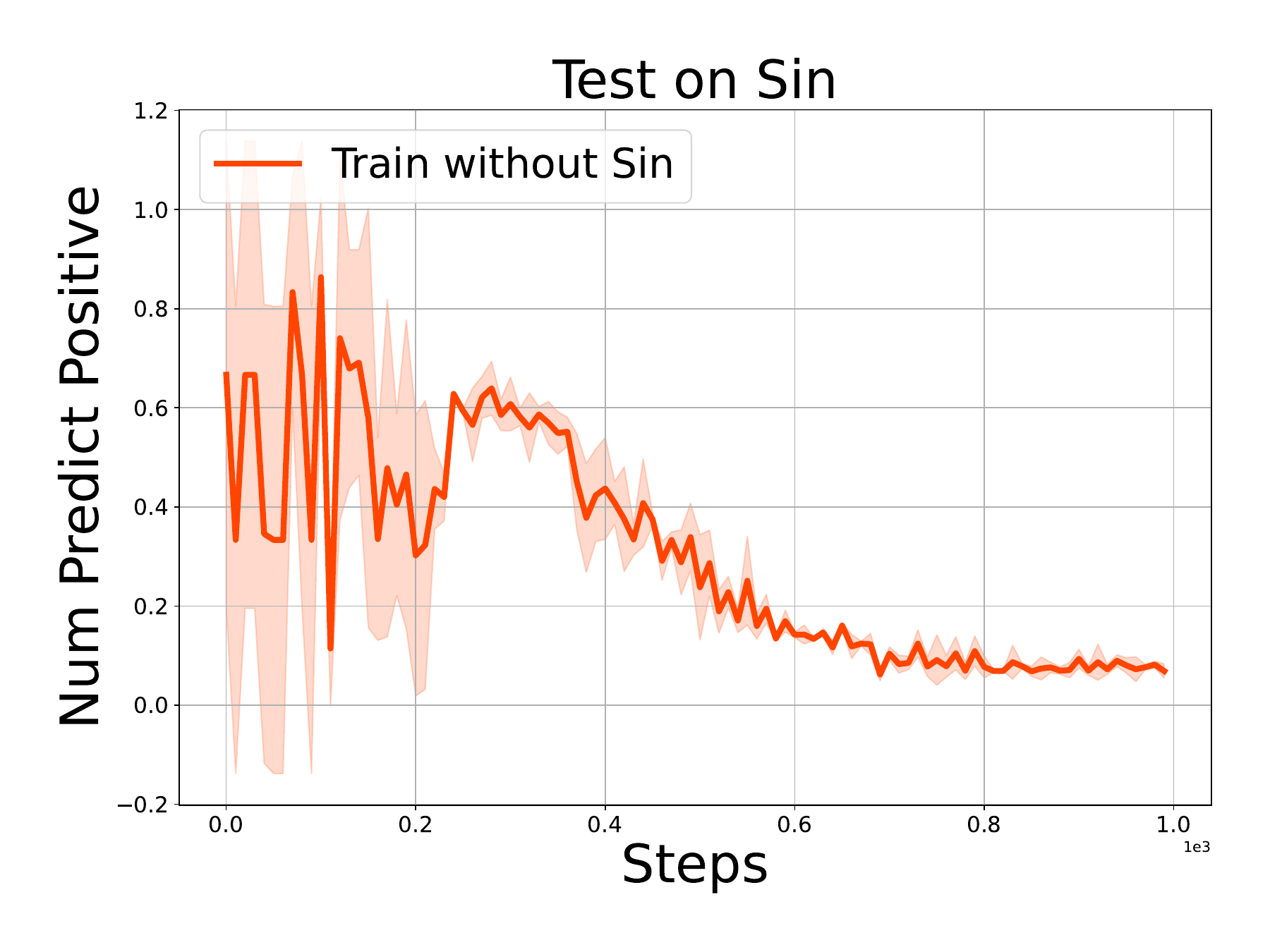}
        \includegraphics[width=0.3\textwidth,height=0.26\textwidth]{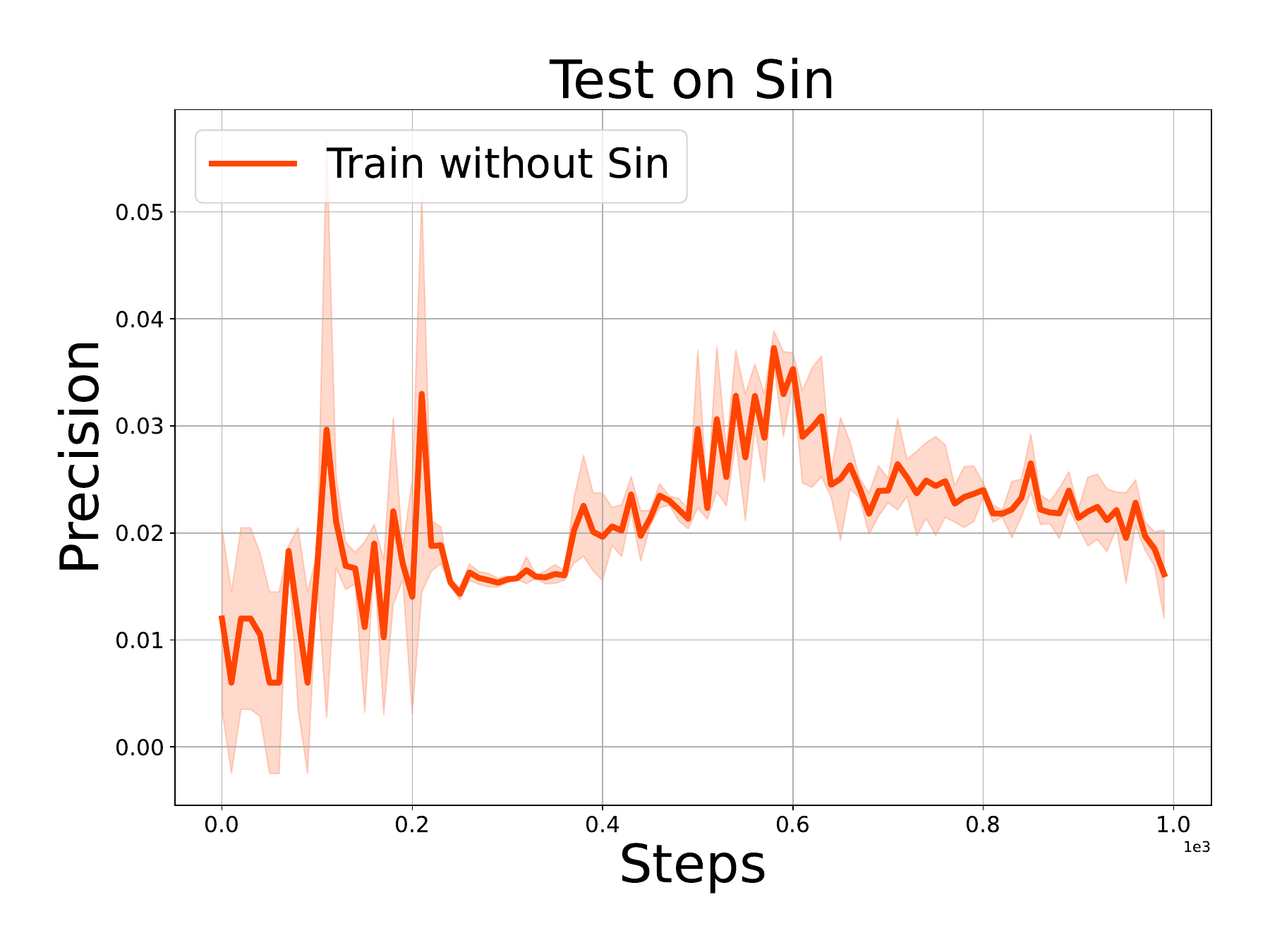}
        \includegraphics[width=0.3\textwidth,height=0.26\textwidth]{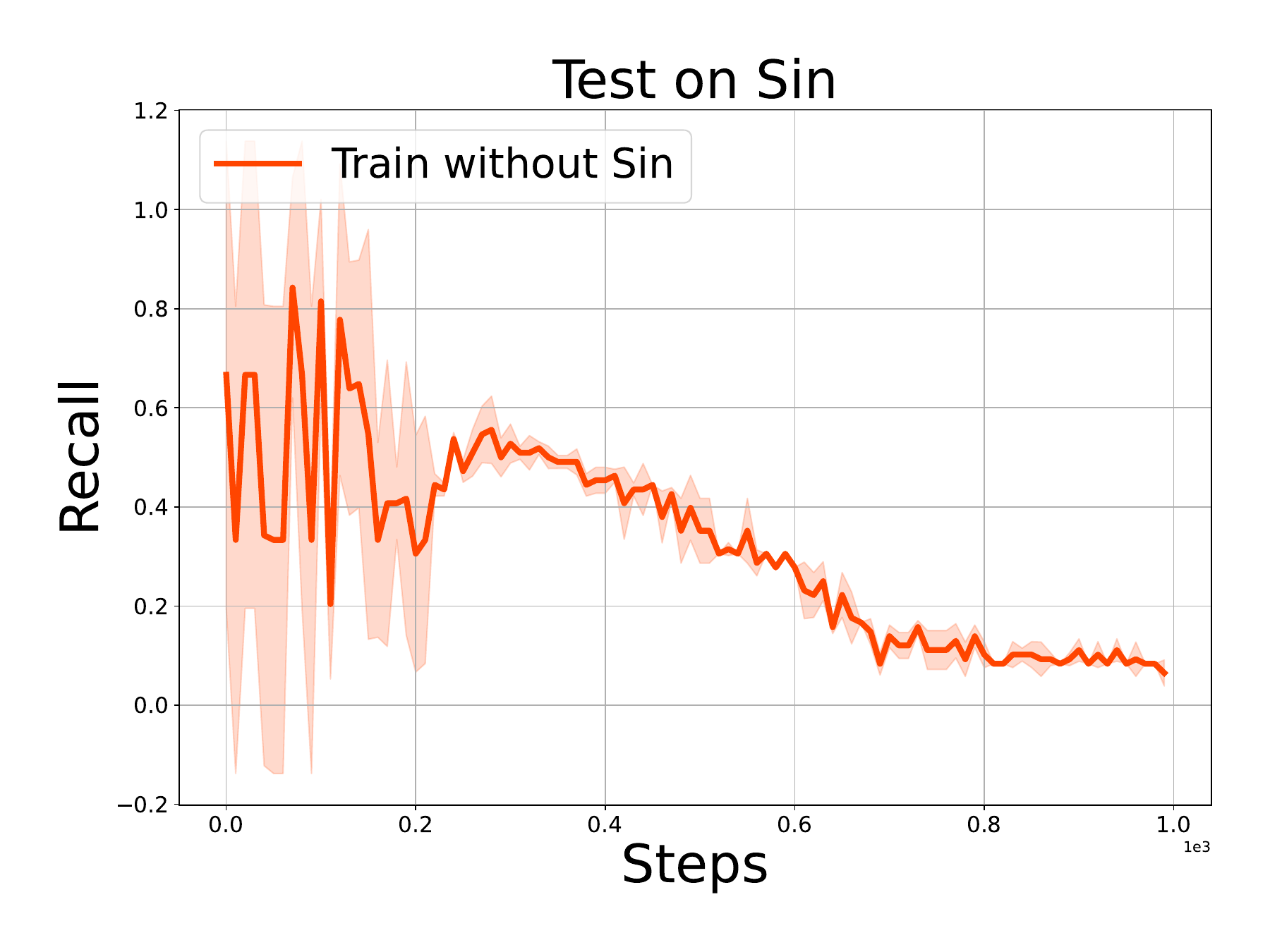}
        \caption{Class Imbalance Problem}
        \label{fig:class_imbalance_problem}
    \end{subfigure}
    \caption{(a) We visualize the data points from different circuits. (b) We train models using EnsembleMLP with a dataset without Sin, and evaluate the models on Sin. The results (Left) show that the number of predicted positive samples is quite small. Moreover, the results (Middle and Right) show that learned models performs poorly in terms of the precision and recall.}
    \label{fig:motivation_nega_bias}
\end{figure*}

\begin{table*}[t]
\caption{The results show that the ratio of the total model time to the runtime of the Mfs2 operator (i.e., Percent) is very low, with an average of $2.66\%$.}
\label{model_inference_time}
\centering
\resizebox{0.99\textwidth}{!}{
\begin{tabular}{@{}ccccccc@{}}
\toprule
\toprule
Circuit & Nodes & Graph construction time (s) & Model inference time (s) & Total model time (s) & Runtime of operator (s) & Percent (\%) \\ \midrule
Log2 & 32060 & 0.48 (0.01) & 2.01 (0.05) & 2.49 & 156.19 (1.66) & 1.59 \\
Hyp & 214335 & 2.56 (0.02) & 6.87 (0.09) & 9.43 & 274.13 (9.90) & 3.44 \\
Vga lcd & 124050 & 1.52 (0.04) & 5.02 (0.33) & 6.54 & 189.10 (3.37) & 3.46 \\
f0089 & 788288 & 10.87 (0.01) & 25.54 (0.26) & 36.41 & 1436.53 (123.54) & 2.53 \\
c8449 & 37051 & 0.57 (0.10) & 2.18 (0.22) & 2.75 & 234.58 (45.54) & 1.17 \\
d3151 & 24778 & 0.41 (0.004) & 2.10 (0.07) & 2.51 & 66.17 (5.25) & 3.79 \\ \bottomrule
\end{tabular}
}
\end{table*}

    \subsection{Discussion on the Class Imbalance Problem}\label{appendix:exp_imbalance}
    We use EnsembleMLP to train models on a dataset without Sin, and evaluate the models on Sin. 
    We use 0.5 as the threshold, and predict the nodes whose model prediction probability is greater than 0.5 as positive samples.   
    As shown in Fig. \ref{fig:class_imbalance_problem}, learned models predict only about 6\% of the samples to be positive on the Sin circuit. This implies that our learned models suffer from serious negative bias problem due to the class imbalance problem. Moreover, the results in Fig. \ref{fig:class_imbalance_problem} show that learned models perform poorly in terms of the precision and recall, which suggests that 0.5 is an inappropriate threshold.

    \subsection{Discussion on the Model Inference Time}\label{appendix:model_inference_time}
        We use multiple representative circuits with node sizes ranging from $20,000$ to $780,000$ to test the bipartite graph construction time and model inference time of our COG. As shown in Table \ref{model_inference_time}, 
        the ratio of the total model time to the runtime of the Mfs2 operator (i.e., Percent) is very low, with an average of $2.66\%$. 
        Note that as the size of the circuit increases, the total model time also increases significantly. For example, on the f0089 circuit with $788,288$ nodes, the total model time takes $36.41$ seconds, which is much longer than that taken on a smaller circuit. The major reason is that our GPU (i.e., a NVidia GeForce GTX 3090 Ti) memory is limited, and 
        we have not implemented multi-card parallel inference.
        That is, our model inference is performed serially and in batches on large-scale circuits, which could greatly increase the model inference time. In contrast, we can implement multi-card parallel inference to further reduce the total model time, especially on large-scale circuits.
    
\section{Details of Datasets Used in This Paper}

\begin{table*}[t]
\caption{A detailed description of circuits from the EPFL benchmark. Nodes denotes the sizes of circuits, and Lev denotes the depths of circuits.}
\label{dataset_description_on_EPFL}
\centering
\resizebox{0.99\textwidth}{!}{
\begin{tabular}{@{}cccccccc@{}}
\toprule
Circuit & PI & PO & Latch & Nodes & Edge & Cube & Lev \\ \midrule
Adder & 256 & 129 & 0 & 1020 & 2040 & 1020 & 255 \\
Barrel shifter & 135 & 128 & 0 & 3336 & 6672 & 3336 & 12 \\
Divisor & 128 & 128 & 0 & 57247 & 114494 & 57247 & 4372 \\
Hypotenuse & 256 & 128 & 0 & 214335 & 428670 & 214335 & 24801 \\
Log2 & 32 & 32 & 0 & 32060 & 64120 & 323060 & 444 \\
Max & 512 & 130 & 0 & 2865 & 5730 & 2865 & 287 \\
Multiplier & 128 & 128 & 0 & 27062 & 54124 & 27062 & 274 \\
Sin & 24 & 25 & 0 & 5416 & 10832 & 5416 & 225 \\
Square-root & 128 & 64 & 0 & 24618 & 49236 & 24618 & 5058 \\
Square & 64 & 128 & 0 & 18486 & 36969 & 18485 & 250 \\
Round-robin ariter & 256 & 129 & 0 & 11839 & 23678 & 11839 & 87 \\
Alu control unit & 7 & 26 & 0 & 175 & 348 & 174 & 10 \\
Coding-cavlc & 10 & 11 & 0 & 693 & 1386 & 693 & 16 \\
Decoder & 8 & 256 & 0 & 304 & 608 & 304 & 3 \\
i2c controller & 147 & 142 & 0 & 1357 & 2698 & 1356 & 20 \\
Int to float converter & 11 & 7 & 0 & 260 & 520 & 260 & 16 \\
Memory controller & 1204 & 1230 & 0 & 47110 & 93945 & 47109 & 114 \\
Priority encoder & 128 & 8 & 0 & 978 & 1956 & 978 & 250 \\
Lookahead XY router & 60 & 30 & 0 & 284 & 514 & 257 & 54 \\
Voter & 1001 & 1 & 0 & 13758 & 27516 & 13758 & 70 \\ \bottomrule
\end{tabular}
}
\end{table*}

\begin{table*}[t]
\caption{A detailed description of circuits from the IWLS benchmark. Nodes denotes the sizes of circuits, and Lev denotes the depths of circuits.}
\label{dataset_description_on_IWLS}
\centering
\resizebox{0.99\textwidth}{!}{
\begin{tabular}{@{}cccccccc@{}}
\toprule
Circuit & PI & PO & latch & nodes & edge & cube & lev \\ \midrule
aes\_core & 259 & 129 & 530 & 20797 & 40645 & 24444 & 28 \\
des\_area & 240 & 64 & 128 & 5005 & 9882 & 5889 & 35 \\
des\_perf & 234 & 64 & 8808 & 98463 & 180542 & 108666 & 28 \\
ethernet & 98 & 115 & 10544 & 46804 & 113378 & 72850 & 37 \\
i2c & 19 & 14 & 128 & 1147 & 2299 & 1375 & 15 \\
mem\_ctrl & 115 & 152 & 1083 & 11508 & 26436 & 14603 & 31 \\
pci\_bridge32 & 162 & 207 & 3359 & 16897 & 34607 & 23130 & 29 \\
pci\_conf\_cyc\_addr\_dec & 32 & 32 & 0 & 109 & 212 & 128 & 6 \\
pci\_spoci\_ctrl & 25 & 13 & 60 & 1271 & 2637 & 1557 & 19 \\
sasc & 16 & 12 & 117 & 552 & 1148 & 766 & 10 \\
simple\_spi & 16 & 12 & 132 & 823 & 1694 & 1089 & 14 \\
spi & 47 & 45 & 229 & 3230 & 6904 & 4054 & 32 \\
steppermotordrive & 4 & 4 & 25 & 228 & 397 & 253 & 11 \\
systemcaes & 260 & 129 & 670 & 7961 & 18236 & 11648 & 44 \\
systemcdes & 132 & 65 & 190 & 3324 & 6304 & 3791 & 33 \\
tv80 & 14 & 32 & 359 & 7166 & 16280 & 9352 & 50 \\
usb\_funct & 128 & 121 & 1746 & 12871 & 27102 & 16378 & 25 \\
usb\_phy & 15 & 18 & 98 & 559 & 1001 & 638 & 12 \\
vga\_lcd & 89 & 109 & 17079 & 124050 & 242332 & 146201 & 25 \\
wb\_conmax & 1130 & 1416 & 770 & 29036 & 77185 & 39619 & 26 \\
wb\_dma & 217 & 215 & 263 & 3495 & 7052 & 4496 & 26 \\ \bottomrule
\end{tabular}
}
\end{table*}

\begin{table*}[t]
\caption{A detailed description of two very large-scale circuits from the EPFL benchmark. Nodes denotes the sizes of circuits, and Lev denotes the depths of circuits.}
\label{dataset_description_on_vlsc_circuits}
\centering
\resizebox{0.99\textwidth}{!}{
\begin{tabular}{@{}cccccc@{}}
\toprule
Circuit & PI & PO & Latch & Nodes & Lev \\ \midrule
twenty & 137 & 60 & 0 & 20732893 & 162 \\
sixteen & 117 & 50 & 0 & 16216836 & 140 \\ \bottomrule
\end{tabular}
}
\end{table*}

\begin{table*}[t]
\caption{A statsical description of 27 industrial circuits (21 training circuits and 6 testing circuits) from Huawei HiSilicon. Nodes denotes the sizes of circuits, and Lev denotes the depths of circuits.}
\label{dataset_description_on_real_circuits}
\centering
\resizebox{0.99\textwidth}{!}{
\begin{tabular}{@{}cccccc@{}}
\toprule
Traning Circuits & PI & PO & Latch & Nodes & Lev \\ \midrule
mean & 8410.5 & 5978.682 & 0 & 104229.4 & 55.95455 \\
max & 59974 & 29721 & 0 & 788288 & 104 \\
min & 41 & 107 & 0 & 2775 & 18 \\ \midrule
Testing Circuits & PI & PO & latch & nodes & lev \\ \midrule
mean & 18540.67 & 18015 & 0 & 356111.2 & 103.3333 \\
max & 42257 & 33849 & 0 & 655243 & 185 \\
min & 523 & 483 & 0 & 24778 & 40 \\ \bottomrule
\end{tabular}
}
\end{table*}

    \subsection{Description of Three Used Benchmarks}\label{appendix:three_benchmarks}
        
        We present detailed statistics of the circuits from the EPFL and IWLS in Tables \ref{dataset_description_on_EPFL} and \ref{dataset_description_on_IWLS}. Moreover, we present detailed statistics of the industrial circuits and very large-scale circuits in Tables \ref{dataset_description_on_real_circuits} and \ref{dataset_description_on_vlsc_circuits}. In general, a circuit is modeled as a directed acyclic graph (DAG), where nodes represent logic gates and directed edges represent wires connecting the gates. The fanins of a node are nodes that are driving this node. The fanouts of a node are nodes driven by the node. The primary inputs (PIs) denote the nodes without fanins. The primary outputs (POs) denote a subset of the nodes of the network. 
        Latches can be considered as specialized nodes within sequential circuits. Cubes refer to logical expressions that represent subsets of input variables in Boolean functions. Lev denotes the depth of the DAG, i.e., the maximum number of edges between PIs and POs.
        
        
    \subsection{Datasets for Evaluation on Open-Source Benchmarks}\label{appendix:datasets_open_source}
    For each circuit and a given X operator, we collect the circuit dataset by applying the X operator to optimizing the circuit and collecting the node features $\{\textbf{x}_i\}_{i=1}^{n}$ and labels $\{y_i\}_{i=1}^{n}$. We found that there are a small number of circuits with no effective nodes. We discard these circuits, as we can directly avoid applying transformation to these circuits without learning a model.  
    
    Specifically, under Evaluation Strategy 1 using the EPFL benchmark, we construct five datasets for evaluation. We use one of the five circuit datasets, i.e., circuit datasets collected from Log2, Hyp, Multiplier, Sin, Square, as the testing dataset, and the other circuit datasets as the training dataset. Furthermore, under Evaluation Strategy 1 using the IWLS benchmark, we construct four datasets for evaluation. We use one of the four circuit datasets, i.e., circuit datasets collected from Vga lcd, Ethernet, Wb conmax, and Des perf, as the testing dataset, and the other circuit datasets as the training dataset.

    Moreover, under Evaluation Strategy 2, we construct a dataset for evaluation. Specifically, we use the five circuit datasets from Log2, Hyp, Multiplier, Sin, Square as the testing dataset, and the other circuit datasets from the EPFL and IWLS as the training dataset. To promote the community of machine learning for LS, we will release these datasets once the paper is accepted to be published.

    \subsection{Datasets for Evaluation on Industrial Circuits and Very Large-Scale Circuits}\label{appendix:datasets_industrial_vlsi}
        In terms of the industrial circuits, we report a statistical description of the training and testing circuits in Table \ref{dataset_description_on_real_circuits}. As shown in Tables \ref{dataset_description_on_real_circuits}, \ref{dataset_description_on_EPFL}, and \ref{dataset_description_on_IWLS}, real industrial circuits are much larger in size than open-source circuits. 
        
        In terms of the very large-scale circuits, we report a detailed description of them in Table \ref{dataset_description_on_vlsc_circuits}. 
        Due to the small number of very large-scale circuits, we evaluate our method using Evaluation Strategy 1 mentioned in Section \ref{results:open_source}. Specifically, we use circuits from the EPFL mentioned in Table \ref{dataset_description_on_EPFL} and the two very large-scale circuits to construct datasets. (1) We set the Sixteen circuit as the testing dataset, and the rest as the training dataset. (2) We set the Twenty circuit as the testing dataset, and the rest as the training dataset. To promote the community of machine learning for LS, we will release the datasets once the paper is accepted to be published.
            
\section{Details of Methods and Experimental Settings}
    \subsection{Details of Experimental Setup}\label{appendix:exp_setup}
    \subsubsection{Optimization Sequence Flows for Collecting Data and Evaluation}\label{appendix:opt_flow} 
    In the industrial setting, we usually apply a sequence of Logic Synthesis (LS) operators to optimizing an input circuit. Thus, we follow the setting throughout all experiments unless mentioned otherwise. Specifically, in terms of the Mfs2 operator, we apply the sequence of operators, i.e., \textit{strash; dch; if -C 12; mfs2 -W 4 -M 5000}, to collect data and evaluate the performance of the Default Mfs2 operator and our PruneX. Note that the optimization sequence flow is a standard academic flow for evaluating the Default Mfs2 operator, which follows previous work \cite{mfs2}. Moreover, in terms of the Resub operator, we apply a common commercial optimization sequence flow widely used in the industrial setting, i.e., \textit{strash; resyn2; resub -K 16 -N 3 -z}, to collect data and evaluate the performance of the Default Resub operator and our PruneX. 
    
    \subsubsection{Optimization Sequence Flows for Evaluating 2PruneX-COG}\label{appendix:opt_flow_2prunex}    
    To apply our PruneX-COG twice, we apply the sequence of operators, i.e., \textit{strash; dch; if -C 12; mfs2 -W 4 -M 5000; strash; if -C 12; mfs2 -W 4 -M 5000}, to evaluate the performance of 2PruneX-COG. Note that the mfs2 operator is a post-mapping optimization operator, whose input DAG is a k-input look-up table graph (K-LUTs).  Moreover, the strash operator transforms the current circuit representation into an And-Inverter Graph (AIG) by one-level structural hashing. Then, the if \cite{technology_mapping} operator maps an AIG into a K-LUTs. Finally, the Mfs2 operator optimizes the input K-LUTs. 

    \subsubsection{Top k Accuracy Metric}\label{appendix_top_k_acc}
    In our implemented PruneX, we sort all the nodes according to the prediction scores given by our learned classifier, 
    and select the top $k$ nodes---that is, the nodes with top $k$ scores. That is, the top $k$ nodes are predicted positive, and the other nodes are predicted negative. Then, the top $k$ accuracy metric is defined by the recall, i.e., the fraction of true positive nodes that are predicted to be
    positive. 
    
    \subsection{Implementation Details of the Focal Loss}\label{appendix:focal_loss}
    The Focal Loss is first introduced to address the class imbalance problem in the object detection scenario \cite{focal_loss}. In this paper, we leverage the Focal Loss to alleviate the class imbalance problem in learning classifiers for logic synthesis. Specifically, the loss function for each data pair $(\textbf{x},y)$ takes the form of 
    \begin{align*}
        l(f(\textbf{x}),y) = & -\alpha y (1-f(\textbf{x}))^{\gamma}\log(f(\textbf{x})) \nonumber \\ 
        & - (1-\alpha)f(\textbf{x})^{\gamma}\log(1-f(\textbf{x})),
    \end{align*}
    where $\alpha\in\left[0,1\right]$ denotes a weighting factor, and $\gamma\geq 0$ denotes a tunable focusing parameter. In this paper, we set $\alpha$ by inverse class frequency, and $\gamma$ by $2$, which follows \cite{focal_loss}. 
    
    \subsection{Implementation Details of EnsembleMLP}\label{appendix:ensemblemlp}
    EnsembleMLP aims to learn a classifier to predict the probability (i.e., score) of a node that it is effective. 
    For a fair comparison, EnsembleMLP uses the same node features as COG in Table \ref{node_features}, and we train EnsembleMLP via the Focal Loss as well. Specifically, we implement the EnsembleMLP with a multi-layer perceptron containing three hidden layers with 1024 units. To enhance the generalization ability, we train the EnsembleMLP with the ensemble learning trick \cite{zhou2021ensemble}. Specifically, we train the EnsembleMLP with 15 ensembles. 

    \subsection{Discussion on Assumption \ref{assumption1}}\label{appendix:discuss_assum1}
    In our setting, for each of $N$ circuits, a set of nodes are obtained by applying an X operator to the circuit. The nodes are then labeled based on whether the corresponding node-level transformation is effective. Thus, the training set consists of heterogeneous samples from several distributions, i.e., nodes from several different circuits. That is, we are given $N$ similar and related distributions, and aim to generalize to a similar but different test distribution.
    Regarding the source of these circuits, each circuit may be a sample of a complex integrated circuit, such as an addition circuit, storage circuit, control circuit, etc. In addition, aggregating circuits with similar functionality is similar to sampling some small building blocks from a complex integrated circuit. Therefore, it is reasonable to assume that the training Circuit Domains and testing Circuit Domains come from the same hyper-distribution. That is, Assumption \ref{assumption1} holds. 
    
    Moreover, we empirically show that the data distributions from $N$ different circuits are $N$ similar but different distributions in Fig. \ref{fig:more_visualization}, which suggests that Assumption \ref{assumption1} holds in practice as well.     

    \subsection{Details on Our Circuit Aggregation Mechanism}\label{appendix:domain_construction_mapping}
    In this paper, we aggregate circuits based on their sample sizes and functionality. In general, we divide the real industry setting into two categories. In the first category, we are given $N$ circuits with their functionality information. In the second category, we are given $N$ circuits without their functionality information for privacy protection. In the first category, we aggregate those circuits with the same high-level functionality. Take the EPFL benchmark as an example, we aggregate circuits with the arithmetic functionality and control functionality into an arithmetic Circuit Domain and a control Circuit Domain, respectively. Thus, the two Circuit Domains contain two large datasets with similar sample sizes, which is beneficial for achieving a small generalization error bound. Moreover, the two Circuit Domains may come from small integrated units from a complex integrated circuit, which follows Assumption \ref{assumption1}. In the second category, we aggregate circuits to make the sample sizes across domains nearly uniform.
    Specifically, we sort the circuits according to their sample sizes, and divide the odd-numbered circuits into one domain, and the even-numbered circuits into another domain. In this paper, we assume that the IWLS and industrial benchmarks fall into the second category. 

    \subsection{Details of the Transformation-Invariant Domain Knowledge}\label{appendix:discuss_transformation_invariant}
    Here we take the node-level transformation in the Resub and Mfs2 operators, i.e., the resubstitution transformation \cite{resub}, as an example. The node-level transformations in the Rewrite \cite{rewrite} and Refactor \cite{refactor} operators are different from but similar to the resubstitution transformation. Please refer to \cite{rewrite} for details. In terms of the resubstitution transformation, we are given a current node called the root node. The resubstitution transformation first uses a heuristic rule to collect a limited number of \textit{candidate input nodes} of the root node. In general, the candidate input nodes include two types of nodes: (a) the nodes that are on the input direction of the root node and distance-m or less from the root node, and (b) the nodes that are on the output direction of the type (a) nodes and with level not exceeding the level of the root node. The resubstitution transformation focuses on the subgraph constructed by the root node and candidate input nodes. Next, the resubstitution transformation aims to find a new set of input nodes from \textit{candidate input nodes} to replace existing input nodes of the current node \textit{without changing its functionality}, thereby reducing the size and/or depth of the circuit. 
    
    Thus, the resubstitution transformation mechanism is invariant across different circuits.
    Moreover, whether a resubstitution transformation can be effective is highly related to the focused subgraph regardless of what the global graph is. Based on the transformation-invariant domain knowledge, we propose to extract the subgraph rooted at a node to learn its node embedding for discriminative classification, which carries the potential to learn domain-invariant representations and thus well generalize to unseen circuits. Please see \cite{resub} for details on the resubstitution. 
    
    \subsection{Details of PruneX and COG}

            
        \subsubsection{Designed Node Features}\label{appendix:node_features}
            As shown in Table \ref{node_features}, we design features for each node to contain the functionality information, structure information, and other possible useful features. 
            As the Resub operator and the Mfs2 operator are logic optimization and post-mapping optimization operators respectively, their corresponding input directed acyclic graphs are different. For the logic optimization operator Resub, the input circuit is usually represented as an And-Inverter Graph (AIG). For the post-mapping optimization operator Mfs2, the input circuit is represented as a k-inputs Look up tables (k-LUTs). Please refer to Section \ref{sec:background} for details on the AIG and k-LUTs. 
            
            Therefore, for the two different types of operators, the specific designed node features are slightly different.
            Specifically, for the Mfs2 operator, we design node features with the truth table of the node (i.e., a 6-input look up table) and other possible useful features, such as the fanin/fanout number and level of the node. Overall, the node feature for the Mfs2 operator is a 69-dimensional vector.
            Moreover, for the Resub operator, we consider features from the node itself and from its two children to capture structural information, which follows \cite{neto2021slap}. We use the two features inv0 and inv1 to capture the functionality information, where inv0/inv1$=1$ denotes that the left/right input edge is inverted. Thus, the node feature consists of the root node feature (8-dimensional vector) and its two children features (two 6-dimensional vectors). Note that its children features do not contain the cut information, i.e., the Leaves number and Cuts number. Overall,  the node feature for the Resub operator is a 20-dimensional vector.
        

\begin{table}[t]
\caption{We provide our manually designed node features.}
\label{node_features}
\centering
\resizebox{0.99\textwidth}{!}{
\begin{tabular}{@{}cccc@{}}
\toprule
\toprule
Operator & \multicolumn{3}{c}{Node features} \\ \midrule
Mfs2 & Functionality information & Structure information & Other features \\ \midrule
Input: 6-LUTs & \multirow{3}{*}{\begin{tabular}[c]{@{}c@{}}Truth table of the   node\\      the node is 6-input LUT\\      (64-dimensional vector)\end{tabular}} & - & Fanin number \\
 &  &  & Fanout number \\
 &  &  & Level \\
 &  &  & LevelR \\
 &  &  & Node ID \\ \midrule
Resub & Functionality information & Structure information & Other features \\ \midrule
Input: AIG & inv0 & Child1 features (6-dimensional vector) & Fanout number \\
 & inv1 & Child2 features (6-dimensional vector) & Level \\
 &  &  & Total Level \\
 &  &  & Node ID \\
 &  &  & Leaves number \\ 
 &  &  & Cuts number \\ \bottomrule
\end{tabular}
}
\end{table}
    
        \subsubsection{Implementation Details of PruneX}\label{appendix:details_prunex}
            As shown in Fig. \ref{fig:paradigm_operators}, our proposed PruneX first extract node features, i.e., constructing bipartite graphs for all nodes as mentioned in Section \ref{method:kdlgr}. PruneX then call the learned classifier to predict the score that a node is effective for all nodes. The learned classifier is operator-specific. That is, we learn an individual classifier for each given operator, as the input circuit representation for different operators are different and thus the designed node features are different. 
            Note that an input circuit is modeled by a directed acylic graph (DAG), and each node on the DAG has a unique identification number (ID). Thus, PruneX sort the node IDs in descending order according to the scores, and select the top $k\%$ node IDs.
            Finally, PruneX only applies node-level transformations in sequence to nodes in the selected node ID set. 
            
            The model inference time in PruneX is low, as PruneX only needs to call the inference model once. 
            However, there may be a marginal shift between the predicted node IDs and the true node IDs. 
            The reason is that when applying node-level transformations in sequence, the structure of the graph may change. This causes the true node IDs to change, while the predicted node IDs are fixed. Fortunately, we found that the shift between the predicted node IDs and the true node IDs has marginal impact on the optimization performance of the operators for the following two reasons. First, the ID set of effective nodes is quite sparse due to the ineffective node-level transformations problem. Therefore, the influence between 
            consecutive effective node-level transformations is slight. Second, the results in Table \ref{oracle_prediction_results} show that only applying transformations to true effective nodes achieves the same optimization performance as the default Mfs2 operator (see Appendix \ref{exps:oracle_results} for details).

        \subsubsection{Discussion on How to Apply PruneX to the Rewrite and Refactor Operators}\label{appendix:discuss_apply_to_rewrite}
        The Rewrite and Refactor operators follow the same paradigm as the Resub and Mfs2 operators. We illustrate the unified paradigm in Fig. \ref{fig:paradigm_operators} (Left) in the main text. Note that the major difference between these operators lies in the node-level transformation. That is, the specific methods for node-level transformation are different between these operators. For example, the Resub operator applies the resubstitution transformation. We provide details on the resubstitution transformation in Appendix \ref{appendix:discuss_transformation_invariant}. As shown in Fig. \ref{fig:paradigm_operators} (Right), our PruneX uses a learned classifier to predict those nodes with effective node-level transformations in advance, and applies node-level transformations to these predicted effective nodes sequentially. Note that the paradigm of our PruneX is independent from the node-level transformations. Thus, our PruneX can also be applied to the Rewrite and Refactor operators.
        
            
        \subsubsection{Implementation Details of COG}\label{appendix:details_cog}
            \noindent\textbf{Hardware Specification} Throughout all experiments, we use a single machine that contains eight GPU devices (Nvidia GeForce GTX 3090 Ti) and two Intel Gold 6246R CPUs.

            \noindent\textbf{Neural Network Architecture} The GCNN model encodes node features into embeddings with dimension 128. The multi-head neural network contains a shared multi-layer perceptron containing three hidden layers with 1024 units with $M$ heads branching off independently. To output classification probabilities for values between 0 and 1, the multi-head neural network uses a Sigmoid activation function at the last layer.

            \noindent\textbf{Training Details} Throughout all experiments, we apply Adam optimizer with learning rate $\alpha = 1e-4$ to optimize our models. Moreover, for training stability, we apply linear learning rate decay with step size 100 and decay rate 0.96. For generalization to very large-scale circuits, we apply Min-Max Normalization to data in training and testing datasets. Due to limited video memory, we set the batch size to 10240. For simplicity, we train models for 3000 epochs, and save the three models at 1000, 2000, and 3000 epochs. Furthermore, we choose the model that performs best on the training set from the three models to evaluate on the testing circuits. To further enhance our method, we can leverage the model selection strategies investigated in \cite{gulrajanisearch}, which we leave as future work.  

\end{document}